\documentclass[10pt]{article} 
\usepackage{epsfig,amssymb}
\usepackage{amsmath}
\usepackage{natbib}
\usepackage{graphics}
\usepackage{graphicx}
\usepackage{dcolumn}
\usepackage{bm}
\usepackage{epsfig}
\usepackage{enumerate}
\usepackage{paralist}

\begin{document}

\begin{center}

\Large{\bf The Study of the Pioneer Anomaly: \\
New Data and Objectives for New Investigation}

\vspace{20pt}

\normalsize

Slava G. Turyshev,$^a$
Viktor T. Toth,$^b$
Larry R. Kellogg,$^c$
Eunice. L. Lau,$^a$ 
and Kyong J. Lee$^a$ 

\normalsize
\vskip 15pt

{\it $^a$Jet Propulsion Laboratory, 
California Institute of Technology, \\
4800 Oak Grove Drive, Pasadena, CA 91109, USA \\
$^b$vttoth.com,  3-575 Old St Patrick St.,  Ottawa ON K1N 9H5, Canada \\
$^c$NASA Ames Research Center, Moffett Field, CA 94035, USA\footnote{Contractor to NASA Ames Research Center employed by Bendix Field Engineering Corporation early on and Orbital Sciences Corporation at the end of the Pioneer missions; presently retired.
}}

\vspace{0.25in}

\end{center}

\begin{abstract}
The Pioneer 10/11 spacecraft yielded the most precise navigation in deep space to date.  However, their radiometric tracking data has consistently indicated the presence of a small, anomalous, Doppler frequency drift. The drift is a blue shift, uniformly changing with a rate of $\sim6\times 10^{-9}$~Hz/s and can be interpreted as a constant sunward acceleration of each particular spacecraft of $a_P  = (8.74 \pm 1.33)\times 10^{-10}$~m/s$^2$ (or, alternatively, a time acceleration of $a_t = (2.92 \pm 0.44)\times 10^{-18}$~s/s$^2$). This signal has become known as the Pioneer anomaly; the nature of this anomaly remains unexplained. We discuss the current state of the efforts to retrieve the entire data sets of the Pioneer 10 and 11 radiometric Doppler data. We also report on the availability of recently recovered telemetry files that may be used to reconstruct the engineering history of both spacecraft using original project documentation and newly developed software tools. We discuss possible ways to further investigate the discovered effect using these telemetry files in conjunction with the  analysis of the much extended Pioneer Doppler data. 
\par 
In preparation for this new upcoming investigation, we summarize the current knowledge of the Pioneer anomaly and review some of the mechanisms proposed for its explanation. We emphasize the main objectives of this new study, namely 
\begin{inparaenum}[i)]
\item analysis of the early data that could yield the true direction of the anomaly and thus, its origin,
\item analysis of planetary encounters, that should tell more about the onset of the anomaly (e.g. Pioneer 11's Saturn flyby),
\item analysis of the entire dataset, that should lead to a better determination of the temporal behavior of the anomaly,
\item comparative analysis of individual anomalous accelerations for the two Pioneers with the data taken from similar heliocentric distances,
\item the detailed study of on-board systematics, and
\item development of a thermal-electric-dynamical model using on-board telemetry.
\end{inparaenum}
The outlined strategy may allow for a higher accuracy solution for the anomalous acceleration of the Pioneer spacecraft and, possibly, will lead to an unambiguous determination of the origin of the Pioneer anomaly.

\end{abstract}


{\small
\tableofcontents
} 
\section{Introduction}
\label{sec:intro}

The Pioneer 10/11 missions were the first spacecraft to explore the outer solar system \citep{Fimmel_etal_74,Hall_74,Fimmel_etal_80,Dyal_Fimmel_82,pioprd}. Their objectives were to conduct, during the 1972-73 Jovian opportunities, exploratory investigation beyond the orbit of Mars of the interplanetary medium, the nature of the asteroid belt, and the environmental and atmospheric characteristics of Jupiter and Saturn (for Pioneer 11). 

Pioneer 10 was launched on 2 March 1972 on top of an Atlas/Centaur/TE364-4 launch vehicle. The launch marked the first use of the Atlas-Centaur as a three-stage launch vehicle. The third stage was required to rocket Pioneer 10 to the speed of 14.39 km/s needed for the flight to Jupiter.

\begin{figure*}[t!]
\centering  
\epsfig{figure=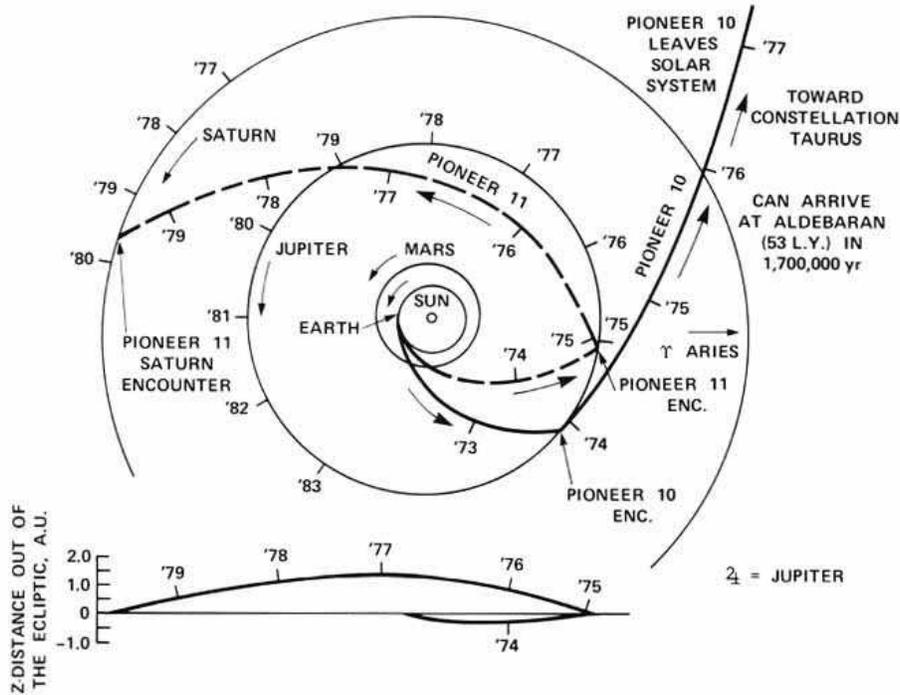,width=122mm}
\caption
{\small
{Trajectories of Pioneer 10 and 11 during their primary missions in the solar system (taken from {\tt http://history.nasa.gov/SP-349/}). The time ticks shown along the trajectories and planetary orbits represent the distance traveled during each year.}} 
     \label{fig:pioneer_inner_path}
\end{figure*}

Its sister craft, Pioneer 11, was launched on 5 April 1973, like Pioneer 10, also on top of an Atlas/Centaur/TE364-4 launch vehicle. After safe passage through the asteroid belt on 19 April 1974, Pioneer 11's thrusters were fired to add another 63.7 m/s to the spacecraft's velocity. This adjusted the aiming point at Jupiter to 43,000 km  above the cloudtops. The close approach also allowed the spacecraft to be accelerated by Jupiter to a velocity of 48.06 km/s -- so that it would be carried across the solar system some 2.4 billion km to Saturn. 

\begin{figure*}[t!]
\centering  
\epsfig{figure=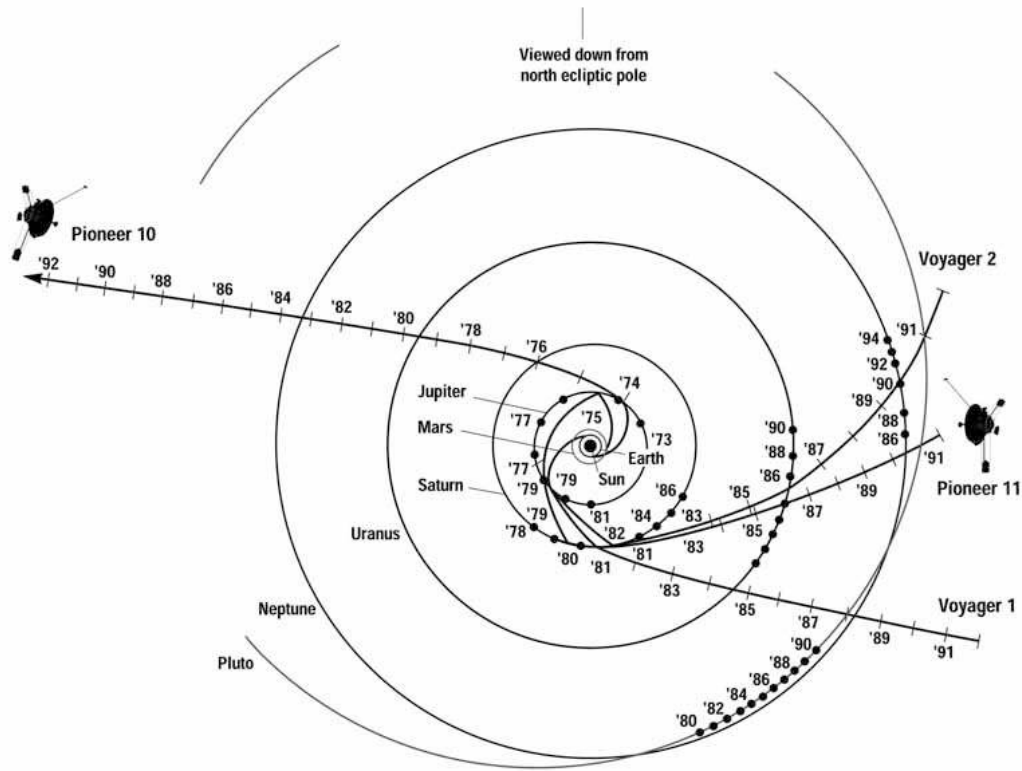,width=140mm}
\caption
{\small
{Ecliptic pole view of the Pioneer 10, Pioneer 11, and Voyager trajectories. Pioneer 10 is traveling in a direction almost opposite to the galactic center, while Pioneer 11 is heading approximately in the shortest direction to the heliopause. The direction of the solar system's motion in the galaxy is approximately towards the top.}} 
     \label{fig:pioneer_path}
\end{figure*}

After the Jupiter and Saturn (for Pioneer 11) encounters (see Figure \ref{fig:pioneer_inner_path}), the craft followed escape hyperbolic orbits near the plane of the ecliptic on opposite sides of the solar system, continuing their extended missions \citep{SP-349,extended}. (See Figure \ref{fig:pioneer_path}.) Pioneer 10 explored the outer regions of the solar system, studying energetic particles from the Sun (solar wind), and cosmic rays entering our portion of the Milky Way.

Pioneer 10 continued to make valuable scientific investigations 
until its science mission ended on March 31, 1997. Since that time, Pioneer 10's weak signal has been tracked by the NASA's Deep Space Network (DSN) as part of an advanced concept study of communication technology in support of NASA's future interstellar probe mission. Pioneer 10 eventually became the first man-made object to leave the solar system. The power source on Pioneer 10 finally degraded to the point where the signal to Earth dropped below the threshold for detection in its latest contact attempt on 7 February, 2003. The previous three contacts had very faint signals with no telemetry received. The last telemetry data point was obtained from Pioneer 10 on 27 April 2002 when the craft was 80~AU from the Sun.

Following its encounter with Saturn, Pioneer 11 explored the outer regions of our solar system, studying the solar wind and cosmic rays. Pioneer 11 sent its last coherent Doppler data on 1 October 1990 while at $\sim30$ AU from the Sun.  In September 1995, Pioneer 11 was at a distance of 6.5 billion km from Earth. At that distance, it takes over 6 hours for the radio signal to reach Earth. However, by September 1995, Pioneer 11 could no longer make any scientific observations. On 30 September 1995, routine daily mission operations were stopped. Intermittent contact continued until November 1995, at which time the last communication with Pioneer 11 took place. There has been no communication with Pioneer 11 since. The Earth's motion has carried our planet out of the view of the spacecraft antenna. The spacecraft cannot be maneuvered to point back at the Earth. It is not known whether the spacecraft is still transmitting a signal. No further tracks of Pioneer 11 are scheduled.

\begin{figure*}[t!]
 \begin{center}
\noindent    
\psfig{figure=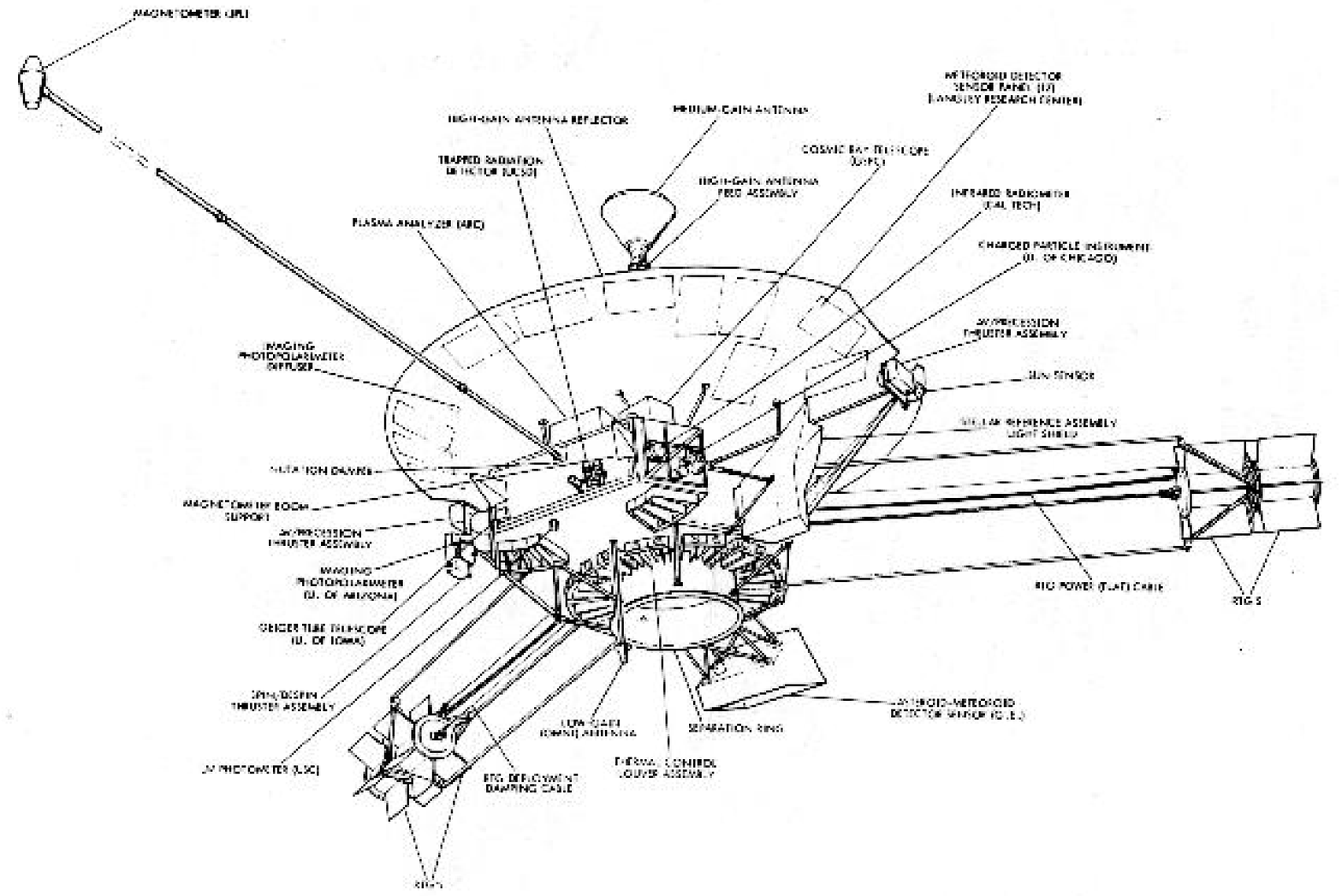,width=160mm}
\end{center}
\vskip -10pt
  \caption{A drawing of the Pioneer spacecraft.  
 \label{fig:pio-craft}}
\end{figure*} 

The Pioneers were excellent craft for the purposes of precision celestial mechanics experiments \citep{pioprl,moriond,pioprd,pio-origin,dust,stanford,problem_set_05,iap-pioneer}.   This was due to a combination of many factors, including their attitude control (spin-stabilized, with a minimum number of attitude correction maneuvers using thrusters), power design (the plutonium-238 powered heat-source RTGs -- Radioisotope Thermoelectric Generators -- being on extended booms aided the stability of craft and also reduced thermal effects on the craft), and precise Doppler tracking (with the accuracy of post-fit Doppler residuals at the level of mHz).  The result was the most precise navigation in deep space to date \citep{null76}. (See Figure~\ref{fig:pio-craft} for a design drawing of the spacecraft.)

Because of the excellent health and navigational capabilities of Pioneer 10, NASA supported a proposal to initiate search for unmodeled accelerations in 1979 (when the spacecraft was at a distance of some 20 AU from the Sun).  The main motivation was the search for Planet X. Eventually both Pioneers were used in the search for trans-Neptunian objects; their Doppler tracking capabilities also yielded the first ever limits on low frequency gravitational radiation \citep{pioprd}.

The Pioneer 10/11 spacecraft had exceptional ``built-in'' acceleration sensitivity naturally allowing them to reach the level of $\sim 10^{-10}$ m/s$^2$.  As indicated by their radiometric data received from heliocentric distances of 20-70 AU, the accuracies of their orbit reconstructions were limited by a small, anomalous, Doppler frequency drift \citep{pioprl,moriond,pioprd}.
By 1980, when Pioneer 10 already had passed a distance of  $\sim$ 20 AU from the Sun and the acceleration contribution from solar radiation pressure on the craft (away from the Sun) had decreased to less than $4\times 10^{-10}$ m/s$^2$, the radiometric data started to show presence of the anomalous acceleration (towards the Sun).
  This blue-shifted drift is uniformly changing with a rate of $(5.99 \pm 0.01)\times 10^{-9}$ Hz/s, which can be interpreted as a sunward constant acceleration of $a_P  = (8.74 \pm 1.33)\times   10^{-10}$ m/s$^2$ \citep{pioprd}. The detailed study of this anomaly  \citep{pioprl,pioprd} led to a better understanding of its properties (summarized in Section~\ref{sec:anomaly}); however, the nature of this anomalous signal is still unknown.  

In this paper we are going to discuss recently recovered set of Pioneer data and objectives for upcoming new investigation of the anomalous  residuals detected in the radiometric Doppler data of the Pioneers 10 and 11. We continue the paper in Section~\ref{sec:anomaly} by discussing the Pioneer anomaly and its known properties. We briefly review the original efforts to understand the signal as well as some recent proposals to explain the Pioneer anomaly. 

Section~\ref{sec:pio-doppler} presents the current status of the effort to retrieve the entire set of the Pioneer radiometric Doppler data. 
Section~\ref{sec:mdr} reports on the newly recovered Pioneer telemetry data in the form of Master Data Records (MDRs). Because of their potentially high value in the investigation of the Pioneer anomaly, we discuss the purpose, means of delivery, storage, processing and possible use of the MDRs.
In Section~\ref{sec:objectives} we present objectives for the upcoming study of the Pioneer anomaly and particularly the use of both the extended set of Doppler data and the MDRs to study the on-board systematics, including the development of an accurate thermal-electric-dynamical model.
In Section~\ref{sec:conclude} we conclude by presenting the next steps for the analysis of the Pioneer anomaly. 

\section{The Pioneer Anomaly and the Search for its Origin}
\label{sec:anomaly}

\subsection{The Pioneer Anomaly}
\label{sec:anomaly_sum}

The analysis of the Pioneer 10 and 11 data (with some support from the Ulysses and Galileo spacecraft) \citep{pioprl,pioprd} demonstrated the presence of an anomalous, Doppler frequency blue shift drift, uniformly changing with a rate of 
$\dot{f}_P \sim 6\times 10^{-9}$~Hz/s \citep{stanford}.
To understand the phenomenology of the effect, consider $f_{\tt obs}$, the frequency of the retransmitted signal observed by a DSN antenna, and $f_{\tt model}$,  the predicted frequency  of that signal modeling conventional forces influencing the spacecraft's motion including both gravitational and non-gravitational in their origin. The observed, two-way (round trip) anomalous effect can be expressed to first order in $v/c$ as 
$\left[f_{\tt obs}(t)- f_{\tt model}(t)\right] =  -2\dot{f}_P\,t.$   

After accounting for the gravitational and other large forces included in standard orbit determination programs this translates to 
\begin{eqnarray}
\left[
f_{\tt obs}(t)- f_{\tt model}(t)
\right]
_{\tt DSN}
= -f_{0}\frac{2a_P~t}{c}. 
\label{eq:delta_nu_syst}
\end{eqnarray}
Here $f_{0}$ is the DSN reference frequency \citep{pioprd,stanford,iap-pioneer} (see discussion of the DSN sign convention in Ref.~[38] of \citep{pioprd}).
 
After accounting for all {\it known} (not modeled) sources of systematic error (discussed in \cite{pioprd,pio-standard}), the conclusion was reached that there exists an anomalous sunward constant acceleration signal of 
\begin{equation}
a_P=(8.74\pm1.33)\times 10^{-10}~~{\rm m/s}^2.
\label{eq:aP} 
\end{equation}
The nature of this anomaly remains unexplained. This signal has become known as the Pioneer anomaly.

Note that there exist a dualism in interpreting the radiometric Doppler data. The anomaly can be due to a force acting on the craft that produces acceleration Eq.~(\ref{eq:aP}), or, alternatively, this signal can also be interpreted as a time deceleration uniformly changing with rate $a_t = (2.92 \pm 0.44)\times 10^{-18}$ s/s$^2$ (see discussion of this possibility in \citep{pioprd,pio-cospar_04,stanford,iap-pioneer}).

By now several studies of the Pioneer Doppler navigational data have demonstrated that the anomaly is unambiguously present for both Pioneer 10 and 11. These studies were performed with four independent (and different!) navigational computer programs 
\citep{pioprl,pioprd,markwardt,Oystein}, namely: 
{}
\begin{itemize}
\item the JPL's Orbit Determination Program (ODP) developed in 1980-2005,
\item The Aerospace Corporation's CHASMP code extended for deep space navigation \citep{pioprl,pioprd}, 
\item code written at the Goddard Space Flight Center \citep{markwardt} that was used to analyze Pioneer 10 data for the period 1987-1994 obtained from the National Space Science Data Center ({\tt http://nssdc.gsfc.nasa.gov/}), and finally 
\item code developed at the Institute of Theoretical Astrophysics, University of Oslo, Norway, that was recently successfully used to analyze the set of the Pioneer 10 data above \citep{Oystein}. 
\end{itemize}

For the most detailed analysis of the Pioneer anomaly to date, \cite{pioprd} used the following Pioneer 10/11 Doppler data: 
\begin{itemize}
\item Pioneer 10: The data set had 20,055 data points obtained between 3 January 1987 and 22 July 1998 and covering heliocentric distances  $\sim40-70.5$~AU.   
\item Pioneer 11: The data set had 19,616 data points obtained between 5 January 1987 to 1 October 1990 and covering heliocentric distances $\sim22.42-31.7$~AU. 
\end{itemize}

The recent analyses of the Pioneer 10 and 11 radiometric data \citep{pioprl,pioprd,markwardt,stanford} have established the following basic properties of the Pioneer anomaly:
{}
\begin{itemize}
\item {\it Distance:} It is unclear how far out the anomaly goes, but the Pioneer 10 data supports its presence at distances up to $\sim$70~AU from the Sun. The Pioneer 11 data shows the presence of the anomaly as close in as $\sim$20 AU. 
\item {\it Direction:} From typical angular uncertainty of Doppler navigation at S-band and spacecraft high gain antennae pointing accuracy set within 3~dB gain bandwidth, $a_P$ behaves as a line-of-sight constant acceleration of the craft generally pointing in the innermost region of the solar system.
\item {\it Constancy:}~Both temporal and spatial variations of the anomaly's magnitude are of order 10\% for each craft, while formal errors are significantly smaller.
\end{itemize}

There are other pieces of information obtained from spot analyses \citep{jda-memo,pioprd,iap-pioneer}; they indicate that:  
\begin{itemize}
\item The anomalous acceleration was present in the Pioneer 11 data at shorter distances, as close in as $\sim 10$ AU (see Figures~6 and 7 in \cite{pioprd}). 
\item The Pioneer 11 data also indicated that the anomaly may be much smaller at distances $<10$~AU. It appears to be amplified (or turned on) at a distance of 
$\sim 10$ AU from the Sun.  This is approximately when the craft flew by Saturn and entered an hyperbolic, escape trajectory.  
\end{itemize}

This information was used as guidance in investigating the applicability of proposals to explain the Pioneer anomaly using both conventional and ``new'' physical mechanisms. In the next section we briefly review these proposals.  

\subsection{Original Efforts to Explain the Anomaly}
\label{sec:origexplain}

Initial announcement of the anomalous acceleration (e.g. \citep{pioprl,moriond}) triggered many proposals that invoked various conventional physics mechanisms, all aimed at explaining the origin of the anomaly.  Finding a systematic origin of the proper magnitude and behavior was the main focus of these proposals, which is yet to be found.  Although the most obvious explanation would be that there is a systematic origin to the effect, perhaps generated by the spacecraft themselves from anisotropic heat rejection or propulsive gas leaks, the analysis performed did not find evidence for either of them; that is, no unambiguous, on-board systematic has been discovered.
This initial search was summarized in \citep{pioprd,pio-standard}, where possible contributions of various mechanisms to the final solution for $a_P$ were given. The entire error budget was subdivided in three main types of effects, namely
\begin{inparaenum}[i)]
\item effects due to sources external to the spacecraft;
\item the contribution of on-board systematics, and
\item computational systematic errors.
\end{inparaenum}

In this section, we present a summary of this earlier work.

\subsubsection{Effects with Sources External to the Spacecraft}

The first group of effects includes those external to the spacecraft, such as the solar radiation pressure, effects of the solar wind, and the effect of the solar corona on the propagation of radio wave signals.  \citep{pioprl,pioprd} also discuss the influences of the Kuiper belt's gravity, galactic gravity, and electromagnetic Lorentz forces.  Errors in the accepted values of the Earth's orientation parameters, precession, and nutation are included in the discussion.  The analysis evaluated the contributions of the mechanical instabilities and the location errors of the DSN antenna structures, the phase stabilities of the DSN antennae and clocks, and effects due to the troposphere and ionosphere. Even though some of these mechanisms are near the limit for contributing to the final error budget, it was found that none of them could explain the behavior of detected signal.  Moreover, some were three orders of magnitude or more too small.  In totality, they were insignificant.

\subsubsection{Study of the On-Board Systematics}
\label{sec:on-board-sytematics}

The second group of effects includes those that originated on-board and are tied to well-known sources; this group, as expected, had the largest impact on the final error.  Among these effects, the radio beam reaction force produced the largest bias to the result, $1.10 \times 10^{-10}$~m/s$^2$.  As the force exerted by the radio beam necessarily points away from the Earth (and thus from the Sun), the correction of the measured data increases the amount of the observed anomalous attractive force and makes the Pioneer effect larger.  Large uncertainties also came from a conjectured differential emissivity of the Radioisotope Thermoelectric Generators (RTGs), radiative cooling of the spacecraft, and propulsive gas leaks from thrusters of the attitude control system: $\pm0.85\times 10^{-10}$~m/s$^2$, $\pm0.48 \times 10^{-10}$~m/s$^2$, and $\pm0.56 \times 10^{-10}$~m/s$^2$, respectively.  The effect due to expelled helium produced within the RTGs was also considered, as well as the small difference in anomaly determinations between the two Pioneers.

The effect of rejected thermal radiation was the second largest bias/uncertainty that has been the most critical systematic bias to quantify.  If heat generated by the on-board power sources was asymmetrically reflected by the body of the craft, a fore/aft acceleration could be produced causing the measured anomaly. The Pioneer spacecraft were powered by SNAP-19 RTGs (Space Nuclear Ancillary Power) mounted on long extended booms (designed to protect the on-board electronics from heat and radiation impact) \citep{pioprl,pioprd,pio-standard}.  In principle, there was more than enough heat available on the craft to cause the anomaly.  However, the spacecraft's spin-stabilized attitude control, special design of the RTGs and the length of the RTG booms that resulted in a relatively small spacecraft surface available for the preferential heat rejection significantly minimized the amount of heat for the mechanism to work.  The analysis of the 11.5 years of Pioneer Doppler data \citep{pioprd} can only support an effect as large as $(-0.55\pm0.55)\times 10^{-10}$~m/s$^2$.  

In summary, although this group represents conceptually the most likely sources for the anomaly, these mechanisms did not gain enough experimental support.  At most one can obtain $\sim$12\% of the discovered effect by employing all of these mechanisms. The possibility exists that a combination of the factors above could amount to the measured effect \citep{scheffer}, but again, the shortness of the data interval used in the analysis, actual spacecraft design and performance data at hand, and also the complexities of modeling thermal radiative processes on the craft made it difficult to unambiguously support this claim \citep{pio-standard}.  

\subsubsection{Computational Systematics }
The third group of effects was composed of contributions from computational errors.  The effects in this group dealt with the numerical stability of least-squares estimations, accuracy of consistency/model tests, mismodeling of maneuvers, and the solar corona model used to describe the propagation of radio waves.  It has also been demonstrated that the influence of annual/diurnal terms seen in the data on the accuracy of the estimates was all small \citep{pioprd}. The total uncorrelated error associated with computational systematics is estimated to be less than $\pm0.35 \times 10^{-10}$~m/s$^2$.

These three groups of effects exhausted all available conventional explanations for the anomaly.  The inability to explain the Pioneer anomaly with conventional physics has led to a significant number of theoretical proposals that use more unusual mechanisms (more details are in \citep{pioprd}).  The Pioneer anomaly is an effect at the limit of what is detectable with radiometric tracking of a deep space probe, but it is huge in physical terms: The anomaly exceeds the corrections to Newtonian motion predicted by general relativity by five orders of magnitude (at 50 AU).  Hence, if the effect is not a result of conventional systematics it would have a considerable impact on our models of fundamental forces, regardless of the anomaly being due to a deceleration of the spacecraft or a blue shift of the radio signal.

\subsection{Recent Efforts to Explain the Anomaly}
\label{sec:recentexplain}

\subsubsection{Search for Independent Confirmation}
\label{sec:experiemnts}

Attempts to verify the anomaly using other spacecraft proved disappointing. This is because the Voyager, Galileo, Ulysses, and Cassini  spacecraft navigation data all have their own individual difficulties for use in an independent test of the anomaly \citep{pio-origin}. In addition, many of the deep space missions that are currently being considered either may not provide the needed navigational accuracy and trajectory stability sensitive to accelerations of under $10^{-10}$~m/s$^2$ or else they have significant on-board systematics that mask the anomaly. A requirement to have an escape hyperbolic trajectory makes a number of other missions \citep{stanford,ESLAB2005_Pioneer,pa-lisa} less able to directly test $a_P$.  Although these missions all have excellent scientific goals and technologies, nevertheless, their orbits lend them a less advantageous position to conduct a precise test of the detected anomaly. 

A number of alternative ground-based verifications of the anomaly have also been considered; for example, using Very Long Baseline Interferometry (VLBI) astrometric observations.  However, the trajectories of spacecraft like the Pioneers, with small proper motions in the sky, make it presently impossible to use VLBI in accurately isolating an anomalous sunward acceleration of the size of $a_P$.

\subsubsection{Conventional Physics Mechanisms}

Efforts to explain the anomaly were originally focused on conventional physics mechanisms generated on-board, such as gas leaks from the propulsion system or a recoil force due to the on-board thermal power inventory. So far, these mechanisms have been found to be either not strong enough to explain the magnitude of the anomaly or else to exhibit significant temporal or spatial variations contradicting the known properties of the anomaly presented above \citep{pioprd,pio-standard}.

A number of other conventional physics possibilities have also been addressed.  In particular, it has been proposed that Kuiper belt objects or dust could explain the anomaly by 
\begin{inparaenum}[i)]
\item a gravitational acceleration,
\item an additional drag force (resistance) and
\item a frequency shift of the radio signals proportional to the distance.
\end{inparaenum}

Of course, one of the most natural mechanisms to generate a putative physical force is the gravitational attraction due to a known mass distribution in the outer solar system; for instance, due to Kuiper belt objects or dust. However, possible density distributions for the Kuiper belt were studied in \citep{pioprd} and found to be incompatible with the discovered properties of the anomaly. Even worse, these distributions cannot circumvent the constraint from the undisturbed orbits of Mars and Jupiter \citep{pioprd}.  The density of dust is not large enough to produce a gravitational acceleration on the order of $a_P$ \citep{pioprd,mmn05,orfeu_paula05} and also it varies greatly within the Kuiper belt, precluding any constant acceleration. Hence, a gravitational attraction by the Kuiper belt can, to a large extent, be ruled out.

Also, the data from the inner parts of the solar system taken by the Pioneer 10/11 dust detectors favors a spherical distribution of dust over a disk in this inner region. Ulysses and Galileo measurements in the inner solar system find very few dust grains in the $10^{-18}-10^{-12}$~kg range \citep{dust}. IR observations rule out more than 0.3 Earth mass from Kuiper belt dust in the trans-Neptunian region. The resistance caused by the interplanetary dust is too small to provide
support for the anomaly \citep{dust}.  Any dust-induced frequency shift of the carrier signal is also ruled out.

Finally we note that, motivated by the numerical coincidence $a_P\simeq cH_0$, where $c$ the speed of light and $H_0$ is the Hubble constant at the present time, there have been many attempts to explain the anomaly in terms of the expansion of the Universe.  \cite{pioprd}  showed that such a mechanism would produce an opposite sign for the effect. A study of the effect of cosmic acceleration on the radio signals rather than on the spacecraft themselves was also undertaken.  This mechanism might be able to overcome the apparent conflict that $a_P$ presents to modern solar system planetary ephemerides \citep{pioprl,pioprd}.

\subsubsection{Possibility for New Physics?} 
\label{sec:new-ph}

The apparent difficulty to explain the anomaly within standard physics became a motivation to look for ``new physics.'' So far these attempts have not produced a clearly viable mechanism for the anomaly.  In particular, the physics of MOND (MOdified Newtonian Dynamics) represents an interesting possibility with its phenomenological long-range modification of gravity invoked to explain the rotation curves of galaxies \citep{Milgrom,Bekenstein}. However, the numerical value for the MONDian acceleration $a_0$ is almost an order of magnitude smaller then $a_P$ and is not likely to be observed on the scales of the solar system.  

There is also an attempt to explain the anomaly in the framework of a nonsymmetric gravitational theory \citep{moffat}. It has been argued that a skew-symmetric field with a suitable potential could account for $a_P$ \citep{moffat05,brownstein-moffat} as well as galaxy and cluster rotation curves. A modification of the gravitational field equations for a metric gravity field, by introducing a general linear relation between the Einstein tensor and the energy-momentum tensor has also been claimed to account for $a_P$ \citep{serge2,mark_serge_05}. Both these approaches are currently being further investigated.

Various distributions of dark matter in the solar system have been proposed to explain the anomaly, e.g., dark matter distributed in the form of a disk in the outer solar system with of a density of $\sim4\times10^{-16}$~kg/m$^3$, yielding the wanted effect. Dark matter in the form of ``mirror matter'' \citep{FootVolkas01}  is one example. However, it would have to be a special smooth distribution of dark matter that is not gravitationally modulated as normal matter so obviously is.

\cite{orfeu04} have shown that a generic scalar field cannot explain $a_P$; on the other hand they proposed that a non-uniformly coupled scalar might produce the wanted effect. Although braneworld models with large extra dimensions offer a richer phenomenology than standard scalar-tensor theories, it is difficult to find a convincing explanation for the Pioneer anomaly \citep{orfeu_jorge05}. Other ideas include Yuka\-wa-like or higher order corrections to the Newtonian potential and a theory of conformal gravity with dynamical mass generation, including the Higgs scalar (see discussion in \citep{pioprd}). 

We must conclude that there are many interesting ideas proposed to explain the physics of the Pioneer anomaly. However, most of them need more work before they can be considered to be viable.
To summarize, the origin of the Pioneer anomaly remains unclear.

\section{Recovery of the Extended Pioneer Doppler Data Set}
\label{sec:pio-doppler}

\cite{stanford,iap-pioneer} have advocated for an analysis of the entire set of existing Pioneer Doppler data, obtained from launch to the last useful data received from the vehicles. This data could yield critical new information about the anomaly, especially during the earlier mission phases \citep{pioprd}. This led to the initiation of an effort to retrieve the early Pioneer data that until recently existed in several places on various obsolete format and media (e.g., 9-track and even 7-track magnetic tapes).

Recovery of radiometric data for a mission operating for more then 30 years is an effort that was never attempted before. Indeed, 30 years is a long time, presenting many unique challenges, including changes in the data formats, navigational software, as well as supporting hardware. Even the DSN configuration had changes -- new stations were built, and some stations moved, upgraded and reassigned. The main asset of the entire mission support -- its people -- changed the most.  By 2005 all the DSN data formats, navigational software used to support Pioneers, all the hardware used to read, write and maintain the data have become obsolete and are no longer operationally supported by existing NASA protocols. 

Despite the anticipated complexities, the transfer of the available Pioneer Doppler data to modern media formats had been initiated at JPL in early June 2005 and, as of December 2005, it is nearing completion. In this section we discuss the current status of this effort to recover the Pioneer 10/11 radiometric Doppler data.

\subsection{Doppler Data}

Doppler data is the measure of the cumulative number of cycles of a spacecraft's carrier frequency received during a user-specified count interval. The exact precision to which these measurements can be carried out is a function of the received signal strength and station electronics, but it is
a small fraction of a cycle. Raw Doppler data is generated at the tracking station and delivered via a DSN interface to customers. 
In order to acquire Doppler data, the user must provide a reference trajectory and information concerning the spacecraft's RF system to JPL's Deep Space Mission System (DSMS), to allow for the generation of pointing and frequency predictions. The user specified count interval can vary from 0.1 sec to 10 minutes, with count times of 10 to 60 seconds being typical \citep{dsms-cat}. The average rate of change of the cycle count over the count interval expresses a measurement of the average velocity of the spacecraft in the line between the antenna and the spacecraft. The accuracy of Doppler data is quoted in terms of how accurate this velocity measurement is over a 60 second count. The accuracy of data improves as the square root of the count interval.

\begin{figure*}[t!]
 \begin{center}
\noindent  \hskip 0pt  
\psfig{figure=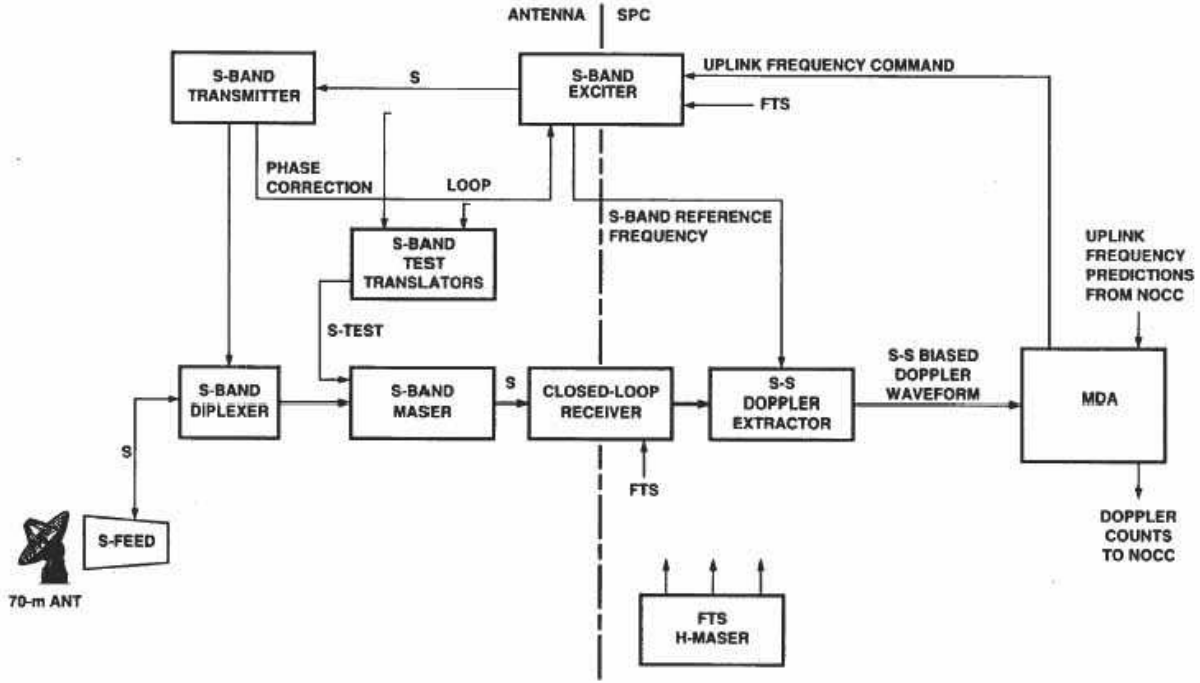,width=162mm}
\end{center}
\vskip -12pt
  \caption{DSN baseline configuration for Pioneer radio science experiments \citep{616-55}. (See more discussion of this chart in Section~III.A of \cite{pioprd}.) 
 \label{fig:pio-radio-science}}
\end{figure*} 

\begin{enumerate}
\item[(1)]  \emph{Non-coherent Doppler Data} (also known as one-way or F1 data) is data received from a spacecraft where the downlink carrier frequency is not based on an uplink signal. The ability of the tracking station to measure the phase of the received signal is the same for non-coherent versus coherent data types, however the uncertainty in the value of the reference frequency used to generate the carrier is generally the dominant error source. Pioneers provided a significant amount of F1 data that is unfortunately not useful for precision orbit determination.
\item[(2)] \emph{Coherent Doppler Data} is received from a spacecraft when the reference frequency of the received carrier signal was based on a transmitted uplink signal from the Earth. This is commonly known as two-way (or F2) data, when the receiving and transmitting ground stations are the same, and three-way (or F3) data, when the transmitting and receiving stations are different. Since the frequency of the original source signal is known, this error source does not affect data accuracy. The accuracy of this data is a function primarily of the carrier frequency, but is affected by transmission media effects. The F2 and F3 data are the primary source for the Pioneer 10 and 11 orbit determination and were the primary focus of the data recovery efforts. 
\end{enumerate}

The Pioneers used S-band ($\sim$2.2 GHz) radio signals to communicate with the NASA Deep Space Network (DSN). The S-band data is available from 26~m, 70~m, and some 34~m antennas of the DSN complex (see baseline DSN configuration in the Figure~\ref{fig:pio-radio-science}). The 1-$\sigma$ accuracy of S-band data is approximately 1 mm/s for a 60 second count interval after being calibrated for transmission media effects. The dominant systematic error that can affect S-band tracking data is ionospheric transmission delays. When the spacecraft is located angularly close to the Sun, with Sun-Earth-spacecraft angles of less than 10 degrees, degradation of the data accuracy will occur. S-band data is generally unusable for Sun-Earth-spacecraft angles of less than 5 degrees. (See more details on the modern capabilities of the DSN, especially its radio science performance for extracting precision Doppler signal demonstrated in \citep{asmar_etal-2004}.)

\begin{figure*}[t!]
 \begin{center}
\noindent    
\psfig{figure=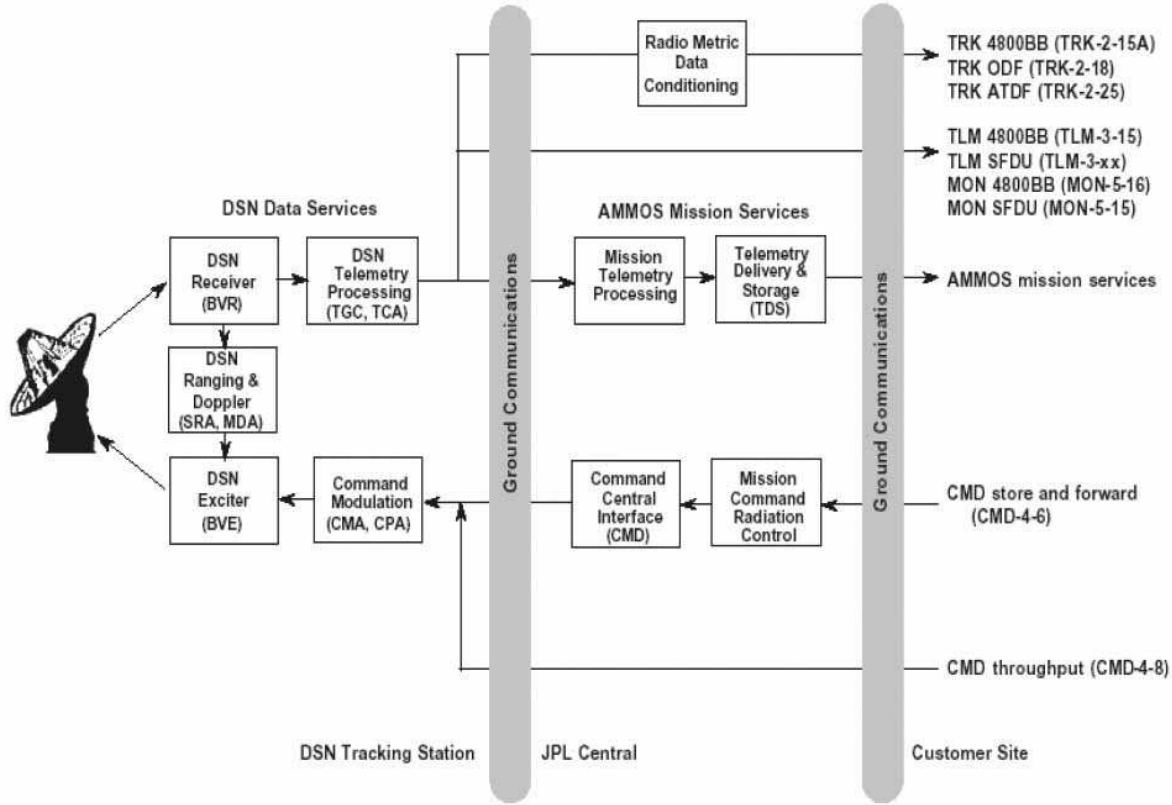,width=160mm}
\end{center}
\vskip -10pt
  \caption{Typical tracking configuration for a Pioneer-class mission
and corresponding data format flow.  
 \label{fig:pio-data-format-flow}}
\end{figure*} 

\subsection{Pioneer Doppler Data Formats}

The Pioneer radiometric data was received by the DSN in ``closed-loop'' mode, i.e., it was tracked with phase lock loop hardware. (``Open loop'' data is recorded to tape but not tracked by phase lock loop hardware.)  There are basically two types of data: Doppler (frequency) and range (time of flight), recorded at the tracking sites of the DSN as a function of UT Ground Received Time \citep{DSN810-5}.  During their missions, the raw radiometric tracking data from Pioneers were received originally in the form of Intermediate Data Record (IDR) tapes, which were then processed into special binary files called ATDFs (Archival Tracking Data Files, see description in \citep{TRK-2-25}), containing Doppler data from the standard DSN tracking receivers (see Figure~\ref{fig:pio-data-format-flow}).\footnote{After the recent DSN upgrade, there is high time resolution Doppler data available from special radio science receivers, so-called ``open loop'' data. This data is stored in files called ODRs (Original Data Records), which became available only at the very end of the Pioneer missions and were not used as a standard format for navigating these missions in deep space.}  Note that the ``closed-loop'' data constitutes the ATDFs that were used in \cite{pioprd}.

After a standard processing at the RMDC (Radio Metric Data Conditioning group) of JPL's Navigation and Mission Design Section,  ATDFs are transformed into ODFs (Orbit Data Files, see description in \citep{TRK-2-18}). A program called {\tt STRIPPER} is used to produce the ODFs that are, at this point, the main product that is distributed to the end users for their orbit determination needs (see more discussion on the conversion process in \citep{pioprd}). At JPL, after yet additional processing, these ODFs are used to produce sequentially formatted input/output files in {\tt NAVIO} format that is used by navigators while working with the JPL Orbit Determination Program. (Note that the {\tt NAVIO} input/output file format is used only at JPL; other orbit determination programs convert ODFs to their particular formats.)   

\subsubsection{ATDF -- Archival Tracking Data File (Format TRK-2-25)}
\label{sec:atdf}

ATDFs are files of radiometric data produced by the Network Operations
Control Center (NOCC) Navigation Subsystem (NAV) (see Figure~\ref{fig:pio-radio-science}).   They are derived
from Intermediate Data Records by NAV and contain all radiometric
measurements received from the DSN station including signal levels
(AGC = automatic gain control in dBm), antenna pointing angles,
frequency (often referred to simply as ``Doppler''), range, and residuals.
Doppler data is often used to infer spacecraft radial motion relative
to the tracking antenna.  Data values in ATDFs are reported at rates no
higher than 10 per second.   During their missions, the Pioneers' 
received frequencies at S-Band were recorded at a nominal sample time of one per either 10, 60, 300, 600 or 1980 seconds.  In addition, for several years from 1980 to 1994 there were tracking passes scheduled for the JPL-led gravity wave experiments that delivered data at a smaller count time of 1 per second. 

Each ATDF consists of all tracking data types used to navigate a particular spacecraft (Pioneers had only Doppler data type) and typically include Doppler, range and angular types (in S-, X-, and L-frequency bands), differenced range versus integrated Doppler, programmed frequency data, pseudo-residuals, and validation data.   (Unfortunately there was no range capability implemented on the Pioneers \citep{pioprl,pioprd}. Early in the mission, JPL successfully simulated range data using modulation of sub-carrier signal and some range data is available, especially for planetary flybys. Later on, this technique of simulating range did was limited by the bandwidth of the sub-carrier signal and did not produce any range data. ) Also, ATDFs contain data for a single spacecraft, for one or more ground receiving stations, and for one or more tracking passes or days. 

The ATDF is described in section TRK-2-25 of \citep{DSN820-13}.  Each ATDF data record contains 117 parameters, stored in records of 288 bytes.  Each ATDF physical record is 2016 32-bit words in length (8064 bytes) and consists of 28 72-word (288-byte) logical records.  Each ATDF contains an integer number of these blocks.  The Tracking Data Table format and content used until early 1997 are described by \citep{DSN820-13}.\footnote{Formats and content used after early 1997 are described in DSN Document SFOC-NAV-2-25.}

The ATDF records are arranged in a sequence that consists of one file identification record, one transponder logical record, tracking data logical records in time order, and software/hardware end-of-file markers. Bit lengths of data fields are variable and cross word boundaries. (Further details of the ATDF content and format are given in \citep{TRK-2-25}.) Each file contains 15 columns per data point specifying a time-tag, S-band, and X-band receiver parameters (the latter not applicable for the Pioneers.)

\subsubsection{ODF -- Orbit Data File (Format TRK-2-18)}

After standard processing at JPL's RMDC with the use of the {\tt STRIPPER} program (see \cite{pioprd} for more details), ATDFs are transformed into ODFs for use in determining  spacecraft trajectories, gravity fields affecting them, and radio propagation conditions.   

ODFs contain radiometric data that has been converted from the ATDF format \citep{TRK-2-18}.  Each ODF physical record is 2016 32-bit words in length and consists of 224 9-word logical records per data block. The ODF records are arranged in a sequence that consists of one file label record, one file identifier logical record, orbit data logical records in time order, ramp data logical records in time order, clock offset data logical records in time order, data summary logical records in time order and software/hardware end-of-file markers. Bit lengths of data fields are variable and cross word boundaries. An ODF usually contains most types of records, but may not have them all. 
The first record in each of the 7 primary groups is a header  record; depending on the group, there may be from zero to many data records following each header.

\subsection{Retrieval of the Pioneer Doppler Data}

The first task was to identify the data storage facilities with potential inventory of the Pioneer Doppler data. In general, we expected to find data in four different  locations, namely:
\begin{enumerate}
\item[1.] Archive of radiometric data at RMDC at JPL:  Our hope was that radiometric data of Pioneers 10 and 11 (in all formats, namely IDR, ATDF, and ODF) may still exist at the various data storage archives at JPL and especially at the RMDC archive.  Furthermore, we expected to find data in all possible media formats, including 7-track and 9-track magnetic tapes and also digital records on the RDC system.

\item[2.] The next logical place was the National Space Science Data Center (NSSDC) at the NASA Goddard Space Flight Center,  {\tt http://nssdc.gsfc.nasa.gov/planetary/pioneer10-11.html}. We knew that NSSDC had  copies of ATDFs on magnetic tapes submitted there by John D. Anderson of JPL, he being the Principal Investigator on the celestial mechanics experiments, as part of his agreement with NASA. The data coverage was expected to be from 1978 to 1994.  In addition to that, we knew that George Null of JPL also submitted Pioneer Doppler data to NSSDC prior to 1978.  It was also know that, in addition to raw Doppler data, he sent some orbital solutions and auxiliary information (maneuvers, spin rates, initial conditions, etc.) that also would be of interest to us. 

\item[3.] We also expected that NASA Ames Research Center (ARC), as the center with the project management responsibilities for all the Pioneer missions, would have some information useful for our investigation. We later realized that no raw orbital information was provided to ARC; instead, the trajectory solutions derived by JPL were used for all project needs.

\item[4.] Several facilities throughout the country were used to provide overflow storage for magnetic tapes in the 1970s and 1980s.  Two of these Federal Data Storage Facilities are located in California: in Long Beach and in Palo Alto. We have not yet tried to approach these data archival centers, but plan to do this in the near future.  
\end{enumerate}

All the sources of radiometric data listed above were used to recover the Pioneer Doppler data.  This data was present in different media formats (magnetic tapes, floptical disks, digital formats on various computer platforms with different hard drive formatting standards) and were written in different data formats (i.e. IDR, ATDF and ODF).  The goal was to recover as much of this data as possible. The next section summarizes the results of our retrieval efforts. 

\subsection{Current Status of the Data Recovery Process}

To recover Pioneer radiometric data we first went to the NSSDC with a request to provide us with copies of all available navigational data (IDR, ATDF, and ODF) for the period 1978-1994 that was previously archived at NSSDC (done by J.\,D. Anderson and E.\,L. Lau of JPL). The NSSDC staff loaded the JPL-supplied IDR and ATDF tapes on their VMS and Unix machines. However, we realized that all the ATDF data files that we were receiving from NSSDC were corrupted. The problem was introduced by the process that was used to transfer the data across multiple computer platforms: an additional byte was inserted at record boundaries. In other words, every 8065th byte is a record marker and is not part of the ATDF stream. This problem was first encountered by C. Markwardt during his study of the Pioneer anomaly \citep{markwardt}. Fortunately,  we were able to recognize the problem early on and were able to fix most of the corrupted files following the procedure suggested by \cite{markwardt-atdf}. When JPL retrieved back the data using FTP, every 8065th byte was removed from the each file record. The resulting data is good to use.

Another data segment retrieved from the NSSDC corresponded to the very beginning of the Pioneer missions, namely the period $\sim$1972-1976 (files were archived by G. Null of JPL). Unfortunately, it appears that not only are these data files corrupted, but they were also written in a format called ``type-66'' (or T66) which has not been used at JPL for more than 25 years, which adds unanticipated delay. This is why  we are still in the process of trying to recover some segments in the hope of adding this data to the new data set. (Note that we were not able to recover any ODF data files from NSSDC due to unknown/unidentified data formats.)

\begin{figure*}[t!]
\hskip -2pt 
\begin{minipage}[b]{.46\linewidth}
\centering \psfig{file=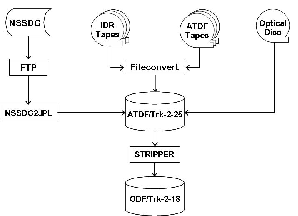, width=78.0mm}
\end{minipage} 
\hskip 15pt
\begin{minipage}[b]{.46\linewidth}
\centering \psfig{file=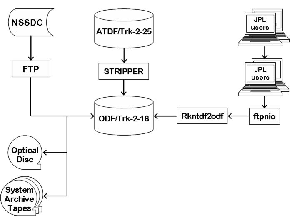, width=78.0mm}
\end{minipage}
\caption{Pioneer 10 and 11 radiometric data retrieval processes. Left: ATDF recovery flow chart. Right: ODF recovery flow chart.}
 \label{fig:recovery-flow-chart}
\vskip -5pt 
\end{figure*}

A significant volume of data (both IDR and ATDF) was found at RMDC on 9-track magnetic tapes (1978-1988) that were read on a still available MiniVAX computer.\footnote{This computer was ready to be decommissioned, but an appeal from \cite{TPS} led to an extension of the operational life of this system by more than 6 months.}  This segment of the data is in a good condition and is ready to use. The retrieval of IDR data was reasonably fast as each tape contained only one file. However, the ATDF 9-track tape had several ATDFs per tape (usually 10-15 files). This increased the time needed to read a single tape on the MiniVAX computer. With regular cleaning of the head of the magnetic tape reader and also cleaning the tapes themselves it usually took almost 90 minutes to read a single tape on this hardware, which was built in the 1980s. With an inventory of more than 200 tapes this was a major effort during the summer of 2005. 

There were a few additional and unexpected finds of Pioneer data at JPL.  Several historical ATDFs for both Pioneer 10 and 11 were found and recovered from the RMDC's archival optical disks.  Also at RMDC, we found SPR  (System Performance Record) data (format TRK-2-15A). The raw SPR data format was used to replace the IDR format, and it was utilized after 1985 when JPL/NAV upgraded the system from UNIVAC to DEC/VMS hardware. We were fortunate to recover several additional historical ATDFs for both Pioneers from the RMDC's System Archive Tapes (SAT) (written on 9-track and 4~mm tapes). These files were also added to the overall file inventory. The resulting ATDF and ODF retrieval methods are shown in Figure~\ref{fig:recovery-flow-chart}.

In addition to the data already recovered, we found and recovered some early ODF data (1973-1974) that was still available in the personal data archives of our JPL colleagues who worked with Pioneer data for other purposes (e.g., development of the solar system ephemerides). In fact, we were able to recover several ODF files written in the {\tt NAVIO} input/output form. These files, obtained from several JPL users, were re-converted back to the ODF data format.  

Finally, we have added the data that was already analyzed by \cite{pioprd} that corresponds to the period of 1987-1998 (for Pioneer 10) and 1987-1990 (for Pioneer 11).  Lastly, we added the most recent data for the period 1998-2002. The resulting 1972-2002 data has some redundancy, but mostly it is a very complete data set assembled for the first time.  After necessary certification at JPL this same set will be available for distribution. 


The currently available expanded data set includes the following data segments for each spacecraft:

\begin{itemize}
\item Pioneer 10: The entire available data set covers mission events from mid-1973 (including Jupiter encounter data) to the last time a Pioneer 10 contact returned telemetry data, 27 April 2002. This interval spanned heliocentric distances  from $\sim4$ AU to 87 AU. The total $\sim$30-year Pioneer 10 data set might have $\sim$60,000 data points and is about 20 GB in size.  We also have most of the information on maneuvers, spin rate, and initial conditions.

\item Pioneer 11:  The entire available data set covers the mission from mid-1974 to late 1994 (including both Jupiter and Saturn encounters). This interval spanned heliocentric distances from $\sim$4 AU to 33~AU. The total  $\sim$20-year Pioneer 11 data set might have $\sim$50,000 data points and is about 15 GB in size. We also have most of the information on maneuvers, spin rate, and initial conditions.

\end{itemize}

To summarize, there exists $\sim$30 years of Pioneer 10 and $\sim$20 years of Pioneer 11 data, most of which had never been well studied for our purposes.  We are still trying to increase the percentage of the F2 and F3 Doppler data in the overal retreived Doppler dataset, as F1 data is useless for our analysis. For this we are working together with RMDC in an attempt to improve the output of the ATDF-to-ODF file conversion effort, which we expect to complete by January 2006.  By the same time we also hope to close the T66 issue and prepare all input information critical for the upcoming re-processing of the Doppler data for the entire Pioneer 10 and 11 trajectories. 

To re-process the complete Pioneer trajectories, one would first have to (re)edit and (re)process the entire data span (from 1972 to 2002), using the same editing strategy, initial conditions, and parameter estimation and noise propagation algorithms. One could also process the high rate Doppler data (i.e., 1 record per second) that previously was used very little.  The high rate data can be used to better determine a spacecraft's spin rate and also to improve the maneuver data file information. Also, the spin rate change was found to be highly correlated with a small but significant spacecraft-generated force, probably from gas leaks \citep{pioprd}. Therefore, one can also use the high-rate data to estimate and/or calibrate valve gas leaks and all the maneuvers. Note that the use of the spacecraft telemetry data may be critical in addressing this issue (see discussion of this possibility in Section~\ref{sec:mass-expulsion}).  

Since the previous analysis \citep{pioprl,pioprd}, physical models for the Earth's interior and the planetary ephemeris have greatly improved. This is due to progress in GPS-, SLR-, LLR- and VLBI-enabling technologies, Doppler spacecraft tracking, and new radio science data processing algorithms. One would have to write and/or update existing orbit determination programs using these latest Earth models (adopted by the IERS) and also  using the latest planetary ephemeris. This will improve the solutions for the DSN ground station locations by two orders of magnitude (1 cm) over that of the previous analysis. Additionally, this will allow a better characterization of not only the constant part of any anomalous acceleration, but also of the annual and diurnal terms detected in the Pioneer 10 and 11 Doppler residuals \citep{moriond,pioprd,Oystein}. 

\section{Pioneer Telemetry and On-Board Systematics}
\label{sec:mdr}  

As pointed out above, we were able to recover almost the entire set of Doppler data obtained from both Pioneers 10 and 11.  We were also able to obtain the entire archive of the spacecraft telemetry data in a form called Master Data Records (MDR) for both craft. Because of their potentially high value in the investigation of the Pioneer anomaly, in this section we discuss the purpose, means of delivery, storage, processing and possible use of the MDRs.

\subsection{Master Data Records}

\begin{figure*}[t!]
 \begin{center}
\noindent   
\psfig{figure=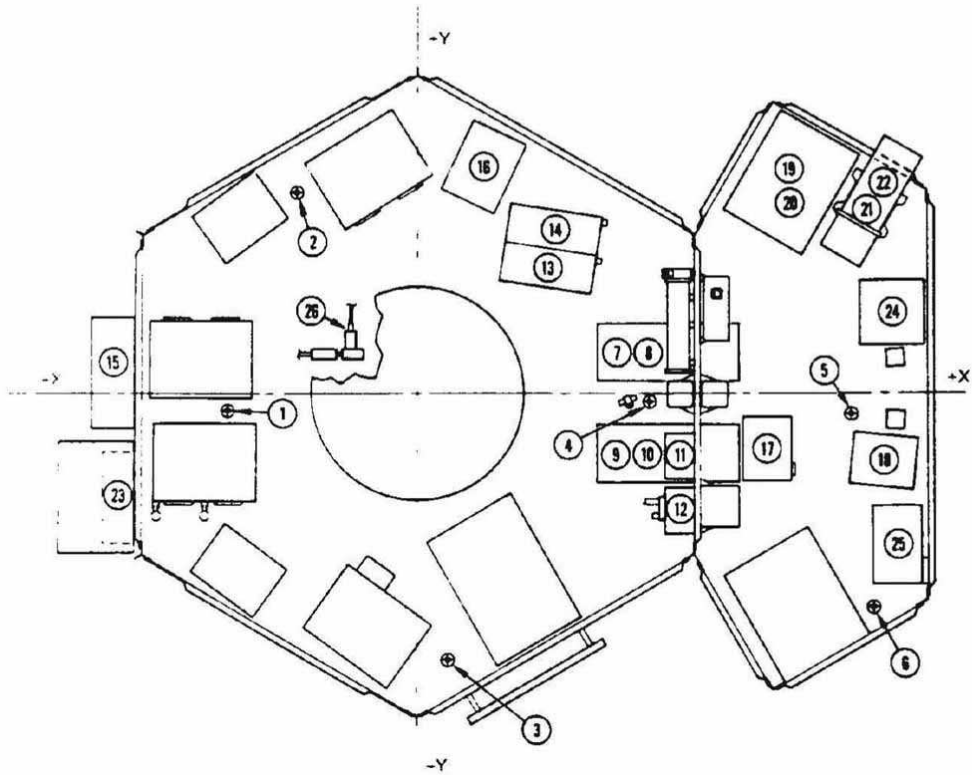,width=130mm}
\end{center}
\vskip -10pt 
  \caption{Location of thermal sensors in the instrument compartment of the Pioneer 10/11 spacecraft (from \cite{PC-202}). Some locations (i.e., end of RTG booms, propellant tank interior, etc.) not shown.
 \label{fig:sensors}}
\vskip -10pt 
\end{figure*}

All data received from the Pioneer 10 and 11 spacecraft by the NASA Deep Space Network (DSN) was initially stored in the form of MDRs. These records contained information about the DSN station and reception characteristics, in addition to the actual, ``raw'' data records themselves, which contained both scientific measurements and engineering telemetry, as received from the spacecraft. (As an example, Figure~\ref{fig:sensors} shows locations of several thermal sensors in the spacecraft instrument compartment.)
 
Normally, MDRs are seen to be of little use once the relevant information is extracted from them. Scientific measurements are extracted, packaged in the appropriate form, and sent to the corresponding experimenters for further processing and evaluation. Engineering telemetry is used by the spacecraft operations team to control and guide the spacecraft. MDRs are not necessarily considered worth preserving once the scientific data has been extracted, and the engineering telemetry is no longer needed for spacecraft operations. Indeed, the MDR retention schedule prescribed that the tapes be destroyed after 7 years.
 
Fortunately, most of the MDRs for the Pioneer 10 and 11 projects have been preserved nevertheless. Long before we (LRK and VTT) heard about the Pioneer anomaly investigation, we recognized the potential value of the MDRs for educational purposes; we envisioned an educational Web site where a visitor could observe the Pioneers as they traveled through the solar system, while viewing telemetry information that showed the health status and operational characteristics of the two spacecraft.
 
For this reason, LRK endeavored to preserve the complete set of MDRs for the two projects and copied the data to modern media. Meanwhile, VTT began developing PC-based tools for processing the MDRs and extracting information.
 
Far beyond the original expectations, this effort is now seen to be of value for the investigation of the Pioneer anomaly, as the MDRs, specifically the telemetry data contained therein, may help in the construction of an accurate model of the spacecraft during their decades long journey, including a precise thermal profile, the time history of propulsion system activation and usage, and many other potential sources of on-board disturbances.

\subsubsection{MDR Media and Data Organization}

In the early days of Pioneer operations, the original MDRs for the Pioneer 10 and 11 missions were copied to magnetic tape at the DSN receiving stations, and then sent to the appropriate NASA office for further processing.  

After more than 20 years of storage, the medium started to show signs of deterioration and all of the recoverable raw data was copied to 128 megabyte magneto-optical (``floptical'') disks.  These disks were readable by a MicroVAX using a Digital Corporation RWZ21 Magneto Optical SCSI disk drive. Digital Corporation is now part of Compaq, the available VAX 4000-300 computer will soon cease to exist, and the SCSI RWZ21 drives will not be usable to read these ``floptical'' disks.  

At present, a near complete set of MDR ``floptical'' disks still exists, along with equipment capable of reading this media. During the course of 2004 and 2005, Kellogg has copied the files from ``floptical'' disks to a modern computer, and also created DVD-ROMs for distribution. The contents of the available MDR disks can be summarized as follows: 
\begin{center}
{\tt Pioneer 10: 155 disks x 128 MB (16.33 GB)\\
Pioneer 11: 217 disks x 128 MB (23.01 GB)}
\end{center}

All MDR files are named using a common naming convention. Each file contains the MDRs for one spacecraft for an entire day. The file name begins with the letter `m', followed by a two-digit spacecraft identifier (23 for Pioneer 10, 24 for Pioneer 11), and a 5-digit date (2 digits for the year, 3-digit DOY.) The filename extension is {\tt .mdr}. Thus, {\tt m2495003.mdr}, for instance, would be the name of the Pioneer 11 file for January 3, 1995.
 
When Kellogg copied the files to modern computers, he established a separate directory for each ``floptical'' disk copied. He also used a consistent naming convention: the directory name began with the two-digit spacecraft identifier, followed by the letter `P', a two-digit year value, and a two-digit sequence number that simply identified the number of the disk for the given year. So for instance, {\tt 23P9602} would be disk 2 for Pioneer 10, for the year 1996.

\subsubsection{Data Integrity and Completeness}

The total amount of data stored in these MDR files is approximately 40~GB. According to the original log sheets that record the transcription from tape to ``floptical'' media, only a few days worth of data is missing, some due to magnetic tape damage.
 
One notable exception is the ``Jupiter encounter period'' of Pioneer 10. According to the transcription log sheets, DOY 332-341 from 1973 were not available at the time the ``floptical'' disks were made. Kellogg's investigation revealed that the tapes in question may have been misplaced long before the transcription has taken place. Unfortunately, it is unlikely that those tapes will ever be located, and even if they're found, chances are that they will no longer be readable due to media deterioration.

Other significant periods of missing data are listed in Table \ref{tb:missing_mdr}. It is not known why these records are not present, except that we know that very few days are missing due to unreadable media (i.e., the cause is missing, not damaged, tapes.)

\begin{table}[h!]
\vskip -12pt
\begin{center}
\caption{Pioneer 10/11 missing MDRs (periods of missing data shorter than 1-2 days not shown.)}
\vskip 8pt
\begin{tabular}{|l|l|l|}\hline
Spacecraft&Year&DOYs\\\hline\hline
Pioneer-10&1972&133-149\\
&1973&004-008, 060-067, 332-341\\
&1974&034-054\\
&1979&025-032, 125-128, 137-157, 171-200\\
&1980&173-182, 187-199, 248-257\\
&1983&329-348\\
&1984&346-359\\\hline
Pioneer-11&1973&056-064, 067-080, 082-086, 088-094\\
&1980&309-330, 337-365\\
&1982&318-365\\
&1983&001-050\\
&1984&343-357\\
&1990&081-096\\\hline
\end{tabular}
\label{tb:missing_mdr}
\end{center}
\vskip -10pt
\end{table}
 
So, the record is fairly complete. But how good is the data? Over forty billion bytes were received by the DSN, processed, copied to tape, copied from tape to ``floptical'' disks, then again copied over a network connection to a personal computer. It is not inconceivable that the occasional byte was corrupted by a transmission or storage error. We do indeed see records that contain what is apparently bogus data, especially from the later years of operation. This, plus the fact that the record structure (e.g., headers, synchronization sequences) is intact suggests reception errors as the spacecraft's signal got weaker due to increasing distance, and not copying and/or storage errors.
 
The MDRs contain no error detection or error correction code, so it is not possible to estimate the error rate. However, it is likely to be reasonably low, since the equipment used for storage and copying is generally considered very reliable. Furthermore, any errors would likely show up as random noise, and not as a systemic bias. In this regard, the data should generally be viewed to be of good quality insofar as the goal of constructing an engineering profile of the spacecraft is considered.

\subsubsection{Interpreting the Data}
MDRs are a useless collection of bits unless information is available about their structure and content. Fortunately, this is the case in the case of the Pioneer 10 and Pioneer 11 MDRs.
 
The MDR record frame, as used in the Pioneer project, is described in detail in \citep{ARC221}. At the top of each MDR record is 10 words (32 bits each) of data including a timestamp, spacecraft identifier, etc. This is followed by four data frames (not all four frames may be used, but they're all present) of 192 bits each. Lastly, an additional 8 words of DSN information completes the record. The total length of an MDR is thus 42 words of 32 bits each.
 
Data frames are organized as 192-bit blocks, usually interpreted as 64 3-bit words or, alternatively, as 32 6-bit words. The Pioneer project used many different data frame formats during the course of the mission. Some formats were dedicated to engineering telemetry (accelerated formats). Other formats are science data formats, but still contain engineering telemetry in the form of a subcommutator: a different engineering telemetry value is transmitted in each frame, and eventually, all telemetry values are cycled through.
 
The Pioneer spacecraft had a total of 128 6-bit words reserved for engineering telemetry. Almost all these values are, in fact, used. A complete specification of the engineering telemetry values can be found in subsection 3.5 (``Data Handling Subsystem'') of \citep{PC-202}. When engineering telemetry was accelerated to the main frame rate, four different record formats (C-1 through C-4) were used to transmit telemetry information. When the science data formats were in use, an area of the record was reserved for a subcommutator identifier and value.
 
The formats are further complicated by the fact that some engineering telemetry values appear only in subcommutators, whereas others only appear at the accelerated (main frame) rate.
 
In the various documentation packages, engineering data words are identified either by mnemonic, by the letter `C' followed by a three-digit number that runs from 1 through 128, or most commonly, by the letter `C' followed by a digit indicating which `C' record (C-1 through C-4) the value appears in, and a two-digit number between 1 and 32: for instance, C-201 means the first engineering word in the C-2 record.

\subsubsection{Decoding Analog Values}
 
Engineering values generally fall into two distinct categories. ``Digital'' values are bit patterns communicating the state of switches or binary sensors on board the spacecraft (i.e., ``On/Off'' type measurements.) ``Analog'' values are digitally encoded readings of sensors such as voltage, temperature, current, etc.  Other binary values include various counters and timers.
 
In order to decode analog values, it is necessary to refer to the original calibration of the analog sensors in question, calibration that was performed prior to launch. These values were available in the form of data files, used in an earlier LabView program that was used in Pioneer operations, and were communicated in \citep{LK031130}. For Pioneer 10, many of the calibration values have been verified against original documentation (\cite{ARC037}). Most values agree, but some minor discrepancies exist that still need to be resolved. No original documentation has yet been located containing Pioneer 11 calibration values.

Analog sensors were calibrated using fifth order polynomials. The following is an example entry, representing the fin root temperature of RTG 1 (parameter C-201):

\begin{verbatim}
{ PRTG1FRTlst }
	+0.1524085978E+03	 +0.2321608606E+01	 +0.1313264518E+00
	-0.4043172698E-02	 +0.3726298780E-04	  0.0
\end{verbatim}
\par
As the trivial case, a value that needs no decoding (i.e., where the binary value is the analog readout) would be represented by a line similar to the following:
\begin{verbatim}
{ NOCOEFlst }
	 0.0	 1.0	 0.0	 0.0	 0.0	 0.0
\end{verbatim}

\vskip -6pt 
\begin{figure*}[ht!]\hskip -4pt
\begin{minipage}[b]{.46\linewidth}
\centering \psfig{file=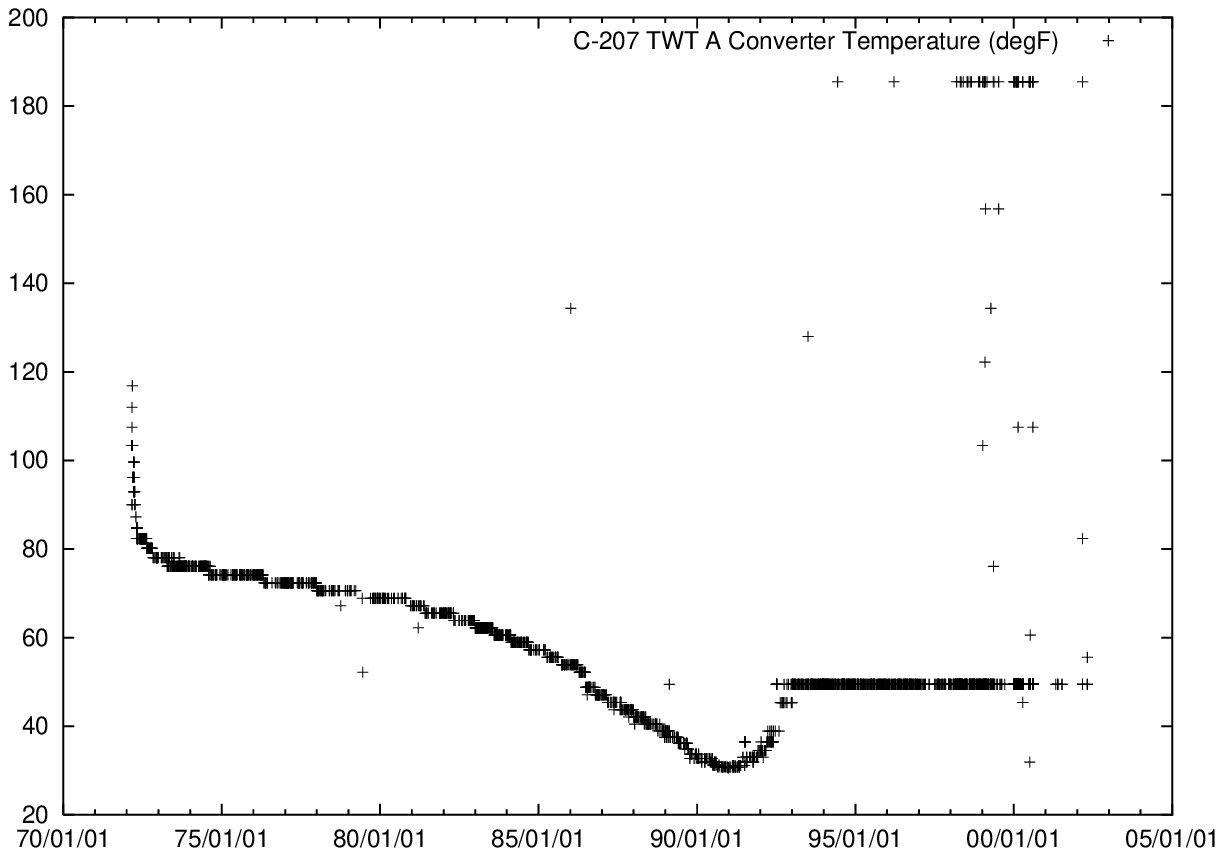, width=80mm}
\end{minipage} 
\hskip 20pt
\begin{minipage}[b]{.46\linewidth}
\centering \psfig{file=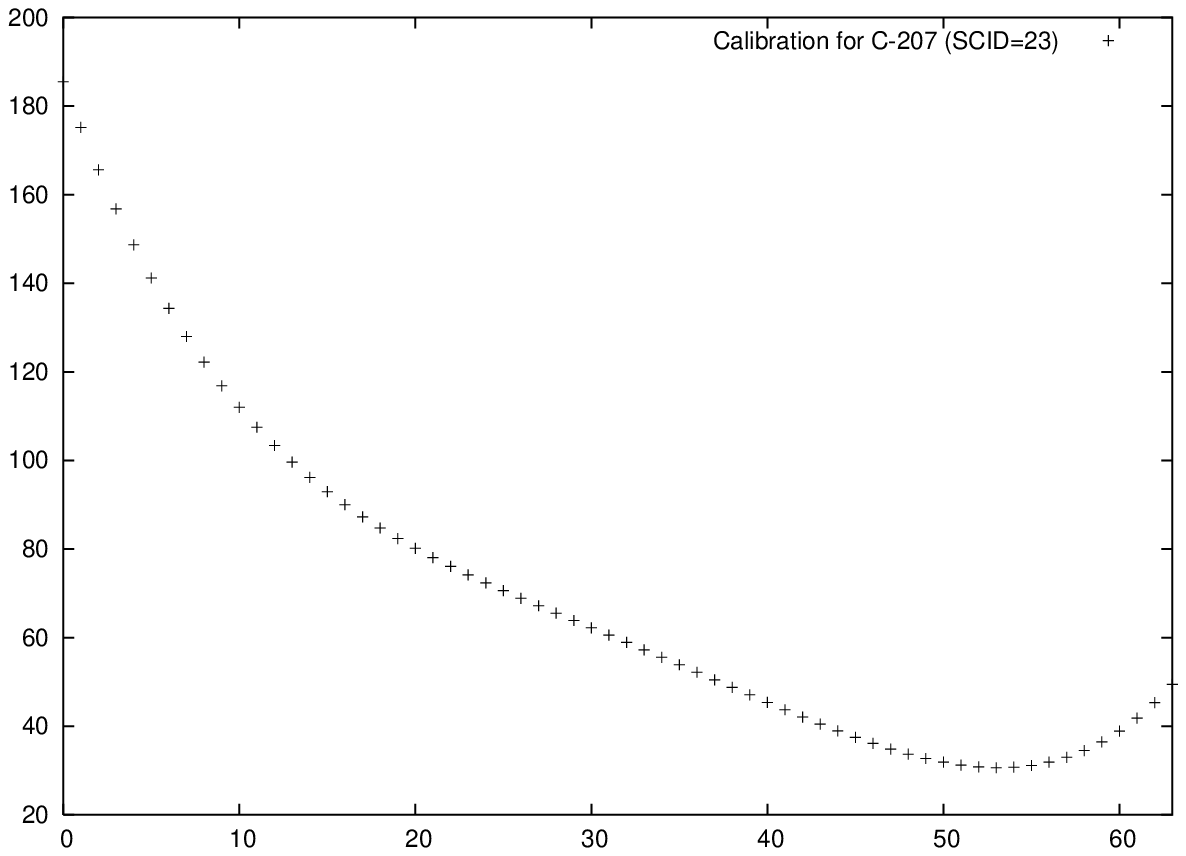, width=80mm}
\end{minipage}
\caption{(Left: TWT A Converter Temperature for Pioneer 10 (in F$^\circ$). The numbers make little sense late in the mission, when the readings are outside the calibrated range, where the calibration polynomial (shown on the right) yields nonsensical values.}
\label{fig:C207_23_calib} 
\end{figure*}
 
Since analog values were encoded in a 6-bit format, this severely limits the resolution of the data. For instance, the resolution of the fin root temperature sensor is approximately 3-4 degrees Fahrenheit, limiting our ability to detect small anomalous temperature changes.
\par
Along with calibration polynomials, the calibrated range for sensor values was also recorded. It is necessary to observe these limits, as in some cases, the calibration polynomial yields invalid results outside the defined range. Consider, for instance, the TWT A Converter Temperature (C-207) reading for Pioneer 10 (plotted on the left in Figure~\ref{fig:C207_23_calib}). At first sight, it may suggest that this temperature unexpectedly began to increase late in the mission. However, if you look at the corresponding calibration table (plotted on the right in Figure~\ref{fig:C207_23_calib}), it becomes evident that the calibration polynomial for this value begins to produce nonsensical readings once the binary value reaches about 53. This is because the calibration limits for this reading are 40 to 125 degrees Fahrenheit, and therefore, the calibration polynomial is only expected to produce meaningful results for values between 8 and 43.

\subsubsection{Software Implementation}

Putting everything together required writing the appropriate software. The data was available thanks to Kellogg's efforts; information required to decode the data was also present. Next came the software implementation.

The initial version of the code was simply a sequential interpreter: reading through the file, it produced human readable output of the records contained therein. Soon, however, it became evident that this form of presentation may not be the most useful for the purposes of data analysis, due to the sheer number of records (nearly a billion telemetry frames in total for the two spacecraft.) For this reason, the second version of the MDR analysis software implemented a different approach: it allowed the user to select a parameter, define timestamps, and then have the software locate the corresponding MDR files and list any values found for the desired parameter that fell between the specified timestamps.

The following demonstrates the (command-line) program in operation:
\begin{verbatim}
$ mdrread -d /data/PIONEER 23 C-201 100501000 100502000
C-201 RTG 1 Fin Root Temperature (deg F)
Pathname/Filename     Pos   Timestamp      DQI  SubCom  Binary  Analog/decoded
23P7301/m2373068.mdr  12    100501272.743  3    Y       43      300.995
23P7301/m2373068.mdr  44    100501464.741  3    Y       42      297.977
23P7301/m2373068.mdr  76    100501656.738  3    Y       42      297.977
23P7301/m2373068.mdr  109   100501848.735  3    Y       43      300.995
\end{verbatim}
This command lists the values for parameter C-201 (RTG 1 Fin Root Temperature) for the period between 4:56:40 AM and 5:13:20 AM on  March 9, 1973. The timestamps are standard UNIX style timestamps, representing the number of seconds elapsed since midnight, January 1, 1970, and supplemented by a decimal part representing thousandths of seconds.

As the output in the example shows, the requested parameter was received four times during this time period, in each case as part of a subcommutator (``Y'' in the fifth column). Both the binary value (as a decimal integer between 0 and 63) and the decoded, calibrated analog value are present in the output.

While this command-line implementation is very efficient when used within script or macro programs, it is a bit cumbersome to use by hand. Therefore, a Web page was also developed that provides a wrapper interface: here, the user can use interactive controls to select the desired parameter and the start/end dates, and output appears in the Web browser. An experimental Web server has been configured that contains a complete copy of the Pioneer MDRs and can be utilized by researchers.

In either case, the output is of a fixed, tab delimited format, suitable for import into spreadsheet programs or other data analysis tools.

So how does the program work? The implementation is fairly straightforward, actually, and can be described as consisting of three distinct parts.

First, several C++ classes are used to define the record format, including the MDR frame, as well as the 192-bit Pioneer data/telemetry records contained therein. Additional structures contain calibration parameters for all decodable values, as well as lookup tables that define which parameter is of what type (analog vs. digital.)

Second, a search logic was implemented that navigates through the directory structures described above, locates the appropriate file(s) that correspond with the requested timestamps, and locates the correct records therein using a fast binary search algorithm.

Third, additional logic is used to decode the records read, identify the record type, and extract the desired telemetry information. If the requested parameter is found, its value is passed to a callback subroutine that is responsible for managing output.

Much of this code has been written in the form of a reusable library, so that it may be utilized in programs yet to be written that implement a different approach towards extracting Pioneer data.

The code at present runs on both Windows and Linux computers. Most of the code is not platform-specific; some portions, notably the directory/file search logic, have two separate implementations, one specific to Windows, while the other a generic UNIX implementation.

\subsection{Available Telemetry Information}
\label{sec:mdr-summary}

To conduct a study of on-board systematics, one can use another critical piece of information that recently became available, namely the Pioneer 10 and 11 spacecraft telemetry data.  The initial studies of the Pioneer anomaly \citep{pioprl,moriond,pioprd,markwardt} and several subsequent papers \citep{pio-standard,stanford,problem_set_05,iap-pioneer} had emphasized the need for a very detailed investigation of the on-board systematics. Other researchers also focused their work on the study of several on-board generated mechanisms that could contribute to an anomalous acceleration of the spacecraft \citep{murphy,katz,scheffer}. 
Most of these investigations of on-board systematics were not very precise. This was due to a set of several reasons, some of them are the shortness of the interval of Doppler data analyzed, the lack of sufficient directional resolution of the anomalous acceleration, and insufficient amount of actual telemetry data from the vehicles. As of 2005, this picture has changed dramatically.  

The growing interest in the Pioneer anomaly helped us to initiate an effort at the NASA Ames Research Center to recover the entire archive of the Pioneer Project documents for the period from 1966 to 2003.  This archive contains all Pioneer 10 and 11 project documents discussing the spacecraft and mission design, fabrication of various components, results of various tests performed during fabrication, assembly, pre-launch, as well as calibrations performed on the vehicles; and also administrative documents including quarterly reports, memoranda, etc. We have also identified most of the maneuver records, spin rate data, significant events of the craft, etc.  

\begin{table}
\begin{center}
\vskip -3pt
\caption{Available parameter set that will be useful for the Pioneer anomaly investigation}
\vskip 3pt
{\small
\begin{tabular}{|l|l|l|}\hline
Parameters&Subsystem&Telemetry words\\\hline\hline
\multicolumn{3}{|l|}{\tt TEMPERATURES}\\\hline
RTG fin root temps&Thermal&$\mathrm{C}_{201}$, $\mathrm{C}_{202}$, $\mathrm{C}_{203}$, $\mathrm{C}_{204}$\\
RTG hot junction temps&Thermal&$\mathrm{C}_{220}$, $\mathrm{C}_{219}$, $\mathrm{C}_{218}$, $\mathrm{C}_{217}$\\
TWT temperatures&Communications&$\mathrm{C}_{205}$, $\mathrm{C}_{206}$, $\mathrm{C}_{207}$, $\mathrm{C}_{228}$, $\mathrm{C}_{223}$, $\mathrm{C}_{221}$\\
Receiver temperatures&Communications&$\mathrm{C}_{222}$, $\mathrm{C}_{227}$\\
Platform temperatures&Thermal&$\mathrm{C}_{301}$, $\mathrm{C}_{302}$, $\mathrm{C}_{304}$, $\mathrm{C}_{318}$, $\mathrm{C}_{319}$, $\mathrm{C}_{320}$\\
PSA temperatures&Thermal&$\mathrm{C}_{225}$, $\mathrm{C}_{226}$\\
Thruster cluster temps&Propulsion&$\mathrm{C}_{309}$, $\mathrm{C}_{326}$, $\mathrm{C}_{310}$, $\mathrm{C}_{311}$, $\mathrm{C}_{312}$, $\mathrm{C}_{328}$, $\mathrm{C}_{325}$\\
SRA/SSA temperatures&ACS&$\mathrm{C}_{303}$, $\mathrm{C}_{317}$\\
Battery temperature&Power&$\mathrm{C}_{115}$\\
Propellant temperature&Propulsion&$\mathrm{C}_{327}$\\
N2 tank temperature&Propulsion&$\mathrm{C}_{130}$\\
Science instr temps&Science&$\mathrm{E}_{101}$, $\mathrm{E}_{102}$, $\mathrm{E}_{109}$, $\mathrm{E}_{110}$, $\mathrm{E}_{117}$, $\mathrm{E}_{118}$, $\mathrm{E}_{125}$, $\mathrm{E}_{128}$, $\mathrm{E}_{201}$, $\mathrm{E}_{209}$, $\mathrm{E}_{213}$, $\mathrm{E}_{221}$\\\hline
\multicolumn{3}{|l|}{\tt VOLTAGES}\\\hline
Calibration voltages&Data handling&$\mathrm{C}_{101}$, $\mathrm{C}_{102}$, $\mathrm{C}_{103}$\\
RTG voltages&Power&$\mathrm{C}_{110}$, $\mathrm{C}_{125}$, $\mathrm{C}_{131}$, $\mathrm{C}_{113}$\\
Battery/Bus voltages&Power&$\mathrm{C}_{106}$, $\mathrm{C}_{107}$, $\mathrm{C}_{117}$, $\mathrm{C}_{118}$, $\mathrm{C}_{119}$\\
TWT voltages&Communications&$\mathrm{C}_{224}$, $\mathrm{C}_{230}$\\
Science instr voltages&Science&$\mathrm{E}_{119}$, $\mathrm{E}_{129}$, $\mathrm{E}_{210}$, $\mathrm{E}_{211}$, $\mathrm{E}_{217}$, $\mathrm{E}_{220}$\\\hline
\multicolumn{3}{|l|}{\tt CURRENTS}\\\hline
RTG currents&Power&$\mathrm{C}_{127}$, $\mathrm{C}_{105}$, $\mathrm{C}_{114}$, $\mathrm{C}_{123}$\\
Battery/Bus currents&Power&$\mathrm{C}_{109}$, $\mathrm{C}_{126}$, $\mathrm{C}_{129}$\\
Shunt current&Power&$\mathrm{C}_{122}$, $\mathrm{C}_{209}$\\
TWT currents&Communications&$\mathrm{C}_{208}$, $\mathrm{C}_{211}$, $\mathrm{C}_{215}$, $\mathrm{C}_{216}$\\
Science instr currents&Science&$\mathrm{E}_{111}$, $\mathrm{E}_{112}$, $\mathrm{E}_{113}$\\\hline
\multicolumn{3}{|l|}{\tt PRESSURE}\\\hline
Propellant pressure&Propulsion&$\mathrm{C}_{210}$\\\hline
\multicolumn{3}{|l|}{\tt OTHER ANALOG}\\\hline
TWT power readings&Communications&$\mathrm{C}_{231}$, $\mathrm{C}_{214}$\\
Receiver readings&Communications&$\mathrm{C}_{111}$, $\mathrm{C}_{212}$, $\mathrm{C}_{232}$, $\mathrm{C}_{121}$, $\mathrm{C}_{229}$, $\mathrm{C}_{213}$\\\hline
\multicolumn{3}{|l|}{\tt BINARY/BIT FIELDS}\\\hline
Conscan&Communications&$\mathrm{C}_{313}$, $\mathrm{C}_{314}$, $\mathrm{C}_{315}$, $\mathrm{C}_{316}$\\
Stored commands&Electrical&$\mathrm{C}_{305}$, $\mathrm{C}_{306}$, $\mathrm{C}_{307}$\\
Thruster pulse counts&Propulsion&$\mathrm{C}_{329}$, $\mathrm{C}_{321}$, $\mathrm{C}_{322}$, $\mathrm{C}_{330}$\\
Status bits&Data handling&$\mathrm{C}_{104}$\\
&Power&$\mathrm{C}_{128}$\\
&Electrical&$\mathrm{C}_{120}$, $\mathrm{C}_{132}$, $\mathrm{C}_{324}$, $\mathrm{C}_{332}$\\
&Communications&$\mathrm{C}_{308}$\\
Power switches&Electrical&$\mathrm{C}_{108}$, $\mathrm{C}_{124}$\\
Roll attitude&Data handling&$\mathrm{C}_{112}$, $\mathrm{C}_{116}$\\
Precession&ACS&$\mathrm{C}_{403}$, $\mathrm{C}_{411}$, $\mathrm{C}_{412}$, $\mathrm{C}_{415}$, $\mathrm{C}_{416}$, $\mathrm{C}_{422}$, $\mathrm{C}_{423}$, $\mathrm{C}_{424}$, $\mathrm{C}_{425}$, $\mathrm{C}_{428}$, $\mathrm{C}_{429}$, $\mathrm{C}_{430}$\\
Spin/roll&Data handling&$\mathrm{C}_{405}$, $\mathrm{C}_{406}$, $\mathrm{C}_{407}$, $\mathrm{C}_{408}$, $\mathrm{C}_{417}$\\
Delta V&ACS&$\mathrm{C}_{413}$, $\mathrm{C}_{414}$, $\mathrm{C}_{426}$\\
ACS status&Propulsion&$\mathrm{C}_{409}$\\
&ACS&$\mathrm{C}_{410}$, $\mathrm{C}_{427}$, $\mathrm{C}_{431}$, $\mathrm{C}_{432}$\\
Star sensor&ACS&$\mathrm{C}_{404}$, $\mathrm{C}_{419}$, $\mathrm{C}_{420}$, $\mathrm{C}_{421}$\\\hline
\multicolumn{3}{|l|}{\tt SCIENCE INSTRUMENTS}\\\hline
Status/housekeeping&Science&$\mathrm{E}_{108}$, $\mathrm{E}_{124}$, $\mathrm{E}_{130}$, $\mathrm{E}_{202}$, $\mathrm{E}_{224}$, $\mathrm{E}_{131}$, $\mathrm{E}_{132}$, $\mathrm{E}_{208}$\\
JPL/HVM readings&Science&$\mathrm{E}_{103}$, $\mathrm{E}_{104}$, $\mathrm{E}_{105}$, $\mathrm{E}_{106}$, $\mathrm{E}_{107}$, $\mathrm{E}_{203}$, $\mathrm{E}_{204}$, $\mathrm{E}_{205}$\\
UC/CPI readings&Science&$\mathrm{E}_{114}$, $\mathrm{E}_{115}$, $\mathrm{E}_{116}$, $\mathrm{E}_{206}$, $\mathrm{E}_{212}$, $\mathrm{E}_{214}$, $\mathrm{E}_{215}$, $\mathrm{E}_{216}$\\
GE/AMD readings&Science&$\mathrm{E}_{122}$, $\mathrm{E}_{123}$, $\mathrm{E}_{222}$, $\mathrm{E}_{223}$\\
GSFC/CRT readings&Science&$\mathrm{E}_{126}$, $\mathrm{E}_{127}$\\
LaRC/MD readings&Science&$\mathrm{E}_{207}$\\\hline
\end{tabular}}
\label{tab:telemetry}
\end{center}
\vskip -10pt
\end{table}

Currently, only a limited subset of the Pioneer 10 and 11 spacecraft science data is available for on-line download from the NSSDC,  {\tt http://nssdc.gsfc.nasa.gov/planetary/pioneer10-11.html}. However, the science data is only a small part of a much larger set of information that also contains engineering telemetry, measured by 114 analog sensors and numerous other devices placed on-board of each spacecraft.  This massive amount of data was sent down to the ground where it was recorded by DSN controllers on 9-track magnetic tape.  In addition to the data on scientific instruments, these sensors provided information on the state of the main spacecraft components, such as data handling, power, electrical distribution, communication, thermal, propulsion, attitude control, and antenna subsystems. This entire set of measurements from various subsystems constitutes the raw data stored as the Master Data Records (MDR) which, until recently, were not available to the public. 
This data set is a unique source of information that was never used in conjunction with spacecraft navigation. The use of the spacecraft telemetry may be one of the most critical element that will help to identify the origin of the Pioneer anomaly. It is our intention to use the set of MDRs together with the Doppler data to precisely adjust the solution for $a_P$ for the effect of the on-board systematics.

Table~\ref{tab:telemetry} summarizes all available telemetry values in the C (engineering) and E (science) telemetry formats. The MDRs also contain a complete set of science readings that were telemetered in the A, B, and D formats; these, however, are unlikely to be of value to us, and in any case, if science readings are required for the analysis, the processed science data at the NSSDC may be of much more use.

As this table demonstrates, telemetry readings can be broadly categorized as temperature, voltage, current readings; other analog readings; various binary counters, values, and bit fields; and readings from science instruments. It is anticipated that temperature and electrical readings will be of the greatest use to us, as they will help us establish a detailed thermal profile of the spacecrafts' major components. Some binary readings may also be useful; for instance, thruster pulse count readings may help us better understand maneuvers and their impact on the spacecrafts' trajectories. It is important to note, however, that some readings may not be available and others may not be trusted. For example, thruster pulse count readings are only telemetered when the spacecraft is commanded to send readings in accelerated engineering formats; since these formats were rarely used late in the mission, we may not have pulse count readings for many maneuvers. Regarding reliability, we know from mission status reports about the failure of the sun and star sensors; these failures invalidate many readings from that point onward. Thus it is important to view telemetry readings in context before utilizing them as source data for our investigation.

\begin{figure*}[t!]
\hskip -6pt 
\begin{minipage}[b]{.46\linewidth}
\centering \psfig{file=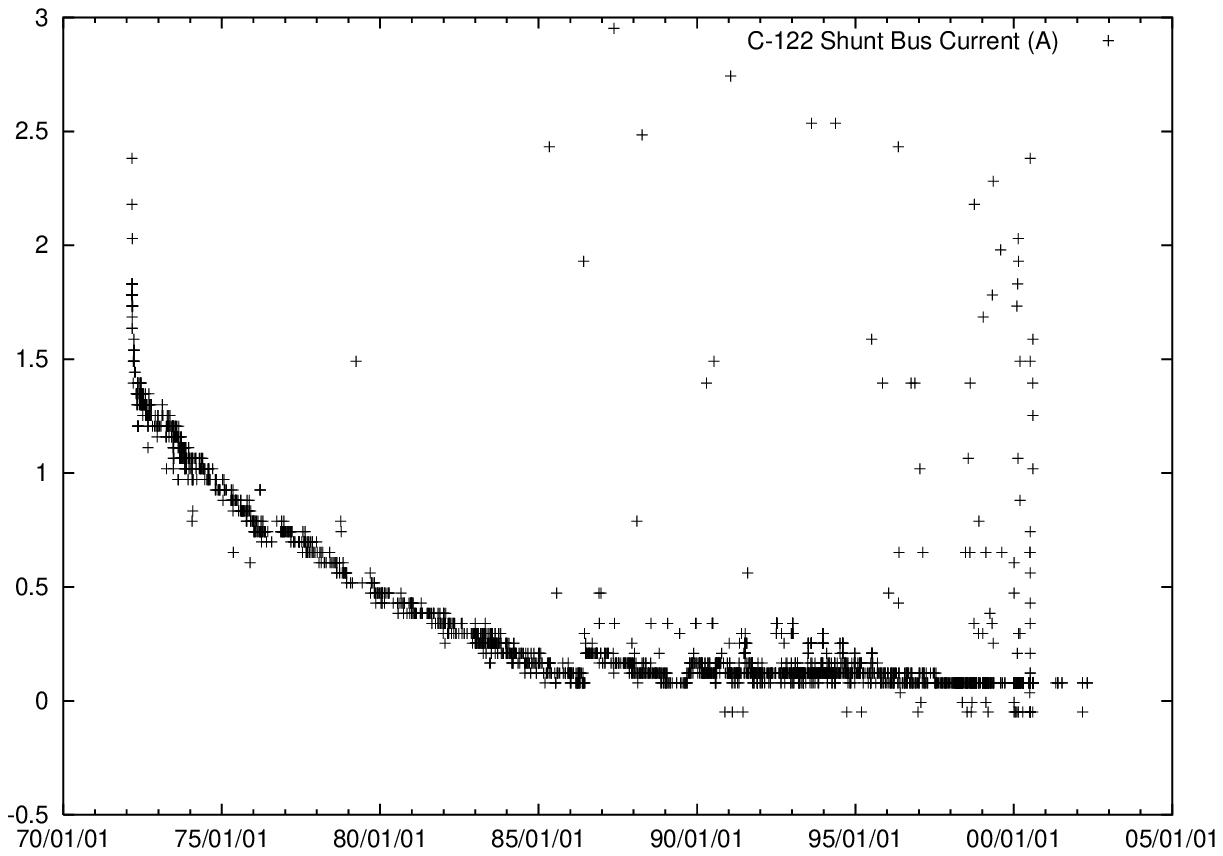, width=80mm}
\end{minipage} 
\hskip 16pt
\begin{minipage}[b]{.46\linewidth}
\centering \psfig{file=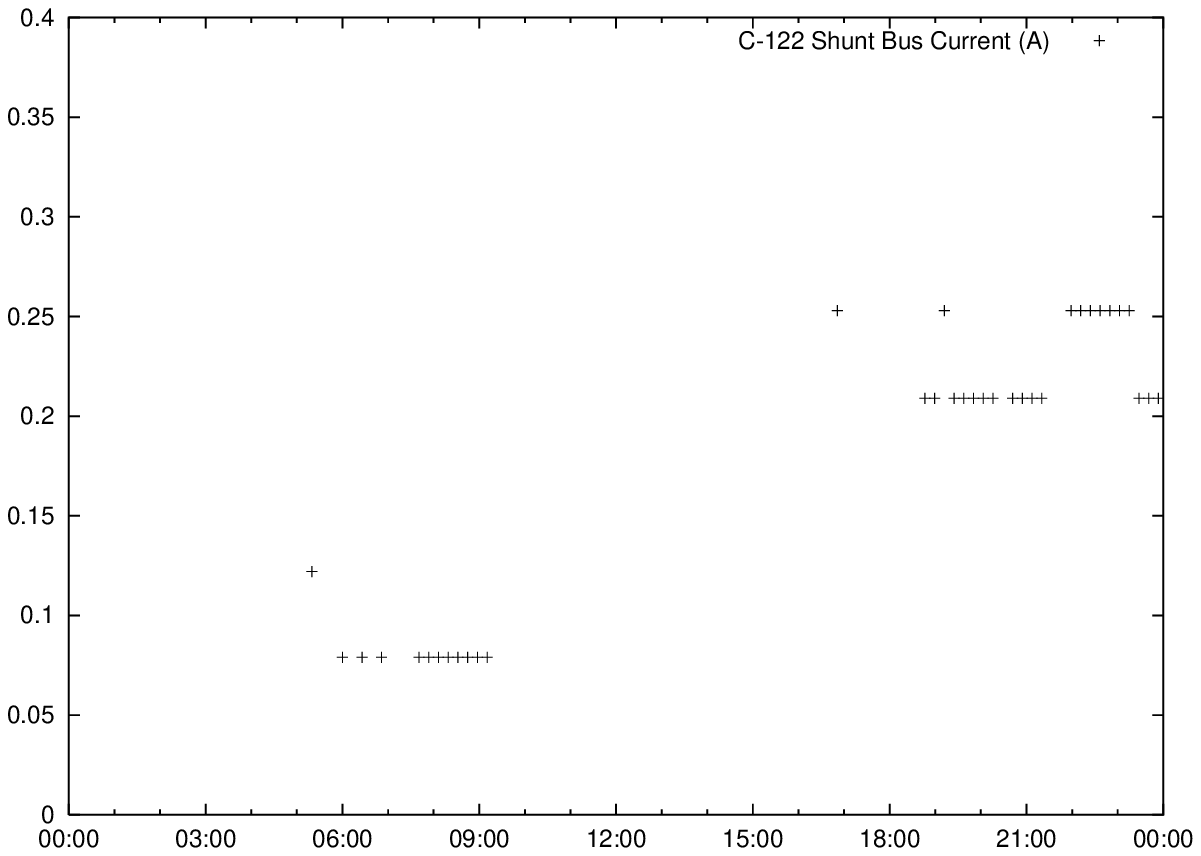, width=80mm}
\end{minipage}
\caption{Left: shunt current history (in A) on Pioneer 10 (telemetry word C-122). Right: change in shunt current  on-board of Pioneer~10 on June 24, 1986, as the HVM instrument is powered down. This change is seen as a major discontinuity near the middle of the left plot.
 \label{fig:C122_HVM}}
\vskip 5pt 
\hskip -6pt 
\begin{minipage}[b]{.46\linewidth}
\centering \psfig{file=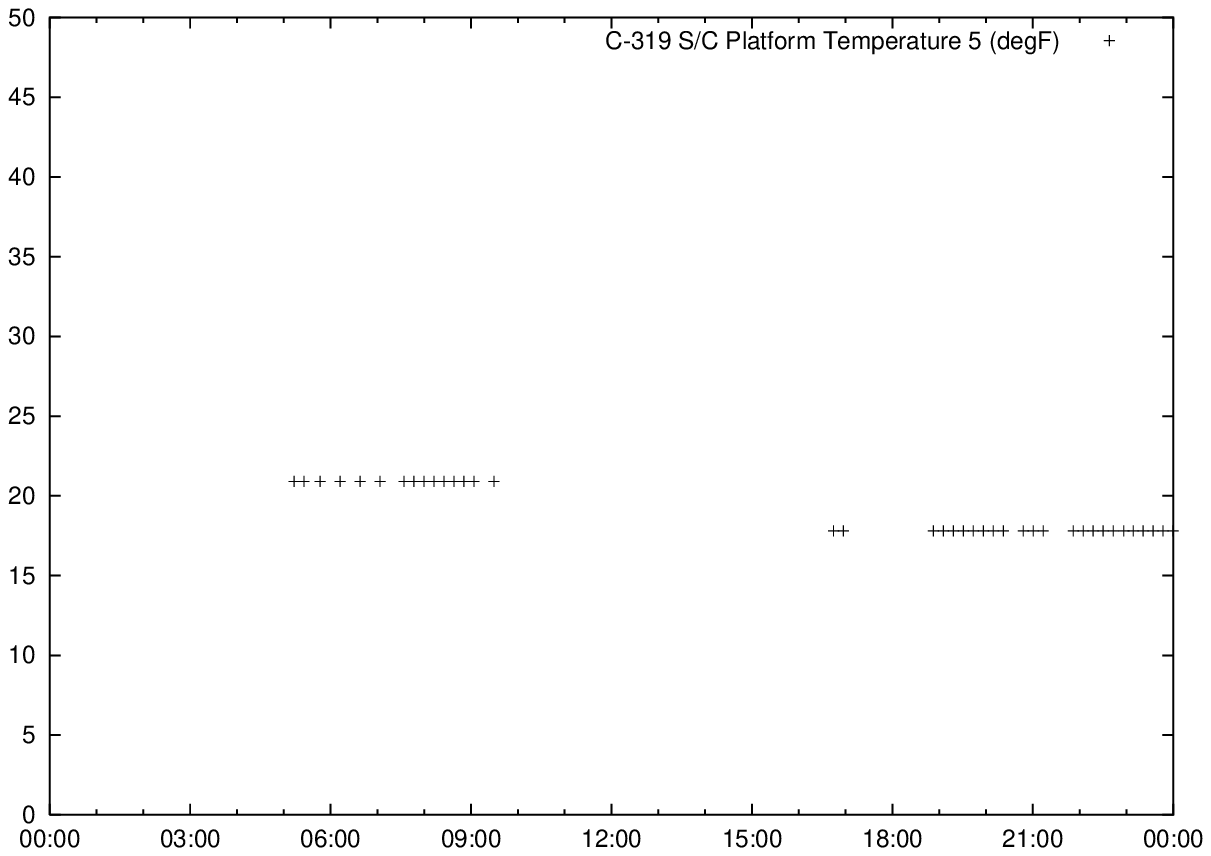, width=80mm}
\end{minipage} 
\hskip 20pt
\begin{minipage}[b]{.46\linewidth}
\centering \psfig{file=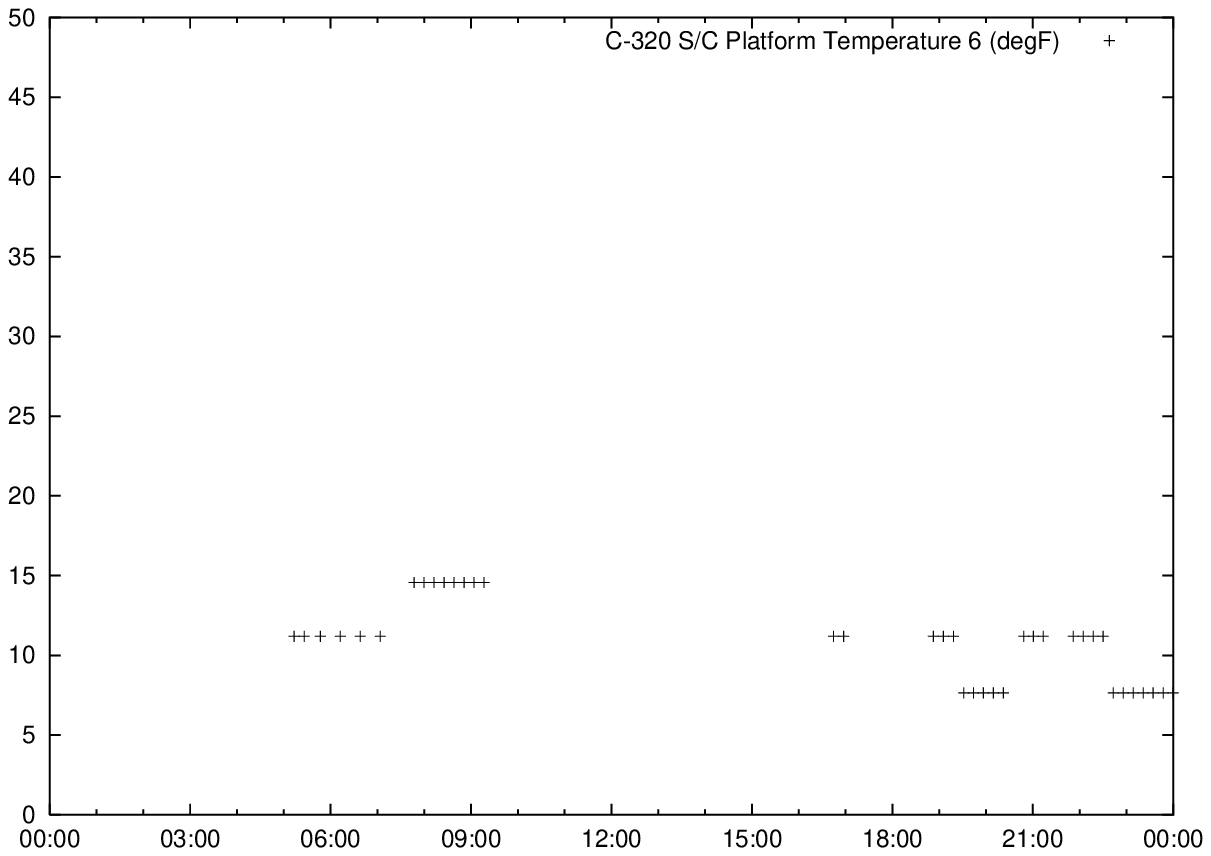, width=80mm}
\end{minipage}
\caption{Temperature changes (in $^\circ$F) inside the science instrument compartment on board Pioneer~10 as the HVM instrument is powered down on June 24, 1986. Both sensors in this compartment (telemetry words C-319, left, and C-320, right) show a decrease in temperature.
 \label{fig:HVMSCI}}
\vskip -5pt 
\end{figure*}
\begin{figure*}[ht!]
\hskip -5pt
\begin{minipage}[b]{.46\linewidth}
\centering \psfig{file=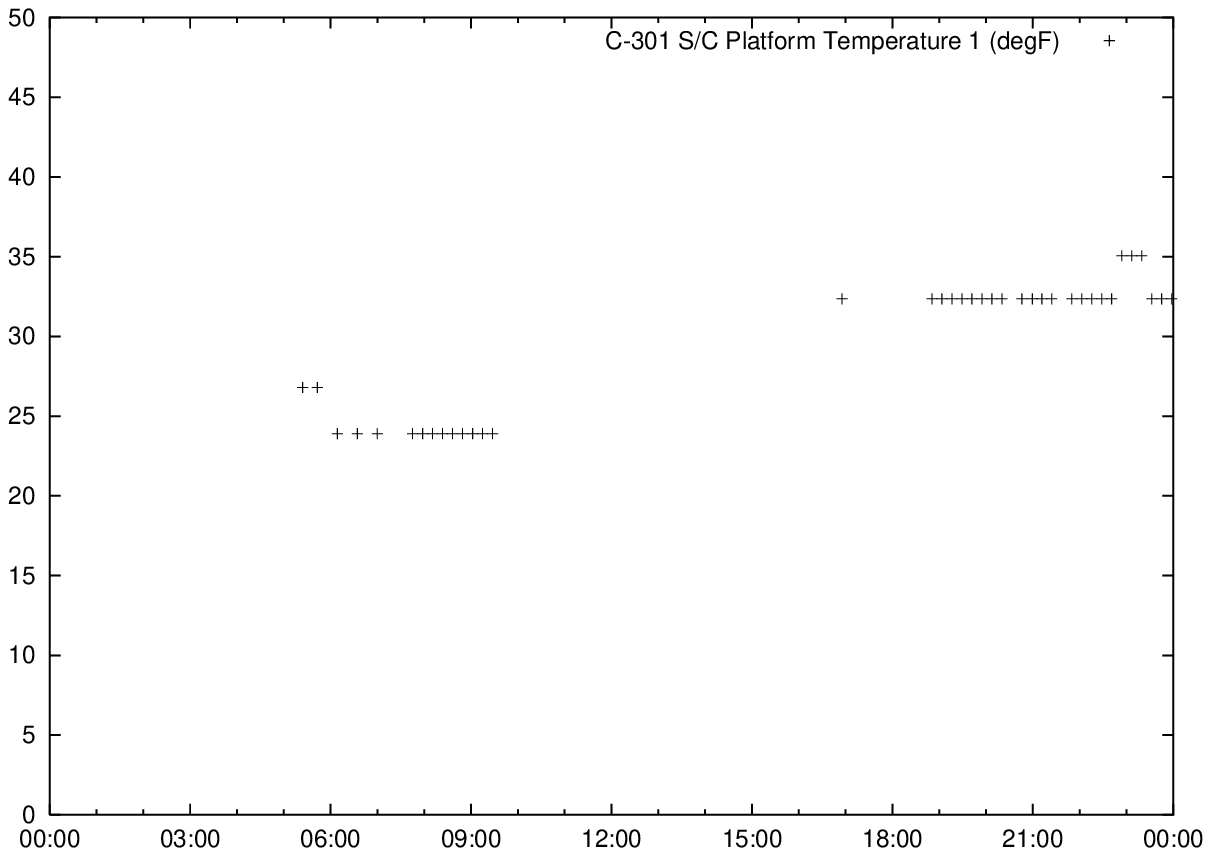, width=79mm}
\end{minipage} 
\hskip 20pt
\begin{minipage}[b]{.46\linewidth}
\centering \psfig{file=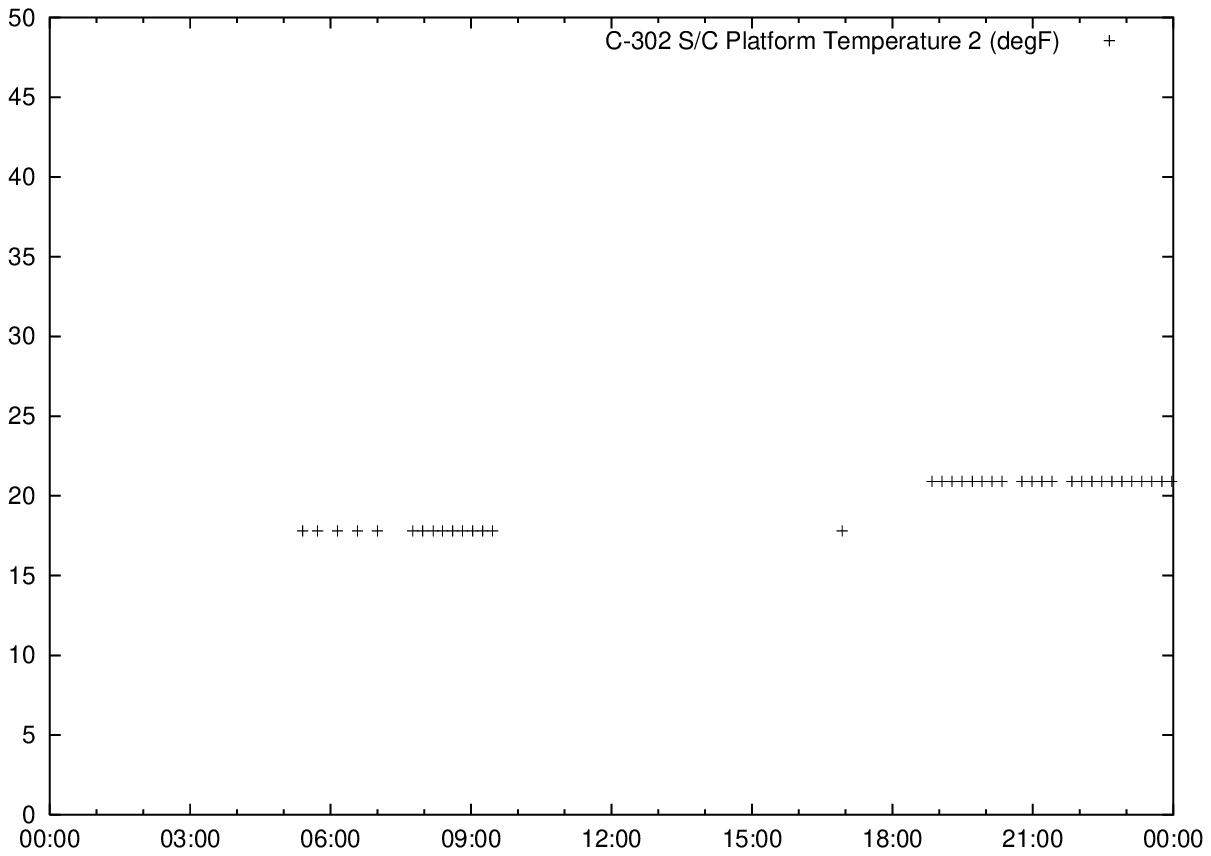, width=79mm}
\end{minipage}
\vskip 10pt\hskip -5pt
\begin{minipage}[b]{.46\linewidth}
\centering \psfig{file=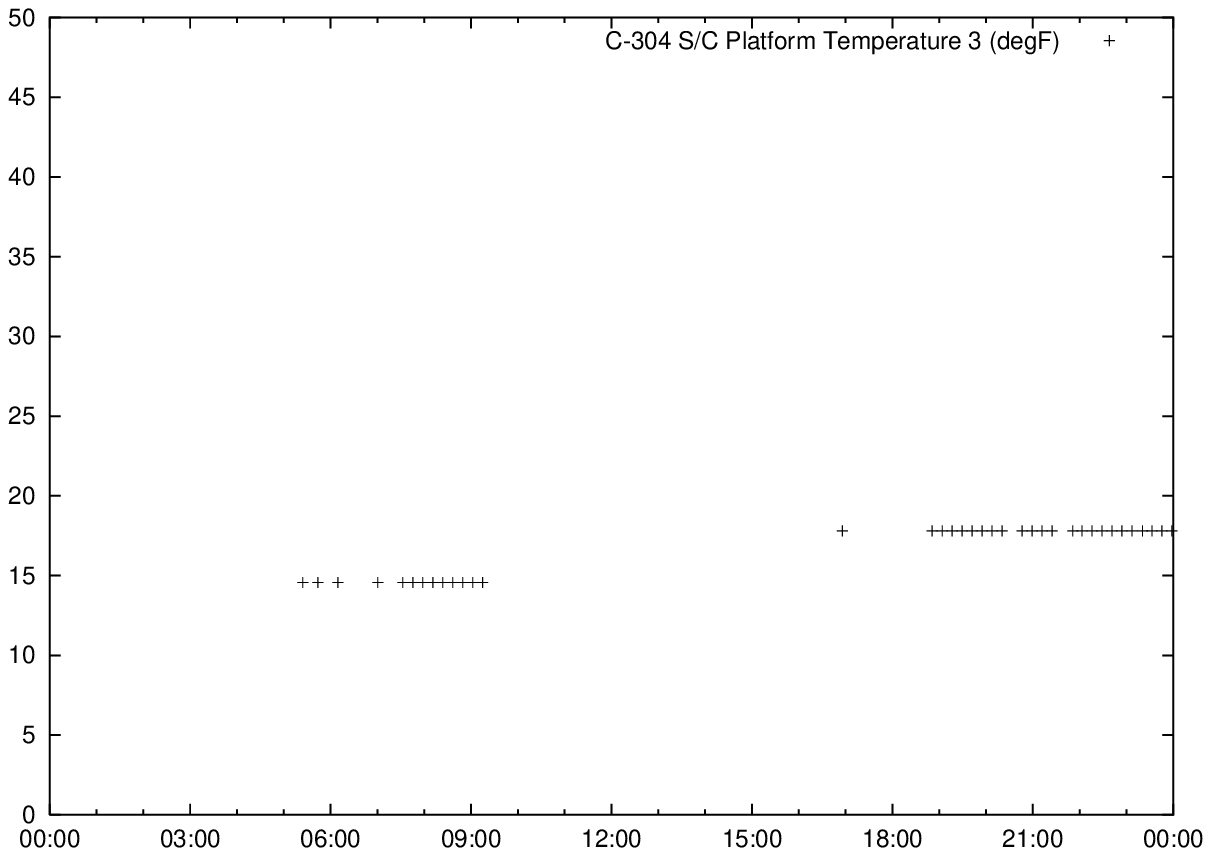, width=79mm}
\end{minipage} 
\hskip 20pt
\begin{minipage}[b]{.46\linewidth}
\centering \psfig{file=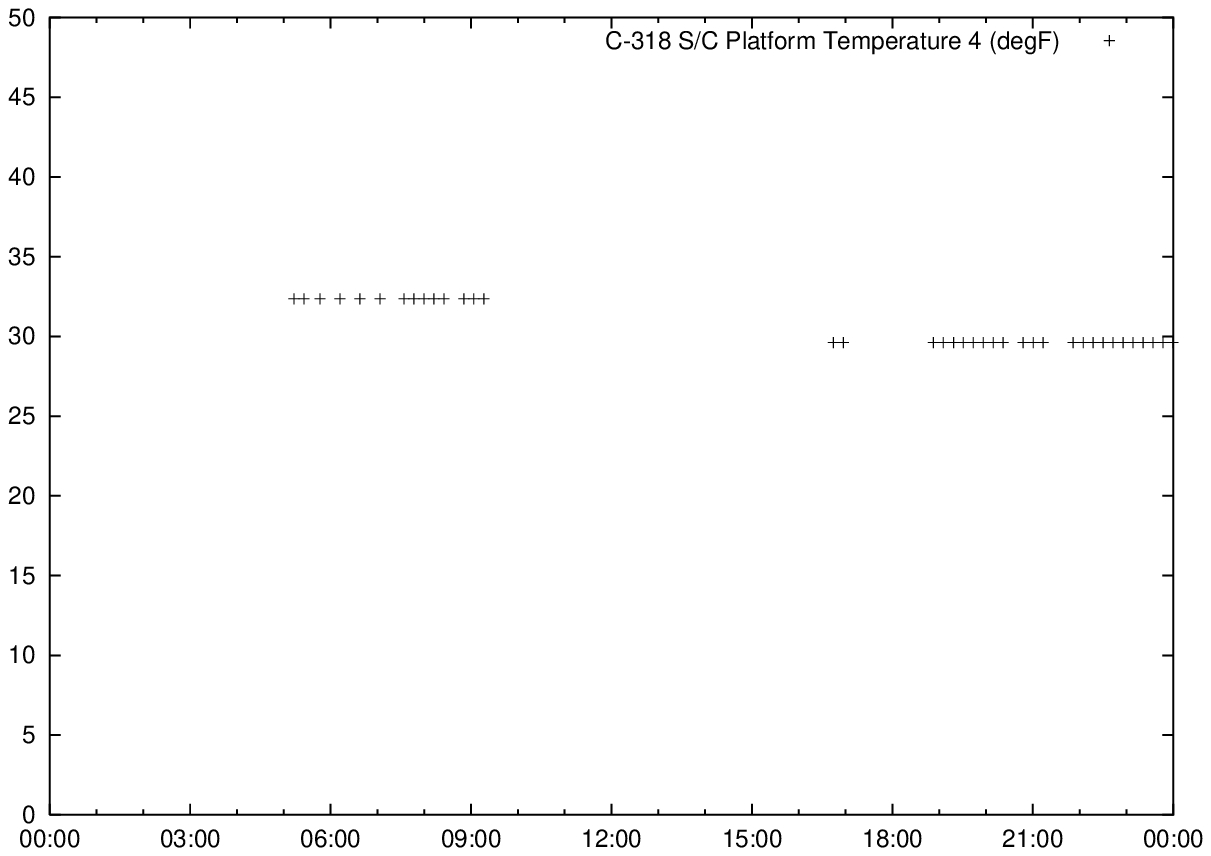, width=79mm}
\end{minipage}
\caption{Temperature changes (in $^\circ$F) inside the main compartment (telemetry words C-301, C-302, C-304, and C-318) on board Pioneer 10 as the HVM instrument is powered down on June 24, 1986. Sensors nearer the power supply electronics show an increase in temperature; the last sensor, farthest away from the power supply electronics and relatively near the science compartment where the powered down instrument is located, shows a decrease.}
\label{fig:HVMMAIN}
\end{figure*}

\subsection{What Can Telemetry Tell Us About the Spacecraft?}

Simply put, telemetry information provides us with a comprehensive and redundant picture of the spacecraft's state. Multiple readings help us not only confirm spacecraft events, but also see the effect of specific events on related spacecraft subsystems.

For example, take June 24, 1986, when the HVM instrument was powered down on board of Pioneer~10. This event can be confirmed from direct telemetry, reading engineering word C-108 that contains several science instrument power states. The first bit represents the HVM instrument:

\begin{verbatim}
$ mdrread -d /data/PIONEER/ 23 C-108 519988907 520015788
C-108 Instrument Power (HVM|PA|CPI|GTT|CRT|0) ((null))
Pathname/Filename       Pos     Timestamp       DQI     SubCom  Binary  Decoded
23P8602/m2386175.mdr    1843    519988908.152   3       Y       62      111110
23P8602/m2386175.mdr    1973    520015788.442   3       Y       30      011110
.end
\end{verbatim}

The timestamps of 519988907 and 520015788 correspond with 09:21:48 and 16:49:48, respectively, on June 24, 1986. (There was, it appears, a gap in the receiver coverage during this event, which is why several hours elapsed between these two telemetry frames.)

Telemetry can tell us a lot more, however, then just the power state of this instrument. First, we would expect that a decrease in spacecraft load by 4\,W should be accompanied by a corresponding increase in the shunt regulator current (excess power available on board was consumed by the shunt circuit.) Indeed, if we look at spacecraft word C-122 (Shunt Bus Current -- see Figure~\ref{fig:C122_HVM}) we see an increase of approximately 0.13\,A, which corresponds with a power increase of approximately 3.6\,W. (The sensor resolution is rather coarse; the shunt current is measured in steps of $\sim0.05\,\mathrm{A}$, which corresponds with steps of $\sim1.4\,\mathrm{W}$ at the nominal 28VDC bus voltage.)

In addition to changes in shunt current, we also expect changes in electronics platform temperatures. Temperatures near the HVM electronics should decrease, while temperatures near the shunt regulator electronics should increase. The HVM electronics box is in the vicinity of sensor location 17 on Figure~\ref{fig:sensors}, so we would expect a decrease in temperature at location 5 and a less significant decrease or no decrease at location 6. This is indeed what the telemetry shows (Figure~\ref{fig:HVMSCI}.) The power supply and control electronics boxes are located at the lower left side of the main compartment in Figure~\ref{fig:sensors}, so we may expect to see an increase in temperatures at sensor locations 1, 2 and 3, with the greatest increase at location 1; at location 4, however, we would expect to see no change or a decrease, as this sensor's location is nearer to the powered down instrument than to the power supply electronics. This is indeed confirmed by on-board telemetry, as seen in Figure~\ref{fig:HVMMAIN}.

\begin{figure*}[ht!]
\hskip -6pt
\begin{minipage}[b]{.46\linewidth} 
\centering \psfig{file=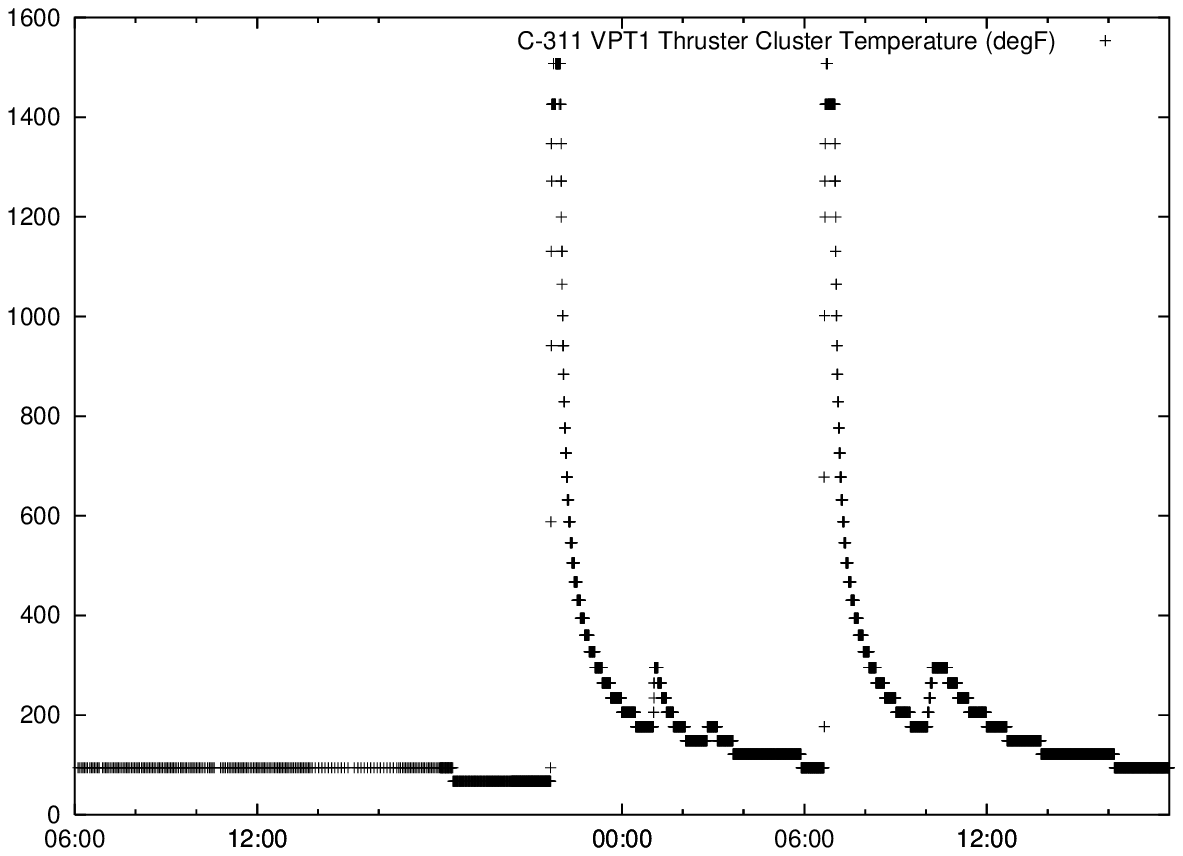, width=80mm}
\end{minipage} 
\hskip 20pt
\begin{minipage}[b]{.46\linewidth}
\centering \psfig{file=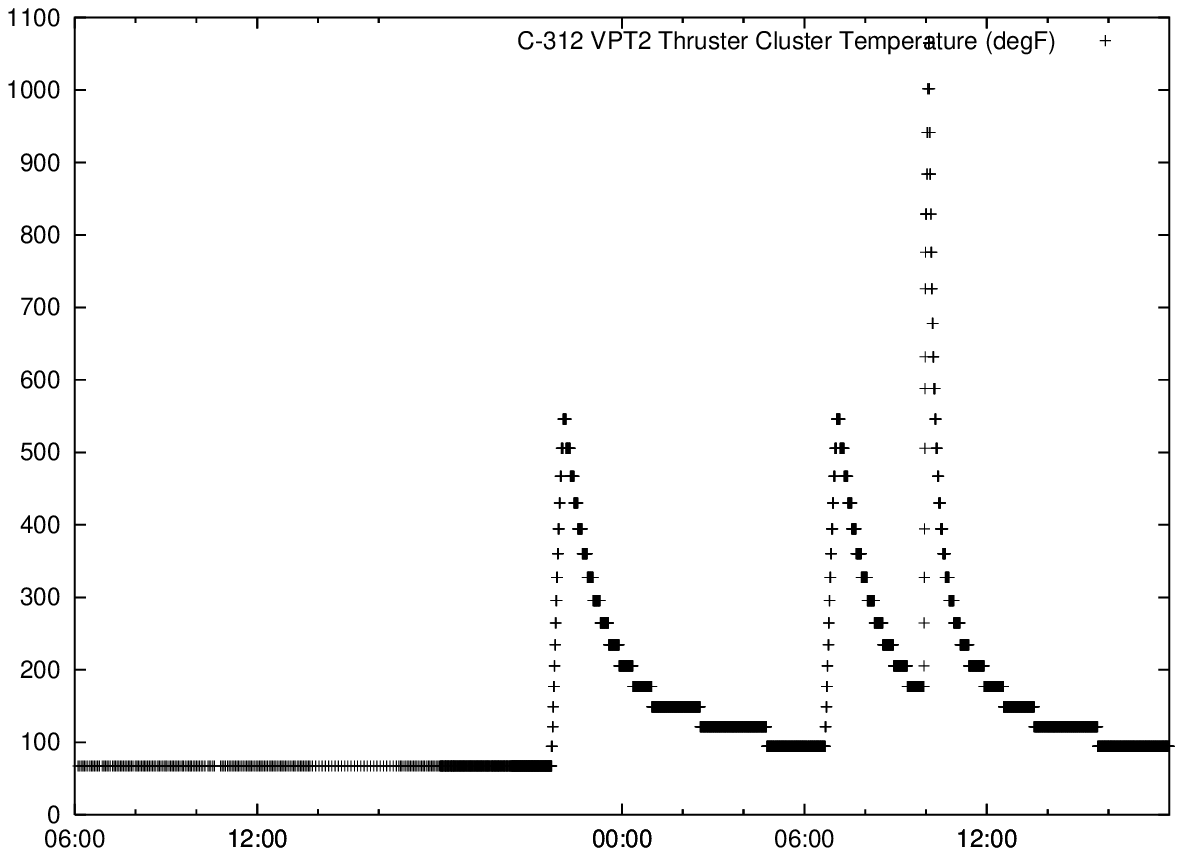, width=80mm}
\end{minipage}
\vskip 10pt\hskip -6pt
\begin{minipage}[b]{.46\linewidth}
\centering \psfig{file=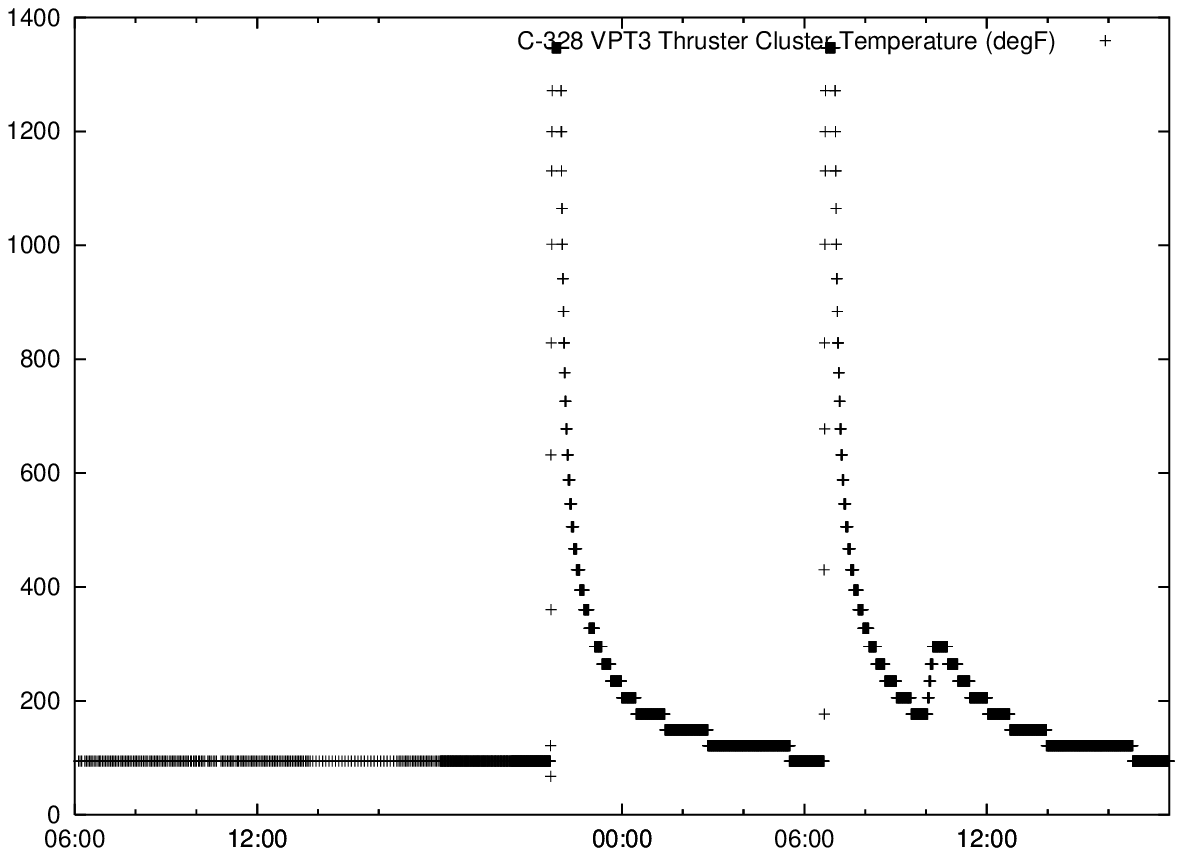, width=80mm}
\end{minipage} 
\hskip 20pt
\begin{minipage}[b]{.46\linewidth}
\centering \psfig{file=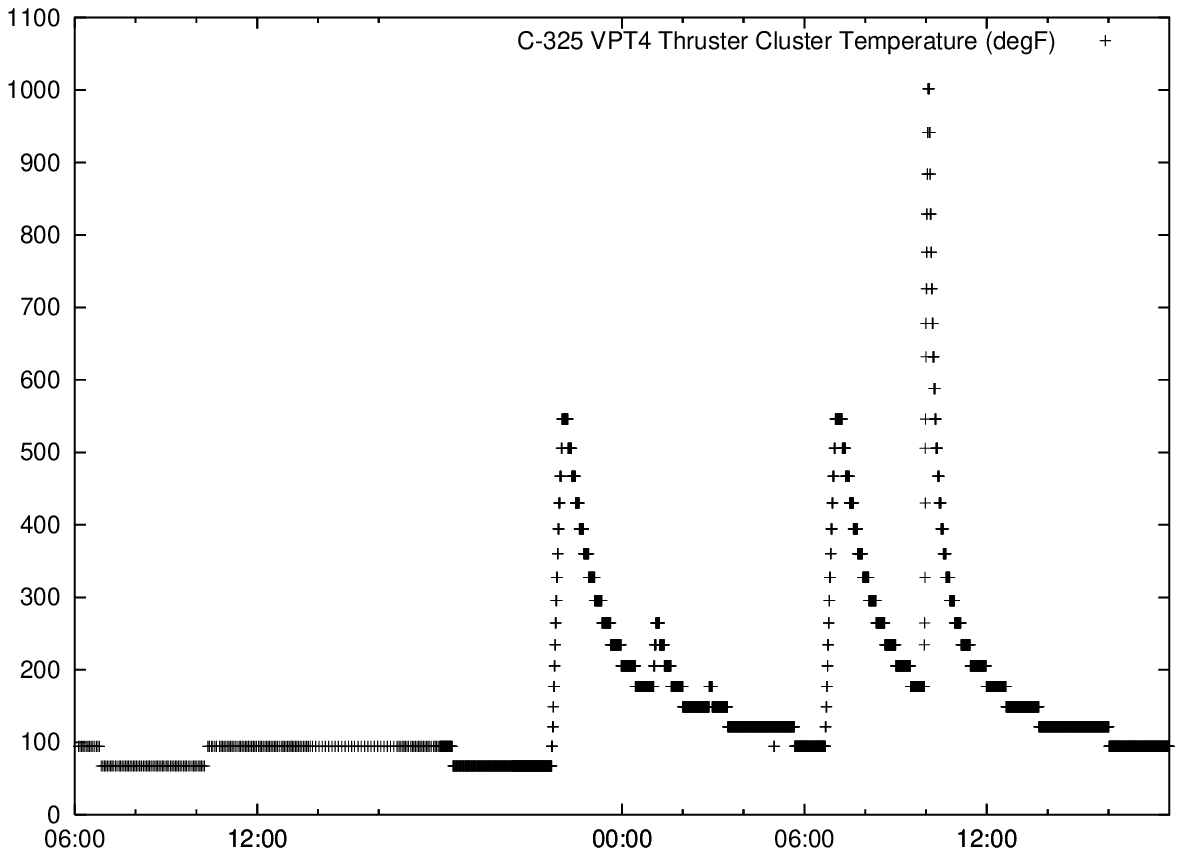, width=80mm}
\end{minipage}
\caption{Changes in Pioneer 11's thruster temperatures (telemetry words C-311, C-312, C-328, and C-325; in $^\circ$F) on April 19-20, 1974, as a result of a major course correction maneuver.}
\label{fig:P11_VPT}
\end{figure*}
\begin{figure*}[ht!]
\hskip -6pt 
\begin{minipage}[b]{.46\linewidth}
\centering \psfig{file=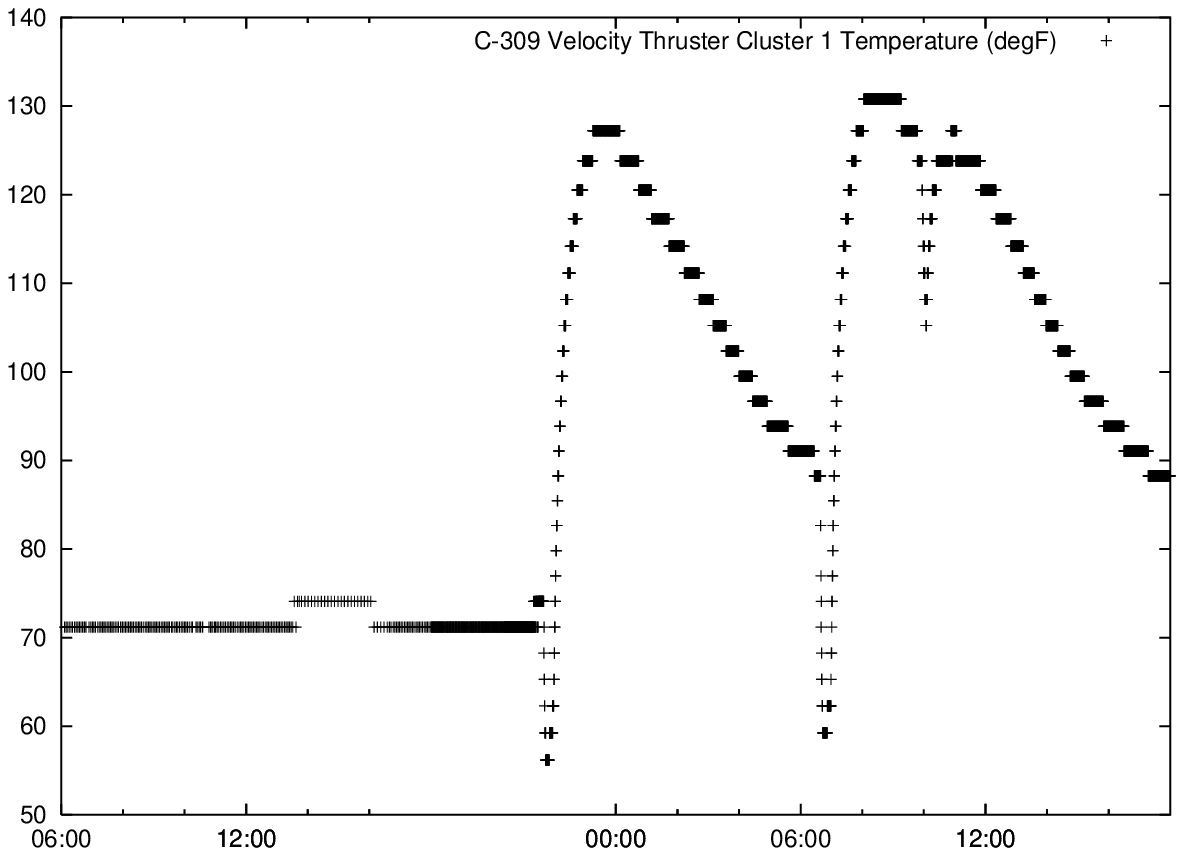, width=80mm}
\end{minipage} 
\hskip 20pt
\begin{minipage}[b]{.46\linewidth}
\centering \psfig{file=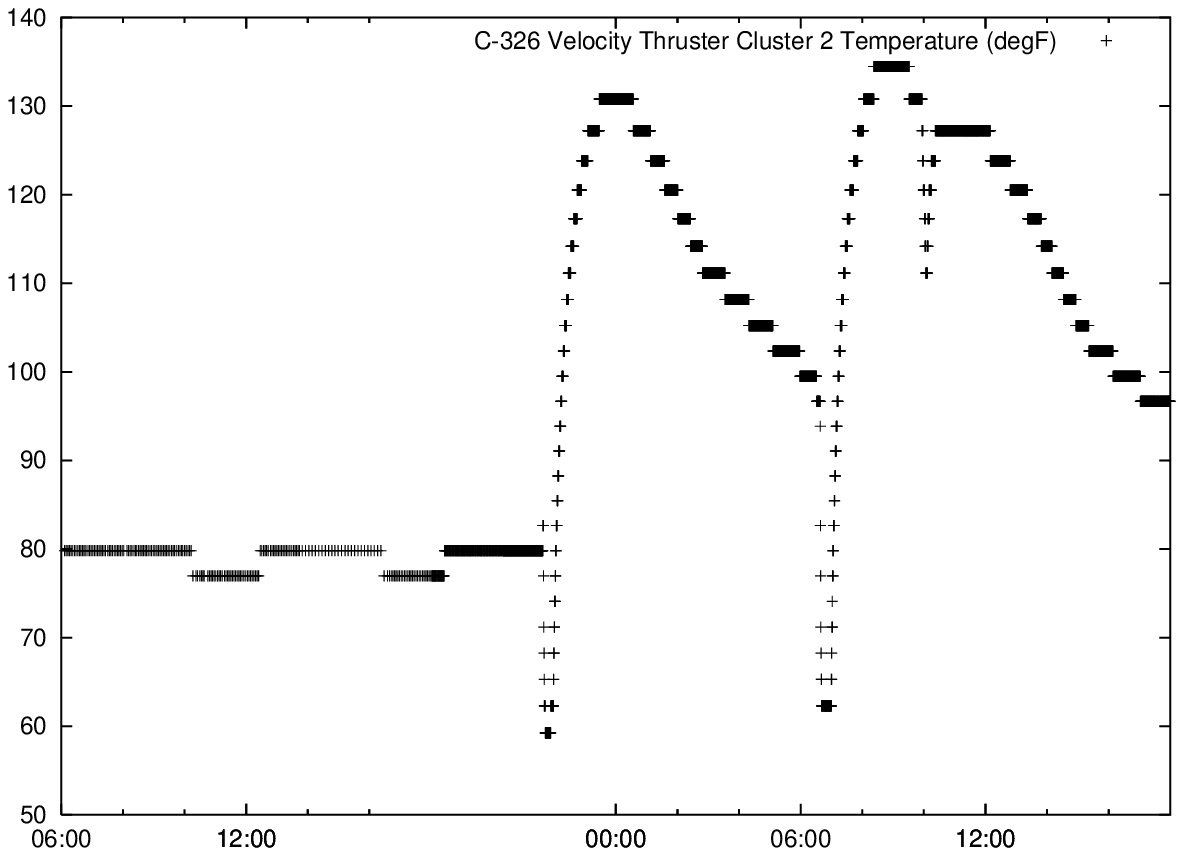, width=80mm}
\end{minipage}
\caption{Temperature changes (telemetry words C-309 and C-326; in $^\circ$F) at Pioneer 11's thruster cluster assemblies on April 19-20, 1974, during a major course correction maneuver.}
 \label{fig:P11_clusters}
\vskip 0pt 
\end{figure*}
\begin{figure*}[ht!]
\hskip -6pt 
\begin{minipage}[b]{.46\linewidth}
\centering \psfig{file=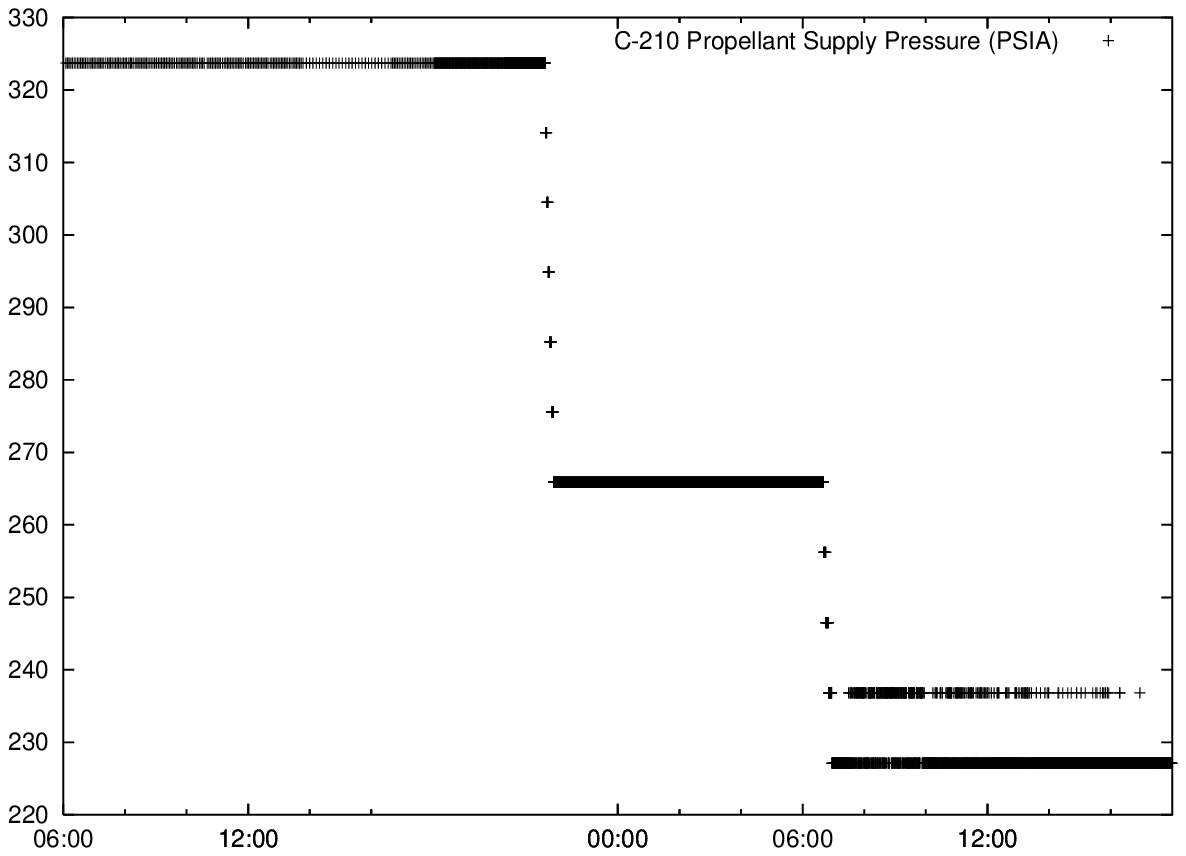, width=80mm}
\end{minipage} 
\hskip 20pt
\begin{minipage}[b]{.46\linewidth}
\centering \psfig{file=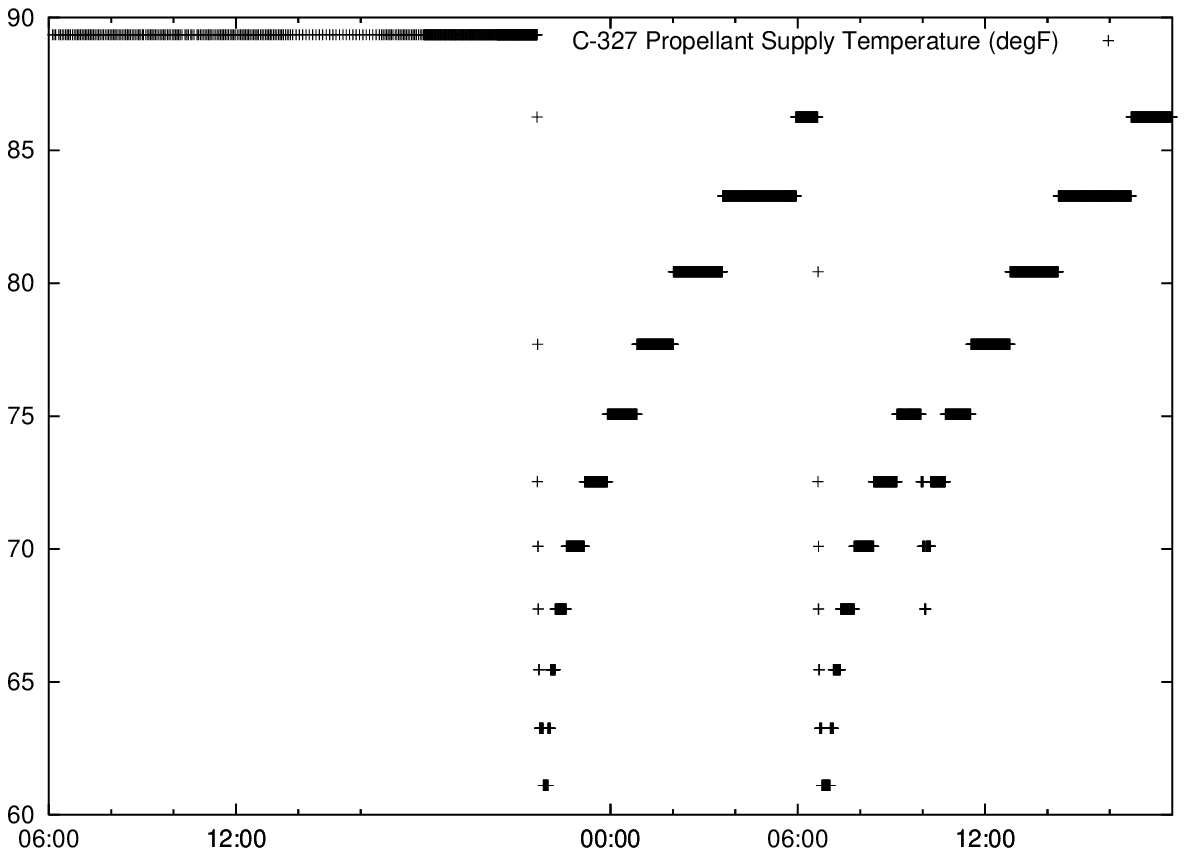, width=80mm}
\end{minipage}
\caption{Changes in Pioneer 11's propellant supply pressure 
(telemetry word C-210; in PSIA, left) and temperature (telemetry word C-327; in $^\circ$F, right) on April 19-20, 1974, during a major course correction maneuver.}
 \label{fig:P11_prop}
\vskip -5pt 
\end{figure*}

As another example, on April 19-20, 1974, Pioneer 11 performed its main course correction maneuver that set it on the appropriate trajectory for a Jupiter swingby to facilitate its eventual encounter with Saturn. Even though thruster pulse counts were not telemetered at this time, we have multiple parameters to confirm this maneuver.

First, we have temperature readings at the individual thrusters: Figure~\ref{fig:P11_VPT} shows the sharp peaks at each of the four individual velocity and precession thrusters (VPTs) on board Pioneer 11.

A smaller corresponding increase can also be seen at the sensors at the two cluster thruster assemblies, as seen in Figure~\ref{fig:P11_clusters}.

Lastly, as this was a major maneuver consuming a significant portion of Pioneer 11's fuel, we would expect to see changes in the readings at the propellant tank. Indeed, we can confirm (Figure~\ref{fig:P11_prop}) that the propellant supply pressure decreased substantially, while the propellant temperature dropped temporarily while the tank was being depleted, only to return to the equilibrium reading a few hours later.

As these examples demonstrate, on-board telemetry not only gives us a detailed picture of the spacecraft and its subsystems, but this picture is redundant: electrical, thermal, logic state and other readings help us examine the same event from a multitude of perspectives.

\section{A Strategy to Find the Origin of the Pioneer Anomaly} 
\label{sec:objectives}

This new extended data set will enable us to investigate the Pioneer anomaly with the entire available Pioneer 10/11 radiometric Doppler data. The objectives of this new investigation of the Pioneer anomaly will be sixfold: 
\begin{enumerate}[(i)]
\item To analyze the early mission data; the goal would be  to determine the true direction of the anomaly and thus, its origin, 
\item To study the physics of the planetary encounters; the goal would be to learn more about the onset of the anomaly (e.g. the Pioneer 11's Saturn flyby),  
\item To study the temporal evolution of $a_P$ with the entire data set; the goal would be a better determination of the temporal behavior of the anomaly, 
\item To perform a comparative analysis of individual anomalous accelerations for the two Pioneers with the data taken from similar heliocentric distances which could highlight properties of $a_P$, and 
\item To investigate the on-board systematics with recently recovered MDRs; the goal here would be to investigate the effect of on-board systematics on the solution for the Pioneer anomaly obtained with the Doppler data, and, finally
\item To build a thermal/electrical/dynamical model of the Pioneer vehicles and verify it with the actual data from the MDRs; the goal here would be to develop a model to be used to calibrate the Doppler anomaly with respect to the on-board sources of dynamical noise.
\end{enumerate}
The objectives presented above not entirely independent from each other; by putting them on this list, we wanted to identify the main areas that we will focus our upcoiming investigation. In this section we will discuss these objectives in more detail.

\begin{figure*}[t!]
 \begin{center}
\noindent   
\psfig{figure=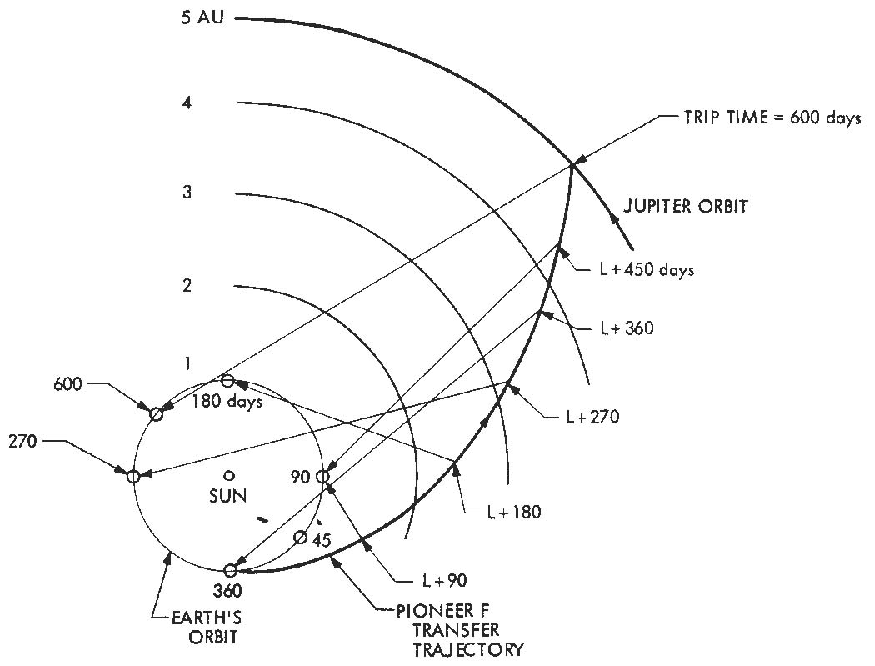,width=105mm}
\end{center}
\vskip -10pt 
  \caption{
Proposed directions (along the spin and antenna axes) from the Pioneer F spacecraft (to become Pioneer 10) toward the Earth.
 \label{fig:antenna-point}}
\vskip -10pt 
%
 \begin{center}
\noindent   
\psfig{figure=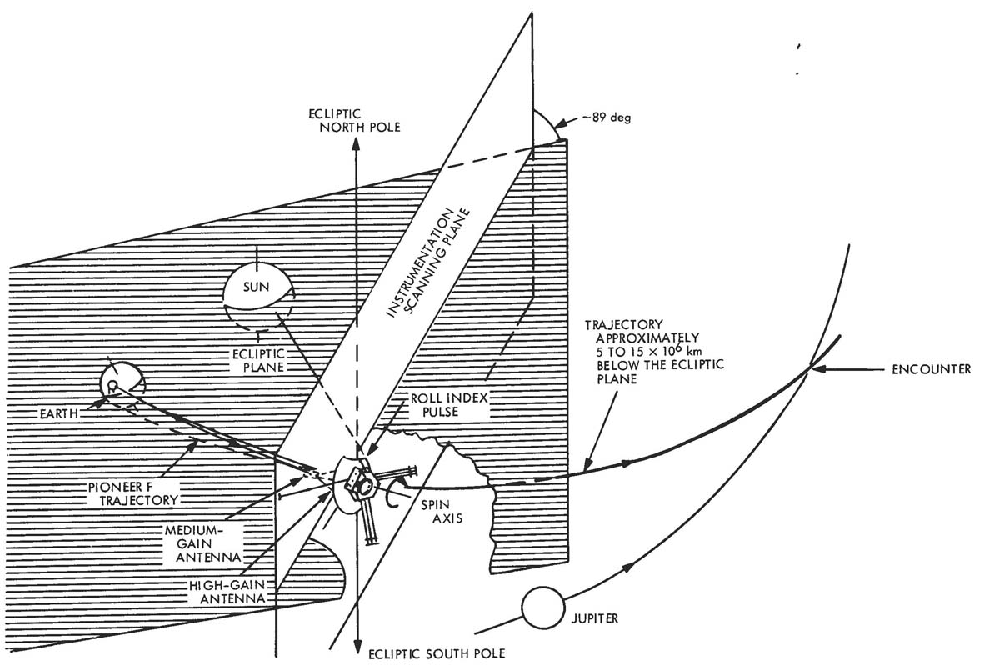,width=150mm}
\end{center}
\vskip -10pt 
\caption{
The earlier part of the Pioneer 10 trajectory before Jupiter encounter, the part of the trajectory when antenna articulation was largest. One can clearly see four possible different directions for the Pioneer anomaly:
(i) toward the Sun,
(ii) toward the Earth,
(iii) along the velocity vector, and
(iv) along the spin axis.
Later in the mission the difference between these directions became very small. 
\label{fig:pio-inner-orient}}
\vskip -10pt 
\end{figure*} 
%

\subsection{Analysis of the Earlier Data} 

We plan to analyze the early parts of the trajectories of the Pioneers with the goal of determining the true direction of the Pioneer anomaly and possibly its origin \citep{pio-origin,stanford,iap-pioneer}. One would expect that the much longer data span will greatly improve the ability to determine the source of the acceleration. In particular, with data from much closer to the Earth and Sun, one should be able to better determine whether the acceleration is 
\begin{enumerate}[(i)]
\item in the sunward direction that would indicate a force originating from the Sun, 
\item in the Earth-pointing direction that would be due to an anomaly in the frequency standards, 
\item in the direction along the velocity vector that would indicate an inertial or drag force, or 
\item along the spin axis direction that would indicate an on-board systematic. 
\end{enumerate} 
Therefore, analysis of the earlier data would be critical in helping to establish a precise 3-dimensional time history of the effect, and therefore to find out whether it is due to a systematic or new physics. 

During the flight in the inner solar system, an Earth-pointing attitude was necessary to enable the narrow beam of the high gain antenna of the spacecraft to illuminate the Earth and to maintain effective communications. Figures~\ref{fig:antenna-point}, \ref{fig:pio-inner-orient} depict the directions along the spacecraft's spin and antenna axes toward the Earth. The geocentric longitude of the craft varied continuously throughout the mission and, therefore, it was necessary to make numerous attitude adjustments. Since the spacecraft high gain antenna has a half-power beamwidth of $\sim3.5^\circ$, many spin axis orientation maneuvers were necessary to compensate for both the movement of the craft relative to the Earth, and also for the precession caused by solar pressure, which was 0.2$^\circ$ per day during the early part of the mission. In addition, to provide the planned encounter trajectories, some adjustments of the velocity vector were performed during the interplanetary flight by generating thrust in a particular direction.  Therefore, with thrusters in a fixed relationship to the spacecraft, reorientations of the spacecraft were also necessary.  Thus, with the early data (where spacecraft reorientation maneuvers were performed much more often than twice per year), we expect to improve the sensitivity of the solutions in the directions perpendicular to the line-of-sight by at least an order of magnitude. 

\begin{figure*}[t!]
 \begin{center}
\noindent   
\psfig{figure=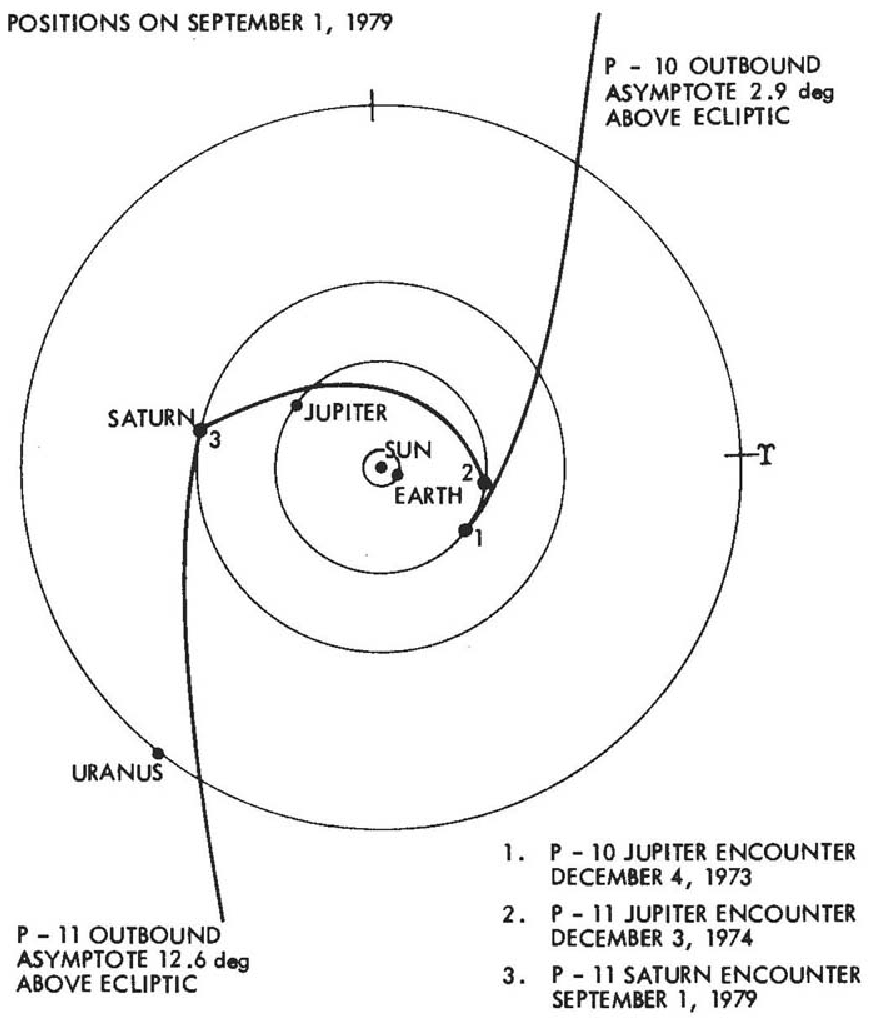,width=86mm}
\end{center}
\vskip -10pt 
  \caption{Heliocentric geometry of Pioneer 10 and 11 trajectories.
 \label{fig:pio-inner-saturn}}
\vskip -10pt 
%
 \begin{center}
\noindent   
\psfig{figure=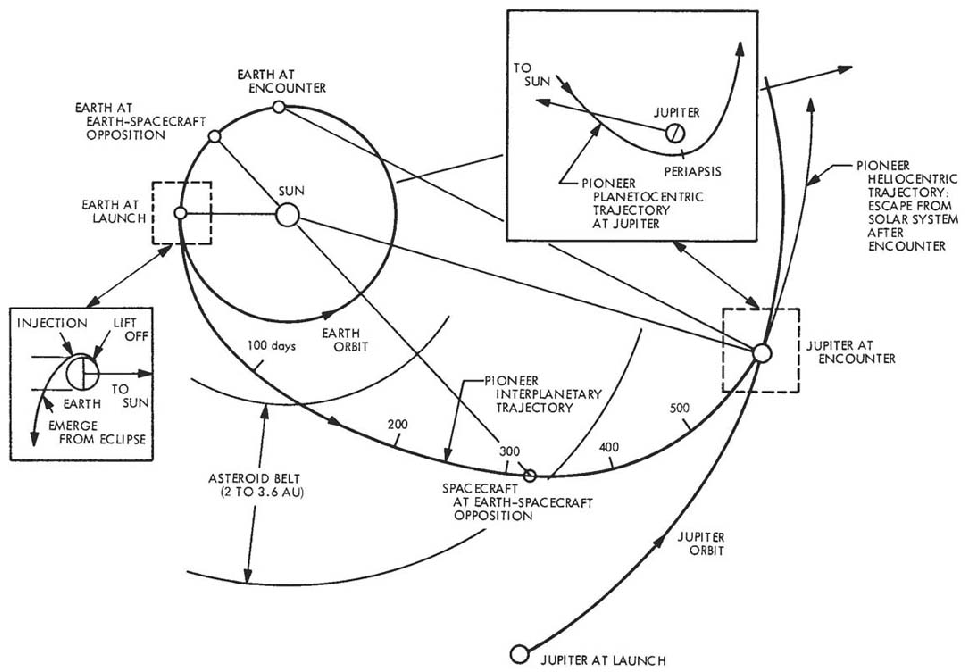,width=148mm}
\end{center}
\vskip -10pt 
  \caption{Heliocentric geometry of the Pioneer 10 Jupiter encounter.
 \label{fig:pio10-inner-jupiter}}
\vskip -10pt 
\end{figure*} 
%

There are strong expectations that a well developed strategy in the analysis of early data could yield a precise direction of the anomaly.  However, we would like to point out some obstacles along the way toward this goal.  It was pointed out \citep{pioprd,pio-origin} that a difficult problem in deep space navigation is precise 3-dimensional orbit determination.  The ``line-of-sight'' component of a velocity is 
much more easily determined by orbit determination codes than motion in the orthogonal directions. Unfortunately there is no range observable on the Pioneer craft which complicates the analysis. There is also the fact that the earlier parts of the trajectory were dominated by solar radiation pressure and attitude control maneuvers which significantly affect the accuracy of the orbit determination at the required level. Nevertheless, at this point there is a hope that these problems can be overcome and the analysis will yield the true direction of the anomaly and, quite possibly, its origin \citep{stanford,iap-pioneer,edata-mmn05}.   
  
\subsection{Study of the Planetary Encounters} 

The early Pioneer 10 and 11 data (before 1987) was never analyzed in detail, especially with a regard for systematics.  However, for Pioneer 10 an approximately constant anomalous acceleration seems to exist in the data as close in as 27 AU from the Sun \citep{pioprl,pioprd}.  Pioneer 11, beginning just after Jupiter flyby, finds a small value for the anomaly during the Jupiter-Saturn cruise phase in the interior of the solar system.  But right at Saturn encounter (see Figure~\ref{fig:pio-inner-saturn}), when the craft passed into an hyperbolic escape orbit,  there was a fast increase in the anomaly whereafter it settled into the canonical value \citep{pioprd,pio-origin,edata-mmn05,iap-pioneer}.

We first plan to study the Saturn encounter for Pioneer 11. We plan to use the data for approximately two years surrounding this event. If successful, we should be able to find more information on the mechanism that led to the onset of the anomaly during the flyby. The Jovian encounters are also of significant interest (see Figure~\ref{fig:pio10-inner-jupiter} for the geometry of the Pioneer 10 encounter). However, they were in the region much too close to the Sun. Thus one expects large contributions from the standard sources of acceleration  noise that exist at heliocentric distances $\sim5$~AU. Nevertheless we plan to use a similar strategy as with the Saturn encounter and will attempt to make full use of the data available.

\subsection{Analysis of the Entire Data Span} 

The same investigation as above could, of course, revisit the question of collimated thermal emission by studying the temporal evolution of the anomaly.  Thus, if the anomaly is due to the on-board nuclear fuel inventory ($^{238}$Pu) present on the vehicles and related heat recoil force, one expects that a decrease in the anomaly's magnitude will be correlated with Pu decay with a half-life of 87.74 years.  The analysis of 11.5 years of data \citep{pioprd,pio-standard} found no support for a thermal mechanism.   However, the now available 30-year interval of data (20 years for Pioneer 11) might demonstrate the effect of a $\sim24$\% reduction in the heat contribution to the craft's acceleration.  At the same time, although not very precise, the analysis performed by \cite{markwardt} indicated a decay in the magnitude of his solution for the anomaly consistent with the temporal change of the on-board thermal inventory.  The much longer data span will help to better determine if there is any signature of an exponential decay of the on-board power source, something not seen with the 11.5 years of data. We plan to analyze the entire set of available data in an attempt to determine whether or not the anomaly is due to the on-board nuclear fuel inventory and related heat radiation or to another mechanism.  

Therefore, the extended data set, augmented by all the ancillary spacecraft data, might not only help to precisely identify the direction of the anomaly but could also help to obtain tighter bounds on its time and distance dependence.  The wealth of additional recently acquired data presents an exciting opportunity to learn more about the Pioneer anomaly in various regimes and, thus, help to determine the nature of the anomalous signal. 

\subsection{Analysis of the Individual Trajectories for Both Pioneers}

The much larger data set for Pioneers 10 and 11 makes it possible to study the properties of the individual solutions for both Pioneers. The data used in the previous analysis \citep{pioprl,pioprd} precluded comparison of the solutions for anomalous accelerations obtained with the data collected from the same heliocentric distances. The new data set would allow such an investigation. Previously, even though we had individual solutions from the two craft, the fact remained that $a_{P10}$ and $a_{P11}$ were obtained from data segments that not only were very different in length (11.5 and 3.75 years), but they were also taken from different heliocentric distances (see discussion in Section \ref{sec:anomaly_sum}). This allowed us to suggest that ``we have only one data point'' in determining the anomalous acceleration. 

In the new analysis we can study the anomalous accelerations derived from the very comparable data sets delivered by the Pioneers. Furthermore, as we shall see in Section~\ref{sec:system-mdr}, based on the information found in the MDRs,  the actual performance of the two craft was different; in fact, the Pioneers had their ``individual signatures'' in their performance.  We anticipate that analysis of their data from similar heliocentric regions could help us to better understand the properties of the anomaly, especially in the case when the anomaly will be attributed to a source of on-board systematics. However, if the anomaly is found to have an extravehicular origin, again the analysis of the individual data would help to calibrate the final solution for the anomaly by properly accounting for the individual properties of the spacecraft.  

\subsection{Investigation of the On-Board Systematics}
\label{sec:system-mdr}

It is believed that the study of the entire Pioneer data set would help to shed light on the role that on-board systematics played in the anomalous acceleration. Therefore, the new analysis of Pioneer Doppler data should be done in parallel with the study of the spacecraft telemetry data. 

In this section, we consider forces that are generated by on-board spacecraft systems and that are thought to contribute to the constant acceleration seen in the analysis of the Pioneer Doppler data. Following \cite{pioprd} the on-board mechanisms discussed are the radio beam reaction force, the RTG heat reflecting off the spacecraft, the differential emissivity of the RTGs, the expelled helium produced within the RTG, and the thruster gas leakage. We specifically focus on the use of MDRs to study these sources of on-board systematics.

\subsubsection{Radio Beam Reaction Force}

The Pioneers have a total nominal emitted radio power of 8\,W. The radiated power was kept constant at all time, independent of the coverage from ground stations. That is, the radio transmitter is always on, even when signals are not received by a ground station. The properties of the Pioneer downlink antenna patterns are well known and the pattern is a reasonably good conical beam \citep{PC-202}. 
\cite{pioprd} found that the corresponding effect of the recoil force due to the emitted radio power on the craft is found to be responsible for an acceleration bias, on the spacecraft away from the Earth. 

The uncertainties in the radiation pattern were estimated to contribute as much as 10\% of uncertainty in this effect.  As a result,  \cite{pioprd} reported the following acceleration contribution of the effect of the recoil force due to the emitted radio power on the craft
\begin{equation}
a_{\rm rp} =(1.10 \pm 0.10)\times 10^{-10}\,{\rm m/s}^2.
\label{eq:rad-beam}
\end{equation}

The estimate above assumed that the antenna was always pointing in the same direction. This is by and large true: throughout their missions, with the exception of (very) brief periods when propulsion maneuvers were performed, both spacecraft aimed their high gain antennae at the Earth with a very small pointing error. As the Earth moved out of the antenna beam due to the combined motion of the Earth and the spacecraft, so-called {\tt CONSCAN} maneuvers \citep{PC-202} were performed to correct the spacecraft attitude. Therefore, the magnitude of the radio beam recoil effect along the line of sight should largely remain constant.

\begin{figure*}[ht!]
\hskip -6pt 
\begin{minipage}[b]{.46\linewidth}
\centering \psfig{file=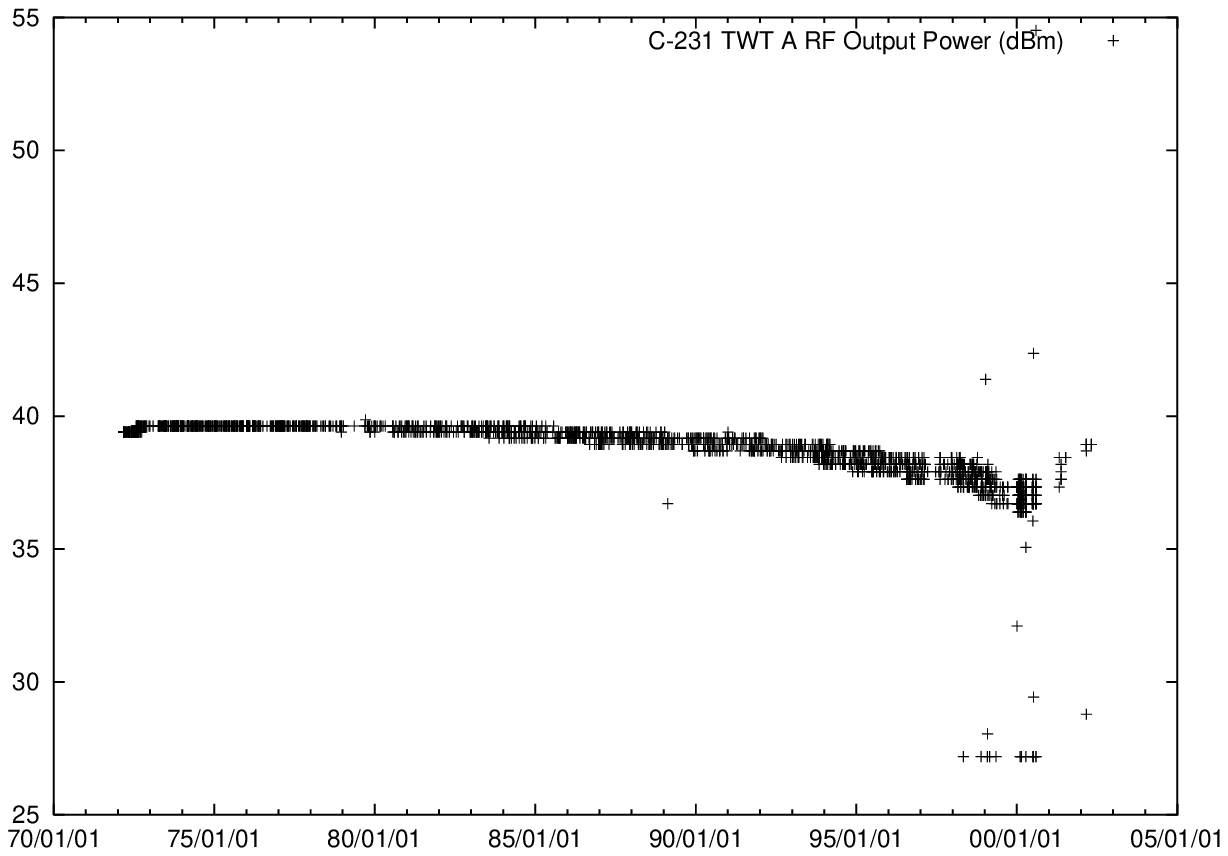, width=80mm}
\end{minipage} 
\hskip 20pt
\begin{minipage}[b]{.46\linewidth}
\centering \psfig{file=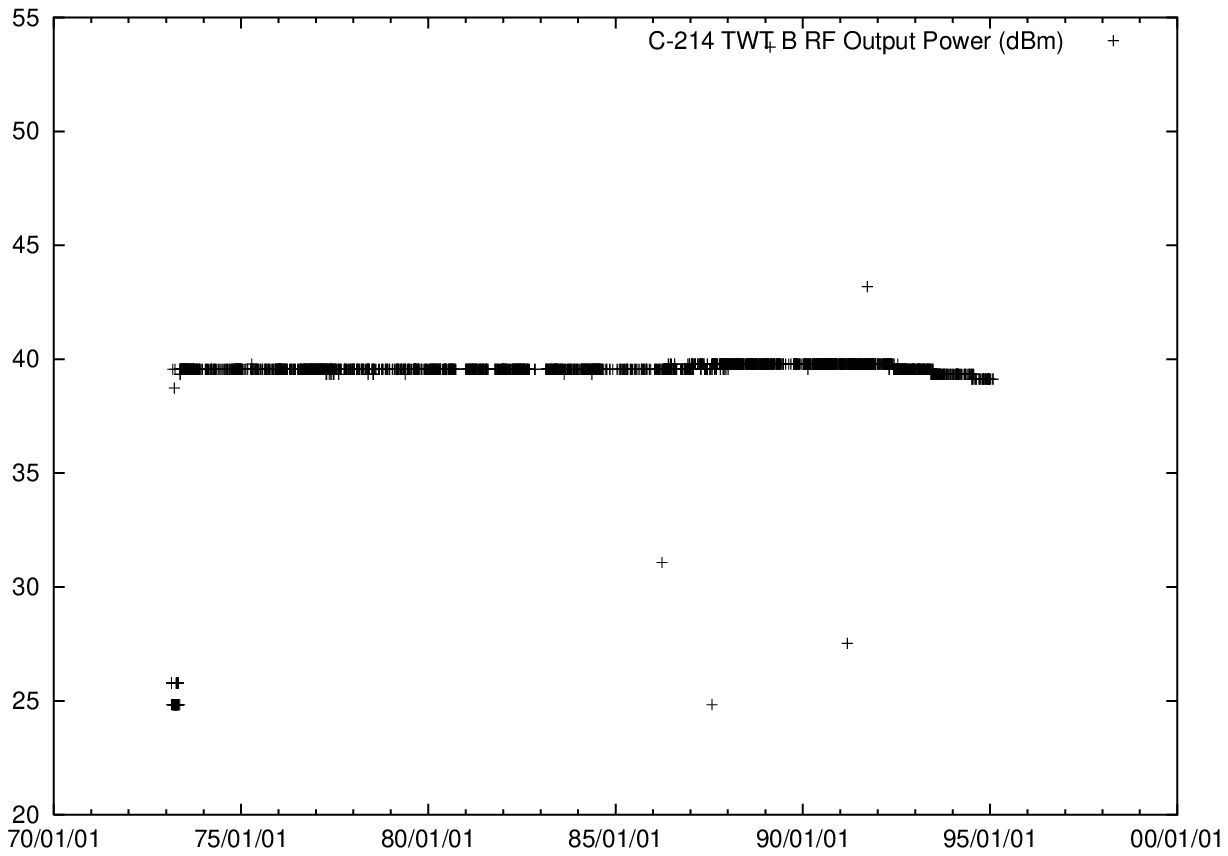, width=80mm}
\end{minipage}
\caption{The emitted power of the traveling wave tube transmitter throughout the mission, as measured by on-board telemetry. Left: Pioneer 10, which used TWT A (telemetry word C-231). Right: Pioneer 11, using TWT B (telemetry word C-214) throughout most of its mission.}
 \label{fig:C231}
\vskip -7pt 
\end{figure*}

The power of the radio beam appears to have varied slowly over time according to on-board telemetry (see Figure \ref{fig:C231}). For Pioneer 10, initially at around 39.4 dBm, corresponding with 8.7 Watts of radiated power, it decreased to around 37 dBm by the year 2000, which corresponds with 5 Watts. After 2000, the on-board telemetry shows an increase, the last measured value being 38.9 dBm (7.8 Watts.) This increase may coincide with the failure of the on-board oscillator, which resulted in the use of coherent mode transmitter operations only; or, it may be entirely a measurement artifact, as during this period, the spacecraft main bus voltage was no longer maintained at the nominal 28VDC level. For Pioneer 11, emitted power was very stable at 39.6-39.8 dBm (9.1-9.5~W) throughout much of the mission, with a slight decrease to 39.1 dBm (8.1~W) towards the end of the mission.

In the upcoming analysis, one should be able to estimate the contribution of this effect to the navigational accuracy of the Pioneers' trajectories with a much better precision. This could be done by correlating the magnitude of the recoil force with changes in the power level of the emitted radio beam. The effect of the non-coherent oscillator failure on board Pioneer 10 (resulting in no radio emissions when the spacecraft was not receiving a signal from Earth) may also be observable during the last years of Pioneer 10's mission.

\subsubsection{Anisotropic Heat Reflection}

It has been argued that the anomalous acceleration may be due to anisotropic reflection of the heat coming from the RTGs off the back of the spacecraft high gain antennae. Realistically, this is the most plausible mechanism to explain the origin of the anomaly and was originally discussed by \cite{pioprl,pioprd}. Note that only $\sim$64~W of directed constant heat is required to explain the anomaly, which certainly is not a great deal of power when the craft has heat sources capable of producing almost 2.5~kW of heat at the beginning of the missions. However, using available information on the spacecraft and RTG designs,  \cite{pioprd} estimated that only 4\,W of directed power could be produced by this mechanism. Adding an uncertainty of the same size, they estimated a contribution to the anomalous acceleration from heat reflection to be
\begin{equation}
a_{\rm hr}= (-0.55 \pm 0.55) \times 10^{-10}\,\mbox{m/s}^2. 
\label{eq:anisat-heat}
\end{equation}

\noindent Furthermore, \cite{katz-reply,pioprd} argued that if this mechanism were the cause of the anomaly, ultimately an unambiguous decrease in the size of $a_P$ should be observed, because the RTGs' radioactively produced radiant heat is decreasing. In fact, one would expect a decrease of about $0.75\times 10^{-10}$\,m/s$^2$ in $a_P$ over the 11.5 year Pioneer 10 data interval if this mechanism were the origin of $a_P$.

Alternative estimates presented by \cite{katz} and later refined by \cite{scheffer} put the magnitude of this effect at $\sim$24\,W or the corresponding value in Eq.~(\ref{eq:anisat-heat}) at the level of $a_{\rm hr}\sim -3.3 \times 10^{-10}\,\mbox{m/s}^2$. These estimates were based on a thermal model built for the Pioneer spacecraft and aided by information on the behavior of the optical properties of the spacecrafts' surfaces subjected to a long exposure in the near-Earth space environment. Without discussing the differences in the two approaches, we comment on the fact that both groups acknowledged that a thermal model for the Pioneer spacecraft is hard to build. However, this seems to be exactly what one would have to do in order to reconcile the differences in analyzing the role of the thermal heat in the formation of the Pioneer anomaly.   

As pointed out by \cite{pio-standard}, any thermal explanation should clarify why either the radioactive decay (if the heat is directly from the RTGs/RHUs) or electrical power decay (if the heat is from the instrument compartment) is not seen. One reason could be that previous analyses used only a limited data set of only 11.5 years when the thermal signature was hard to disentangle from the Doppler residuals or the fact that the actual data on the performance of the thermal and electrical systems was not complete or unavailable at the time the analyses were performed.

The present situation is very different. Not only do we have a much longer Doppler data segment for both spacecraft, we also have the actual telemetry data on thermal and electric power subsystems for both Pioneers for the entire lengths of their missions.  The upcoming analysis would be able to unambiguously determine weather or not the anomaly is due to the on-board generated systematics (especially heat related ones) or, at least, it will be able to properly calibrate the solution for the anomaly for the presence of these systematic effects. 

\begin{figure*}[ht!]
\hskip -6pt 
\begin{minipage}[b]{.46\linewidth}
\centering \psfig{file=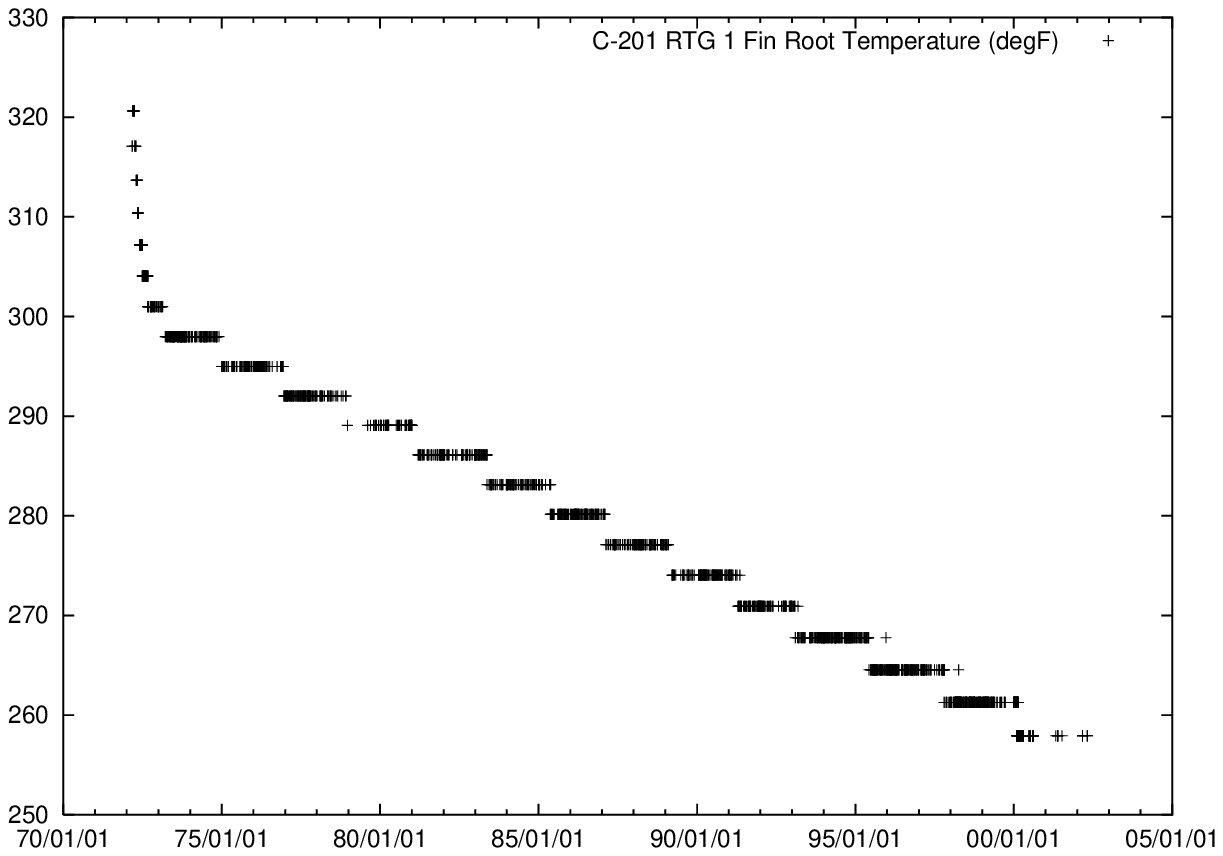, width=80mm}
\end{minipage} 
\hskip 20pt
\begin{minipage}[b]{.46\linewidth}
\centering \psfig{file=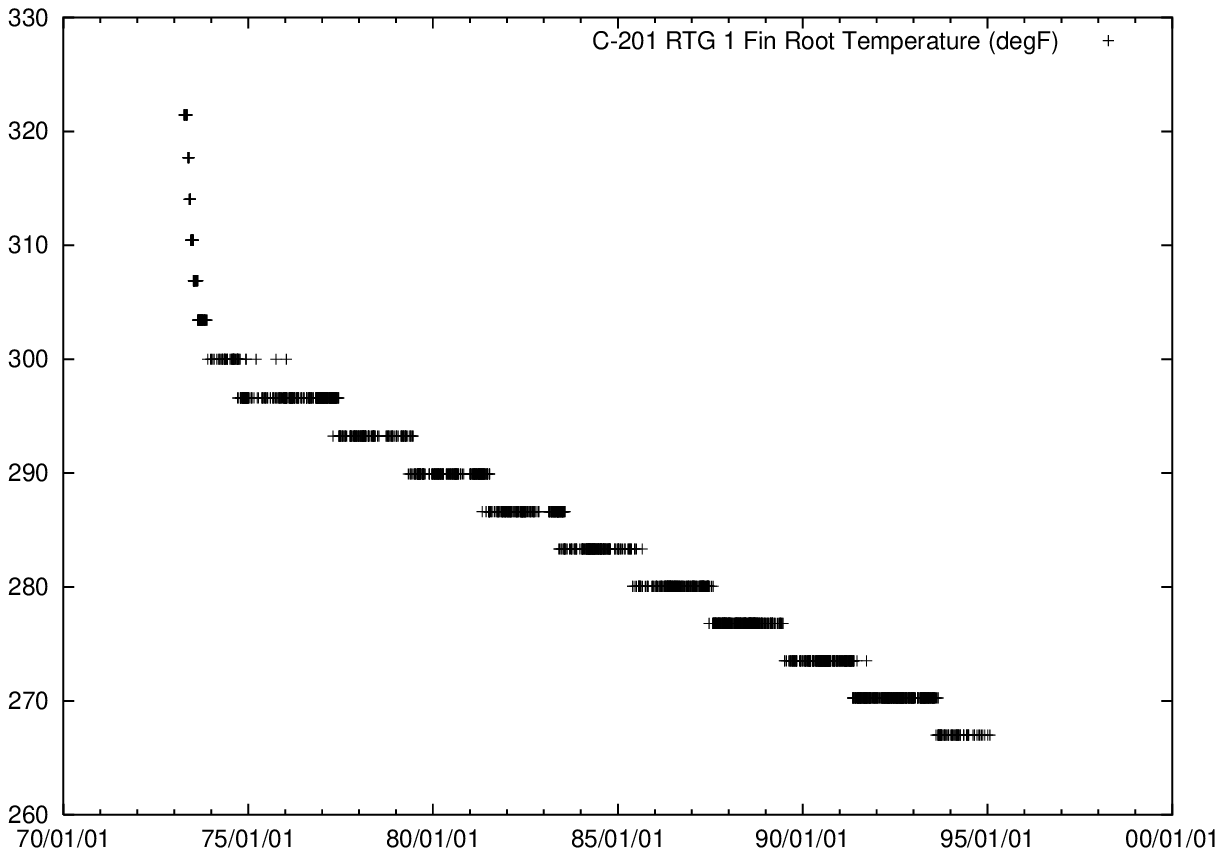, width=80mm}
\end{minipage}
\vskip -5pt 
\caption{RTG 1 fin root temperatures (telemetry word C-201; in $^\circ$F) for Pioneer 10 (left) and 11 (right). The behavior of the other RTGs was nearly identical.}
 \label{fig:root-fin}
\end{figure*}

Relevant on-board telemetry falls into two categories: temperature and electrical measurements. In the first category, we have data from several temperature sensors on-board, most notably the fin root temperature readings for all four RTGs. Figure~\ref{fig:sensors} shows the location of most temperature sensors on board for which readings are available. Other temperature sensors are located at the RTGs and inside the propellant tank. (Figure~\ref{fig:root-fin}  shows data from the ``RTG fin root temperature'' sensors for Pioneer 10 and 11 correspondingly, as an example of the volume and quality of the available information.\footnote{Becasue of the large volume of data, the following sampling values were used to generate all full-mission-duration telemetry plots used in this paper: for years 1972-1975 every 5000th record, for 1976-1982 every 1000th, for 1983-1989 every 200th, for 1990-1997 every 100th, for 1998-1999 every 10th, and for years after 2000 we used all available records.})

The electrical power profile of the spacecraft can be reconstructed to a reasonable degree of accuracy using electrical telemetry measurements. Briefly, the only electrical power source on board are the four RTGs; taking cable and inverter losses into account, we find the total available AC power on board; this supplied two separate circuits, one (called CTRF, or Central Transformer Rectifier Filter) providing low voltages to various components, while the other providing the 28VDC main bus voltage that supplied primary power to most subsystems, with excess power directed to the shunt radiator. We have individual voltage and current readings for the RTGs; we have readings on the main bus voltage and current, as well as the shunt current; we know the power on/off state of most spacecraft subsystems. From these and other readings, the complete electrical profile of the spacecraft can be calculated, as shown in Table~\ref{tb:power}.

\begin{table}[t]
\begin{center}
\caption{Reconstruction of the Pioneer 10 and 11 electrical power profile from the available telemetry.}
\vskip 8pt
\begin{tabular}[1]{lp{4.8in}}\hline
&\\[-8pt]
Subsystem&Power\\
&\\[-10pt]
\hline\hline\\[-4pt]
RTGs\footnotemark&
$P_{\tt RTG}=I_{\tt RTG1}U_{\tt RTG1}+I_{\tt RTG2}U_{\tt RTG2}+I_{\tt RTG3}U_{\tt RTG3}+I_{\tt RTG4}U_{\tt RTG4}$\\[6pt]
RTG cable losses&
$P_{\tt cable}=0.017\left(I_{\tt RTG1}^2+I_{\tt RTG3}^2\right)+0.021\left(I_{\tt RTG2}^2+I_{\tt RTG4}^2\right)$\\[6pt]
Inverter loss&
$\left(100\%-e_{\tt inv}\right)\left(P_{\tt RTG}-P_{\tt cable}\right)$\\[6pt]
CTRF loss&
$\left(100\%-e_{\tt CTRF}\right)P_{\tt RTG}/P_{\tt cable}$\\[6pt]
CTRF load&
$P_{\tt CTRF}=8.941+P_{\tt DDU}+P_{\tt DSU}+P_{\tt CEA}+P_{\tt CDU}+P_{\tt CONSCAN}$\\[6pt]
Nominal values&
Nominal values are available for $P_{\tt DDU}$, $P_{\tt DSU}$, $P_{\tt CEA}$, and $P_{\tt CONSCAN}$ as a function of their power and logic states.\\[6pt]
TRF loss&
$\left(100\%-e_{\tt TRF}\right)P_{\tt TRF}/e_{\tt TRF}$\\[6pt]
TRF load&
$P_{\tt TRF}=\left(P_{\tt sub}+P_{\tt exp}+P_{\tt TWT}+P_{\tt PCU}+P_{\tt BL}\right)U_{\tt bus}^2/784+P_{\tt shunt}$\\[6pt]
Nominal values&
Nominal values (@ 28VDC) for experiments ($P_{\tt exp}$), subsystems ($P_{\tt sub}$), the TWT amplifier ($P_{\tt TWT}$), the Power Control Unit ($P_{\tt PCU}$), and the battery loss ($P_{\tt BL}$) are available as a function of their power and logic states.\\[6pt]
Shunt power&
$P_{\tt shunt}=I_{\tt shunt}U_{\tt bus}$\\[6pt]
Direct telemetry&
Telemetry values are available for RTG voltages ($U_{\tt RTGn}$), currents ($I_{\tt RTGn}$), the main bus voltage ($U_{\tt bus}$), and the shunt current ($I_{\tt shunt}$). The main bus current ($I_{\tt bus}$) is also known, and can be used to verify the computed value of $P_{\tt TRF}$ above.\\[6pt]
Efficiencies&
A linear fit applied to nominal efficiency values yields $e_{\tt inv}=(100-P_{\tt RTG})/37+89\%$ and $e_{\tt CTRF}=0.94P_{\tt CTRF}+47.6\%$. The 28VDC converter efficiency is assumed to be constant: $e_{\tt TRF}=89\%$.\\\hline
\end{tabular}
\label{tb:power}
\vskip -10pt
\end{center}
\end{table}
\footnotetext{Some documentation (e.g., \cite{PC-202}) suggest that the telemetry readings $U_{{\tt RTG}n}$ measure voltages at the inverters, 
i.e.,~$\sum{I_{{\tt RTG}n}U_{{\tt RTF}n}}=P_{\tt RTG}-P_{\tt cable}$.  Actual readings, however, strongly suggest that this is not the case, and that the RTG voltage sensors are calibrated to measure the actual RTG voltages.}

To utilize this data for our investigation, one would have to reconstruct the direction of heat flow: absorption and re-emission by, and reflection off the craft surfaces.  We see this as the main challenge of the upcoming analysis.  How to convincingly use a reading at several points on the craft to reconstruct the heat dynamics? (Similar questions exist for gas leaks and all other parameters.) If we don't see their signature in the Doppler signal, we should be able to put an upper limit on the potential contribution from these sources to the Pioneer anomaly. Similarly, the temperature data in the MDRs must put an upper limit on the changes of the optical surfaces of the RTGs (i.e. their emissivity and reflectivity).

Because of the complexities involved, clearly more resources are needed; we invite researchers to join us for this part of the upcoming study.

\subsubsection{Differential Change of the RTG's Radiant Emissivity}

Another suggestion related to the RTGs is based on the idea that during the early parts of the missions, there might have been a differential change of the radiant emissivity of the Sun-pointing sides of the RTGs as compared to the sides facing deep space. Note that, especially closer to the Sun, the inner sides were subjected to the solar wind. On the other hand, the outer sides were sweeping through the solar system dust cloud. These two processes could result in different levels of degradation of the optical surfaces of the RTGs. In turn this degradation could result in asymmetric patterns of heat radiation away from the RTGs in the fore/aft directions along the spin axis. Therefore, it can be argued that such an anisotropy may have caused the anomaly.
Using the actual design information on the RTGs and a sophisticated model of the fin structure, \cite{pioprd} estimate an upper limit of 6.12\,W to be the uncertainty in the contribution from the differential emissivity of the RTGs to the anomalous acceleration of the craft.  This value, in turn, results in an acceleration uncertainty of 
\begin{equation}
\sigma_{\rm de} = 0.85 \times 10^{-10}~{\mathrm{m/s}}^2.
\end{equation}
The authors also point out the significance of radioactive decay for this mechanism and especially its expected contribution to the decrease of the magnitude of $a_P$, which has not been observed. 

The effect of the radioactive decay on the thermal output of the RTGs can be directly observed from telemetry. All RTGs had a temperature sensor located at the root of one of the radiating fins (``fin root temperature''). Preliminary analysis suggests that these readings are in agreement with what would be expected from the radioactive decay of the plutonium fuel (see Figure~\ref{fig:root-fin}).

One can try to detect changes in the measurements of fin root temperature during each of the three planetary encounters and see if the RTGs had developed an anisotropic radiating pattern. (Note that the same test can be conducted for the electrical energy dissipation.) Thus, if the temperatures at the fins change during the fly-bys, this may indirectly indicate an alteration of the optical properties of the RTGs during the planetary encounters.  This should allow us to ask the question: ``If the system survived flawlessly than why would it change long after the encounter?''  One can also look at the different segments of the thermal history in an attempt to identify different heat decay rates.  However, the planetary flybys might be the strongest test for the system's design and operational longevity.  

\subsubsection{Constant Electrical Heat Radiation as the Source}

If the mechanisms above were treated as separate constituents to the anomalous accelerations, one can ask what if a combination of them is responsible for the effect. It has been recently suggested that most, if not all, of the unmodeled acceleration  of  Pioneer 10 and 11 is due to an  essentially constant supply of heat coming from the central compartment, directed out the front of the  craft through the closed louvers \citep{scheffer}.  This is a more subtle version of an earlier proposal  \citep{murphy} calling on the total electrical power as a mechanism. That proposal was argued against because of the observed lack of decay  of the acceleration with time \citep{murphy-reply,pioprd}.   

For this mechanism to work, one essentially needs to find a constant supply of heat radiated off the back of the spacecraft that would produce a thermal recoil force with the well established properties (discussed in Section~\ref{sec:anomaly}). However, the assumption of constancy came against the actual design of the Pioneers and experimental data on the performance of the 
vehicles \citep{pio-standard}.  

In fact, the lack of constancy of heat dissipated during the longest Doppler segment analyses (i.e. 11.5 years of the Pioneer 10 data)  invalidated the hypothesis. However, this claim can and will be further investigated with the newly recovered data, both Doppler and telemetry (see Tables~\ref{tab:telemetry},\ref{tb:power}).  

The newly acquired data (both Doppler and telemetry) can significantly contribute to addressing this possibility. For one, we also have a much larger segment of Doppler data that will be used to analyze these heat dissipation processes on the vehicles. In addition, we now have the actual design, fabrication, testing, pre- and in-flight calibration data that characterize the  Pioneer craft performance for the duration of their missions.  This data can tell precisely at what time the louvers were open and closed, when a certain instrument was powered on and off, what was the performance of the battery, shunt current and all electric parts of the spacecraft. This is truly a unique data set that will be used to reconstruct the physics of the anomalous acceleration, if it is due to on-board systematics.  

In fact we have all the detailed information on properties of the spacecraft and the data needed to reconstruct the behavior of its major components, including electrical power and thermal subsystems. This information may be used in developing a thermodynamical model of the spacecraft that would help us to establish the true thermal and electrical power dissipation history of the vehicles and also correlate major events on the Pioneers (such as powering ``on'' or ``off'' certain instruments or performing a maneuver) with the available Doppler data; this possibility is addressed in Section~\ref{sec:thermal-model}.

\subsubsection{Helium Expulsion from the RTGs}

Another possible on-board systematic error may be coming from the expulsion of the He being created in the RTGs from the $\alpha$-decay of $^{238}$Pu (see discussion in \citep{pioprd}). The Pioneer RTGs were designed so that the He pressure is not totally contained within the Pioneer heat source over the life of the RTGs. Instead, the Pioneer heat source contains a pressure relief device that allows the generated He to vent out of the heat source and into the thermoelectric converter. The thermoelectric converter housing-to-power output receptacle interface is sealed with a viton O-ring. The O-ring allows the helium gas within the converter to be released by permeation into the space environment throughout the mission life of the Pioneer RTGs. 

For this mechanism to work, what one needs is all He produced to preferentially leave the RTGs in one direction at a rate of 0.77\,g/yr. Assuming a single elastic reflection, \cite{pioprd} ruled out helium permeating through the O-rings as the cause of $a_P$. They also give the estimate for acceleration contribution in the direction away from the Sun due to this systematic expulsion
\begin{equation}
a_{\rm He} = (0.15 \pm 0.16)\times 10^{-10}~\mathrm{m/s}^2.
\end{equation} 

It is highly unlikely that this mechanism is responsible for the Pioneer anomaly seen on both spacecraft; the use of MDRs can further support this conclusion. 

\subsubsection{Propulsive Mass Expulsion}
\label{sec:mass-expulsion}

The possible use of a similar strategy to the above to study the effect of propulsive mass expulsion due to gas leakage has to be assessed. It is known that, although this effect is largely unpredictable, many spacecraft have experienced gas leaks producing accelerations on the order of $10^{-9}$\,m/s$^2$. Gas leaks generally behave differently after each maneuver. The leakage often decreases with time and becomes negligible. (See more discussion of these effects in \citep{pioprd}.)

\begin{figure*}[h!]
\hskip -6pt 
\begin{minipage}[b]{.46\linewidth}
\centering \psfig{file=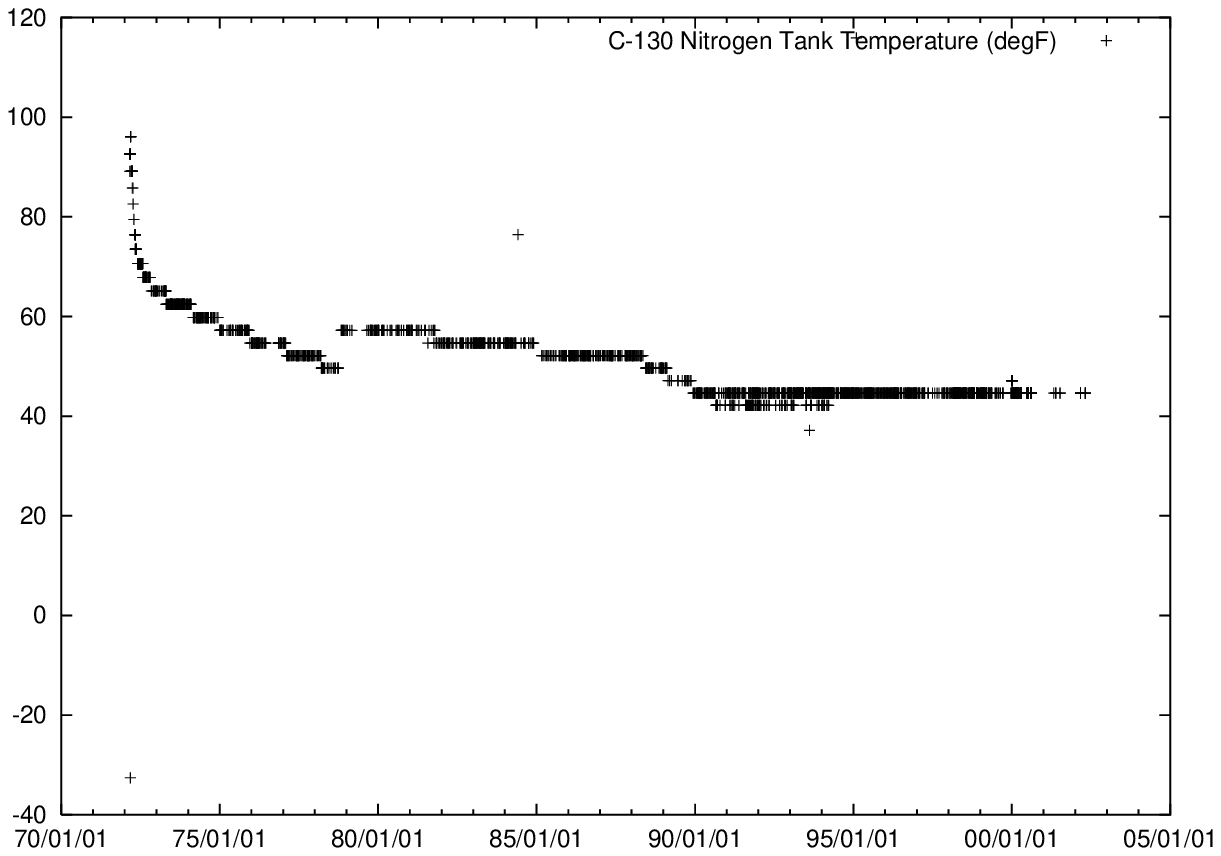, width=80mm}
\end{minipage} 
\hskip 20pt
\begin{minipage}[b]{.46\linewidth}
\centering \psfig{file=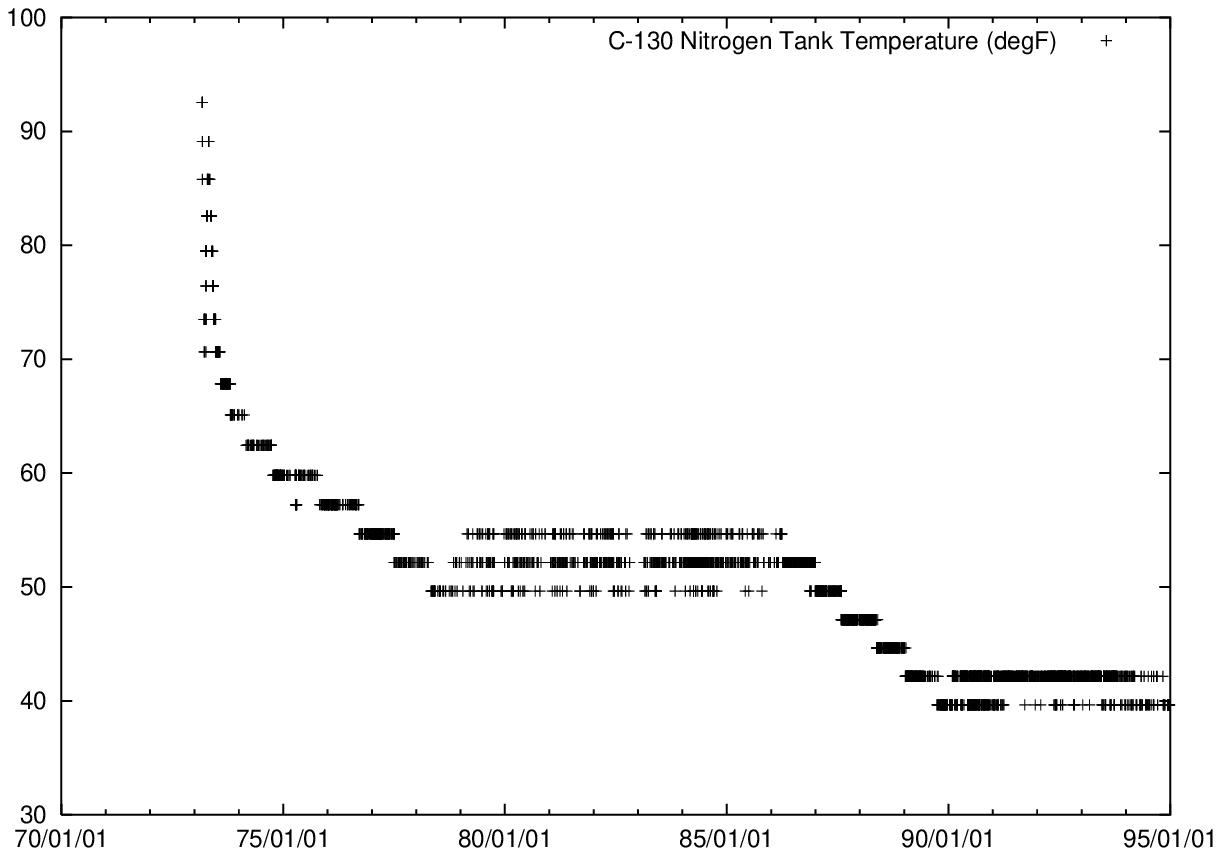, width=80mm}
\end{minipage}
\vskip -5pt 
\caption{Nitrogen tank temperature (telemetry word C-130; in F$^\circ$). Left: Pioneer 10; right: Pioneer~11.}
 \label{fig:thrusters-C-130}
\end{figure*}
\begin{figure*}[ht!]
\hskip -6pt 
\begin{minipage}[b]{.46\linewidth}
\centering \psfig{file=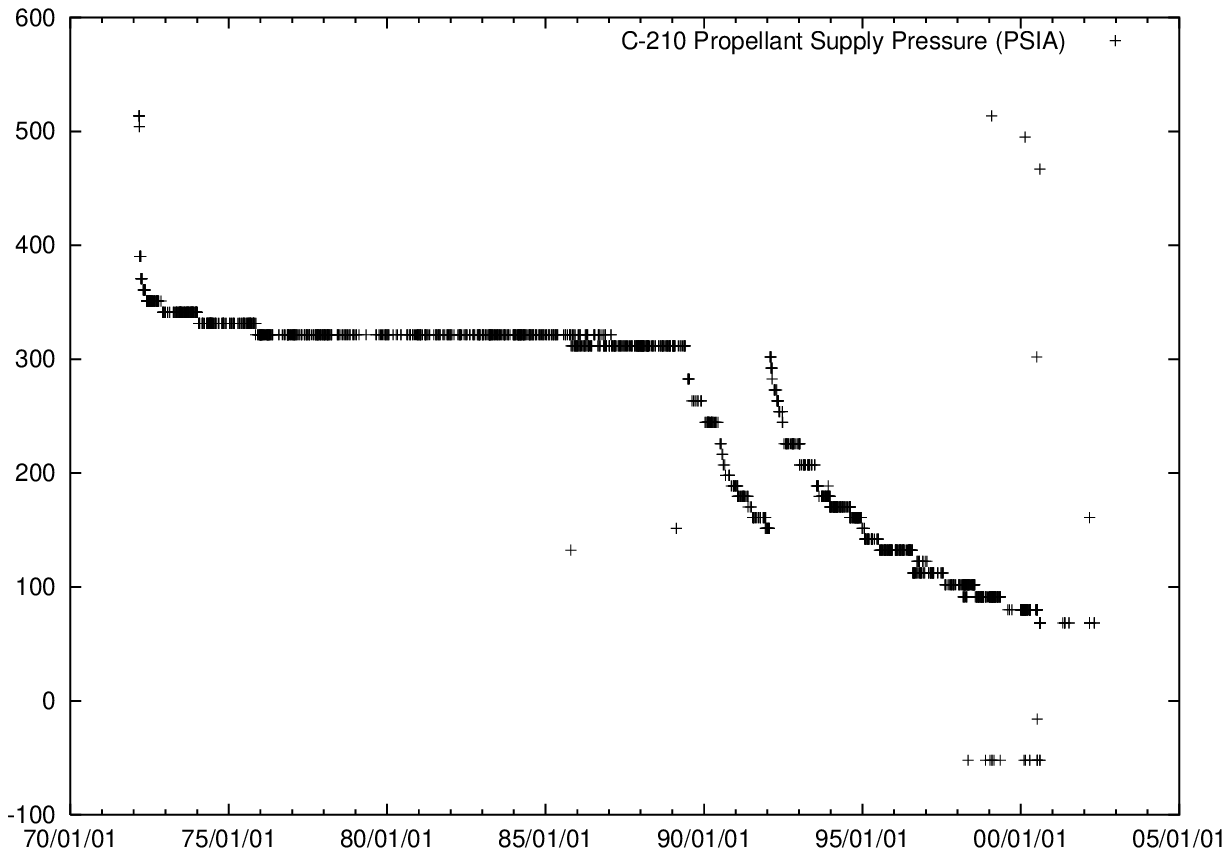, width=80mm}
\end{minipage} 
\hskip 20pt
\begin{minipage}[b]{.46\linewidth}
\centering \psfig{file=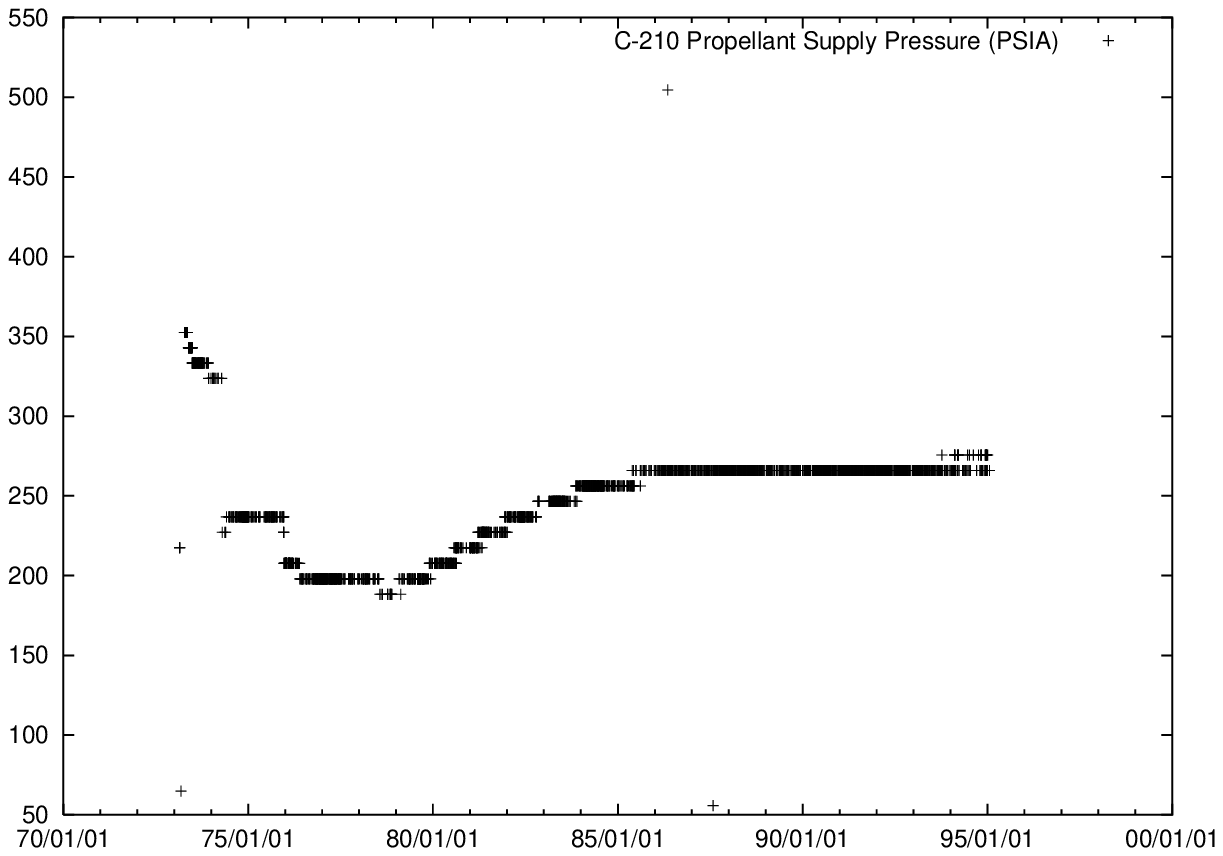, width=80mm}
\end{minipage}
\vskip -5pt 
\caption{Propellant supply pressure (telemetry word C-210). Left: Pioneer 10; right: Pioneer 11.}
 \label{fig:thrusters-C-210}
\vskip 10pt
\hskip -6pt 
\begin{minipage}[b]{.46\linewidth}
\centering \psfig{file=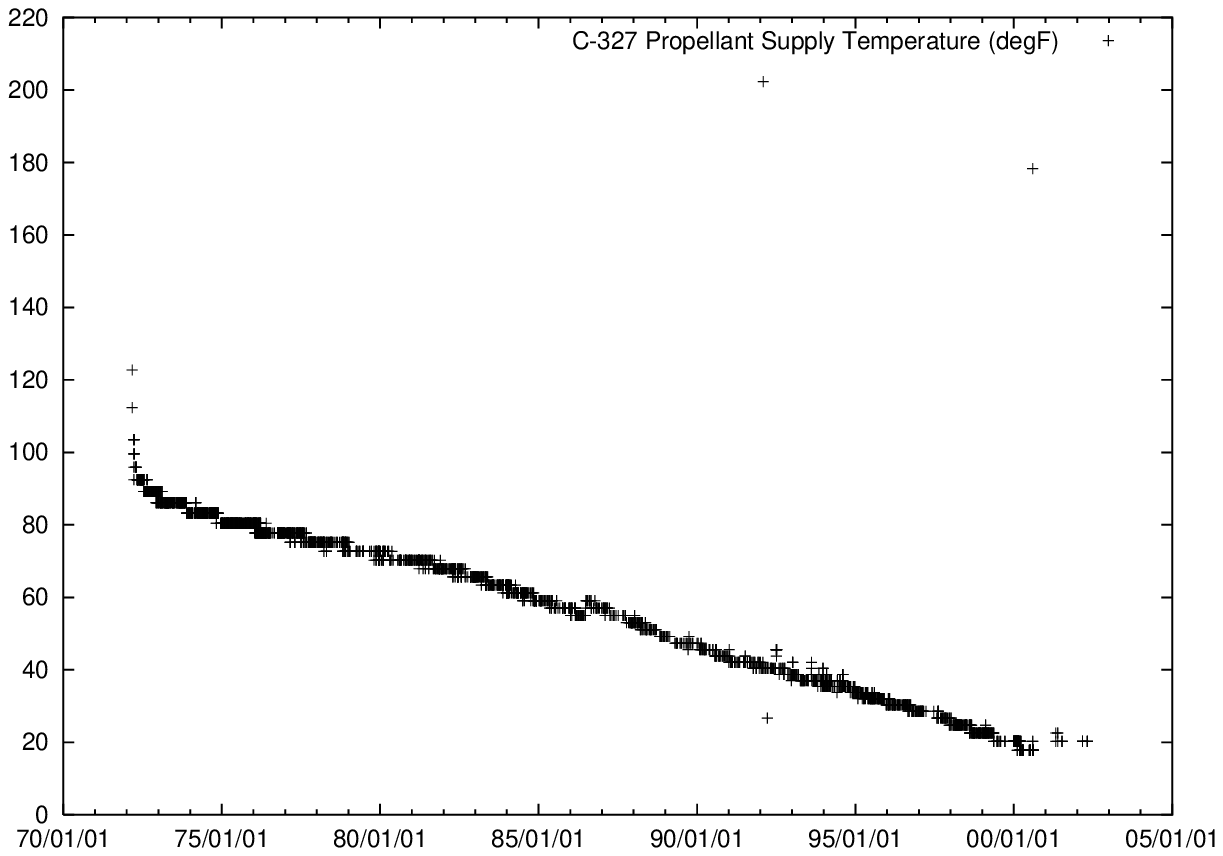, width=79mm}
\end{minipage} 
\hskip 20pt
\begin{minipage}[b]{.46\linewidth}
\centering \psfig{file=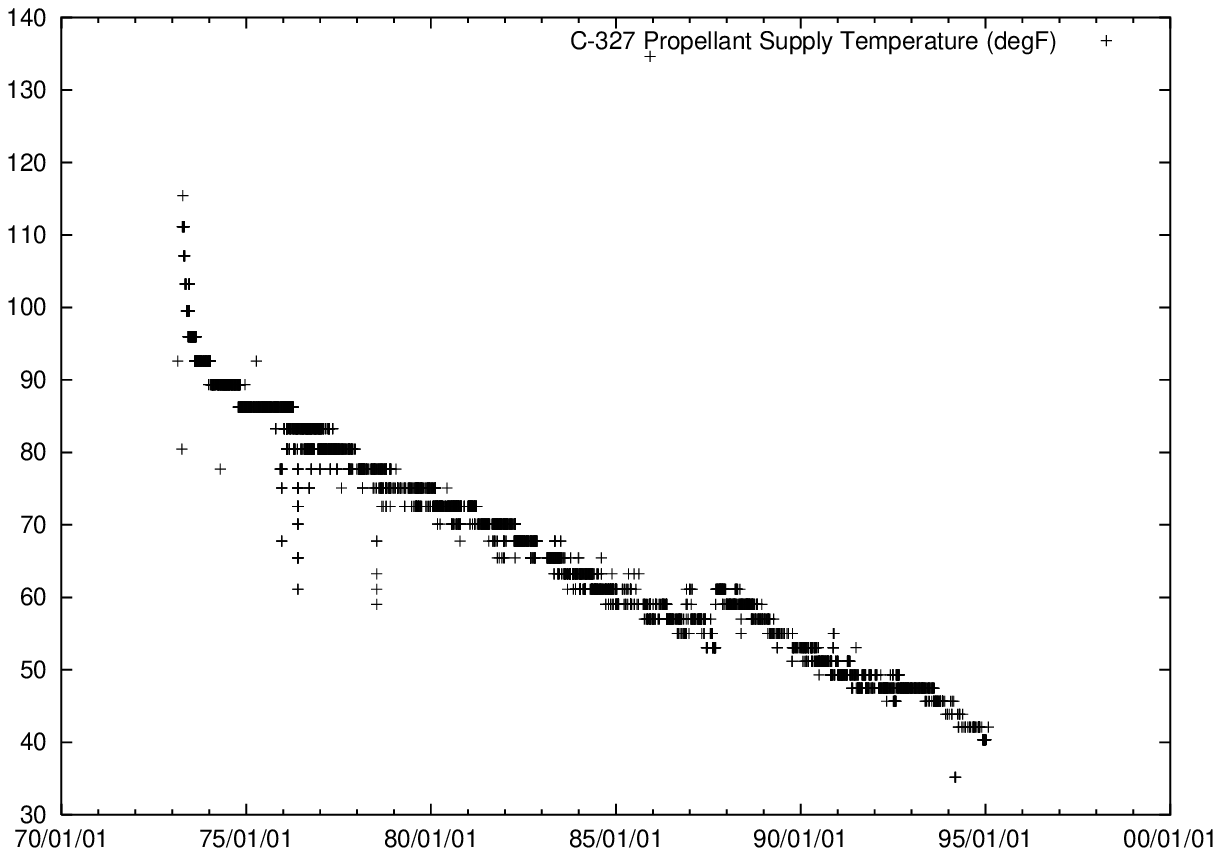, width=79mm}
\end{minipage}
\vskip -5pt 
\caption{Propellant supply temperature (telemetry word C-327). Left: Pioneer 10; right: Pioneer 11.}
 \label{fig:thrusters-C-327}
\vskip 3pt 
\end{figure*}

Gas leaks can originate from the Pioneers' propulsion system, which is used for mid-course trajectory maneuvers, for spinning up or down the
spacecraft, and for orientation of the spinning spacecraft. (Consult \citep{PC-202} for design of the Pioneer propulsion and attitude control systems.)  By studying time histories of the spin rates of the Pioneers, \cite{pioprd} estimated the effect of the gas leakage mechanism on the anomalous acceleration. This mechanism was found to produce an acceleration uncertainty of 
\begin{equation}
\sigma_{\rm gl}= \pm 0.56\times10^{-10}~{\rm m/s}^2.
\end{equation}
\noindent Furthermore, they concluded that the gas leak mechanism is very unlikely to be the explanation for the anomalous acceleration.  This is because it is difficult to understand why this effect, being a stochastic variable obeying a Poisson distribution, would affect Pioneers 10 and 11 by the same amount \citep{pioprd}. 

The new analysis can refine the conclusions above. We have three readings from the propellant tank on board that can help establish the propellant loss history of the two spacecraft, namely propellant tank temperature (Figure~\ref{fig:thrusters-C-130}), propellant supply pressure (Figure~\ref{fig:thrusters-C-210}), and propellant supply temperature (Figure~\ref{fig:thrusters-C-327}). We also have several temperature readings from the thruster clusters and individual thrusters. Not only may this new information be used to precisely establish the time of the maneuvers, it can also provide some insight as to why the two identically designed systems behaved so differently, especially when changes to their spin rates are concerned (see Figures 11 and 12 of \cite{pioprd}.) The corresponding analysis had been initiated and results will be reported.

\subsection{Building a Thermal/Electrical/Dynamical Model for the Pioneers}
\label{sec:thermal-model}

In this section we will argue that, based on the information provided by the MDRs, one can develop a high accuracy thermal/electrical/dynamical model of the Pioneer spacecraft.  Such a model can be used to further improve our understanding of the anomalous  acceleration and especially to study contribution from the on-board thermal environment to the anomaly.  In this section we will comment on our preliminary evaluation of what is needed to develop such an accurate model.

It is clear that a thermal model for the Pioneers spacecraft would have to account for all heat radiation within the spacecraft and out to deep space.  Specifically, this model would have start from the RTGs and include description of the heat radiated, absorbed, reradiated, and conducted by each and every instrument, all cables, support structures and etc.  The objective is to develop a broadband (IR and optical) thermal map of the spacecraft's exterior, accurate to few degrees Kelvin.  As input, this model will use electrical telemetry readings from the on-board sensors that were installed on the spacecraft, while temperature readings (see Figure~\ref{fig:sensors}) can be used for its calibration.  We will then attempt to verify this model by using the radiometric Doppler data from the attitude maneuvers performed at the earlier mission phases.   

It is worth noting that no other mission in the past used telemetry data to improve its navigational capabilities; what we are going to implement with the Pioneers is a novel navigational technique. Pioneer 10/11 may be the first missions to benefit from this technique, designed to reduce the impact of on-board sources to non-gravitational accelerations on the accuracy of trajectory reconstruction. 
 
Our proposal uses the fact that the Pioneers already have very good ``built-in'' navigational capabilities (i.e. guidance, navigation and control); our goal is to reach the absolute limit allowed by these capabilities.  As a quantitative objective, we hope to improve acceleration contribution due to on-board systematic sources to the level below $1 \times 10^{-10}$ m/s$^2$. To reach the stated level of accuracy, one would have to adjust navigational data by using the developed thermal/electrical/dynamical model of the spacecraft. To achieve this we will utilize the following available information: 

\begin{enumerate}
\item 	Dynamical properties of the spacecraft, i.e. geometry, dimensions, and weights of all the components of the spacecraft, including cabling, fuel and thermal blankets.  In short, we will need to have precise mass and power budget data for the entire craft.  Ideally, we would prefer to use not only design numbers, but rather actual values established during the final pre-launch tests at the Kennedy Space Center, Florida (KSC). This information is still to be identified and retrieved.

\item	Thermal, electric and optical (infrared) properties of the spacecraft materials used to fabricate all major components of the spacecraft, including thermal blankets and cabling.  Again, we would prefer to have these values from the tests at the KSC. Note that we will have access to the initial values of these parameters; their degradation during the 30 years in space is unknown.  There is very limited information available from long-term exposure tests.  Furthermore, for obvious reasons (tests conducted in a different environment, e.g., in low Earth orbit (LOE), for different durations), even these results will not be directly applicable to our model.

\item	We will benefit from the design information of the spacecraft, especially information concerning their thermal, power, propulsion, and communication subsystems.  We will use the knowledge on the location of each and every telemetry sensor placed on the spacecraft and use all the recently retrieved Pioneer telemetry data.  This information will be used to build and calibrate the thermal model of the spacecraft.  In particular, the following telemetry information will be critical for the upcoming investigation:
\begin{enumerate}[a).]
\item RTG fin root temperature, as an indicator of RTG thermal output
\item RTG current and voltage, providing a measure for total electrical power
\item All thermal data from sensors on the spacecraft and instrument compartment
\item Temperatures and pulse counts from thruster cluster assemblies
\item Battery current and temperature
\item Shunt current
\item Radio beam gain and transmitted power, TWT amplifier temperature
\item Propellant tank and supply lines temperature
\end{enumerate}

\item We would have to analyze the design and thermal properties of the SNAP-19 RTGs used in the Pioneers.  Much information is already available, but additional data on their thermal behavior will be very helpful in building a 3-dimensional thermal radiation pattern. 

\end{enumerate}

Once the thermal model is completed, it must be calibrated for in-flight conditions.  For this, we may need to analyze all the attitude control (i.e.{\tt~CONSCAN}) maneuvers in the earlier mission phases \citep{PC-202,pioprd}.  The earlier phases had significant variations in the Sun-Earth line to spin axis angle.  These variations will help us to calibrate the effect of thermal recoil forces on the radiometric data received from the craft and, thus, to precisely evaluate the magnitude of the thermal recoil force. The closer the craft to the Earth is, the better the ultimate accuracy of the model will be.  This is due to the fact that we know the properties of the solar radiation and thermal environment in the inner solar system to a much better accuracy than in the deep space.  In fact, it will be hard to do the required calibration after the Jupiter flyby, and it will be impossible at distances greater than 10 AU.  Clearly we will benefit from the larger size of the Earth-probe-Sun angle, which is getting much smaller after 10 AU.  After 20 AU, both the Sun and the Earth will be in the same field of view, making it difficult to decouple any thermally induced force from other physical effects.   

However, given the significant uncertainties on the degradation of the thermodynamical and optical properties of the materials used to build the Pioneers, the outcome of this model is difficult to predict.  Nevertheless, we expect that, given the simplicity of the Pioneers' design, one should be able to build a model with an uncertainty of no more than 15\%. We have initiated this work with anticipated involvement of several navigational and thermal engineers at JPL; results of this work  will be reported.

The plan outlined here is ambitious. Before we embark on this complex exercise, we shall also attempt to build a simpler (but enthusiastically not ``simplistic'') thermal model of the Pioneers based on available telemetry data. Such a model would necessarily utilize simplifying assumptions on the spacecraft's geometry, and ignore contributions from small components (e.g., structural elements, struts, cables as heat sources). Nevertheless, it is possible that even such a model may lead to a much better understanding of the contribution of spacecraft systematics to the anomalous acceleration, and may also guide us to better direct our efforts as we carry out a more elaborate effort.


\section{Conclusion}
\label{sec:conclude}

On March 2, 2002 NASA's DSN made the last contact with Pioneer 10 and confirmed that the spacecraft was still operational thirty years after its launch on March 3, 1972 (UT). The uplink signal was transmitted on March 1 from the DSN's Goldstone, California facility and a downlink response was received twenty-two hours later by the 70-meter antenna at Madrid, Spain. At this time the spacecraft was 11.9 billion kilometers from Earth at about 79.9 AU from the Sun and heading outward into interstellar space in the general direction of Aldebaran at a distance of about 68 light years from the Earth, and a travel time of two million years. Although science operations were officially terminated on March 31, 1997, Pioneer's onboard RTGs still provided just enough electrical power for operations of the radio transmitter and receiver systems and for the Geiger Tube Telescope experiment of the University of Iowa. After 1997, NASA ARC continued to support occasional tracking of Pioneer 10 for development and testing of communications technology that might be used for a future interstellar probe mission.\footnote{\tt http://spaceprojects.arc.nasa.gov/Space\_Projects/pioneer/PNStat.html} Pioneer 10's signal was also used as a standard test source for SETI radio telescope studies (i.e. Project Phoenix, see details at {\tt http://www.seti.org/}).

By 2005, the existence of the Pioneer anomaly is no longer in doubt.  Further, after much understandable hesitancy, a steadily growing part of the community has concluded that the anomaly should be subject to further investigation and interpretation.  The results of the investigation of the Pioneer anomaly would be win-win; improved navigational protocols for deep space at the least, exciting new physics at the best.  A strong international collaboration (e.g., {\tt http://www.issi.unibe.ch/teams/Pioneer/}) that extends in broader areas of fundamental physics, technology and mission design is an additional outcome of the discussed program of the study of the Pioneer anomaly \citep{ESLAB2005_Pioneer}.

Concluding, we would like to emphasize one immediate impact of the study of the Pioneer anomaly, this time on the Pioneer 10 mission itself.  Recently it became clear that there exists one last opportunity to contact Pioneer 10 in the deep space.  Specifically, in February-March 2006, the mutual motions of the Earth and the spacecraft will again put the Earth in the field of view of Pioneer 10's antenna, thus making it possible to establish a radio contact with the craft. At this time, Pioneer 10 will be at a heliocentric distance of $\sim$90.1~AU, moving at a speed of nearly 12.1 km/s with round trip light time from the Earth of almost 25 hours (thus, the same DSN antenna, DSS 14 at Goldstone, is planned for the operations.) This would be the last time when DSN will initiate contact with Pioneer 10, as the on-board power system on the craft is at its limits. This intriguing possibility to re-acquire a coherent Doppler signal from Pioneer 10 is currently being investigated at JPL and results will be reported.

\section*{\large Acknowledgments}

We would like to express our gratitude to our many colleagues who have either collaborated with us on this manuscript or given us their wisdom.
Firstly we  must acknowledge the many people who have helped us with
suggestions,  comments, and constructive criticisms.  We specifically thank John~D. Anderson,
Sami Asmar, and Timothy P. McElrath,
who provided us with very valuable comments while this manuscript was in preparation.  Invaluable information on the history, spacecraft design and mission operations of the Pioneers 10 and 11, as well as the structure of their telemetry data, came from Lawrence Lasher and David Lozier of the NASA Ames Research Center. We also thank Belinda Arroyo, Peter J. Breckheimer, John E. Ekelund, Jordan Ellis, Scott E. Fullner, Gene L. Goltz, Olga King, George D. Lewis, Robert A. Jacobson, Margaret Medina,  Neil Mottinger, Teresa Thomas of JPL for their help in obtaining, understanding and conditiononing of the Pioneer Doppler data.  We thank Moustafa T. Chahine, William Folkner, Ulf E. Israelsson, Tomas J. Martin-Mur, Thomas A. Prince, and Michael M. Watkins of JPL for encouragement and stimulating discussions regarding the Pioneer Doppler data retrieval effort. John F. Cooper and Sharlene Rhodes of NSSDC provided pivotal help in the retrieval and understanding of the historical Pioneer Doppler data archived with NSSDC. Louis K. Scheffer of Cadence Design Systems contributed with useful observations on the thermal modeling and analysis. 

We especially thank The Planetary Society for support and, in particular, Louis D. Freidman, Charlene M. Anderson, and Bruce Betts for their interest, stimulating conversations and encouragement. 
This work was partially performed at the International Space Science Institute (ISSI), Bern, Switzerland, when two of us (SGT and VTT) visited ISSI as part of an International Team program. 
In this respect we would like to thank Roger M. Bonnet, Vittorio Manno, Brigitte Fasler and Saliba F. Saliba of ISSI for their hospitality and support. The work of SGT was supported in part by the Office of the JPL Chief Scientist under the Research and Technology Development program and carried out at the Jet Propulsion Laboratory, California Institute of Technology, under a contract with the National Aeronautics and Space Administration. 


\end{document}